\documentclass[apj]{emulateapj}
\slugcomment{{\sc Accepted to ApJS:} January 6, 2009}

\usepackage{graphicx}
\usepackage{epstopdf}
\usepackage{stmaryrd}
\usepackage{amsmath}
\usepackage{ccaption}
\usepackage{multirow}
\usepackage{bm}
\usepackage{nicefrac}
\usepackage{comment}
\usepackage{lscape}

\newcommand{\solm}{M_{\odot}}

\newcommand{\bdm}{\begin{displaymath}}
\newcommand{\edm}{\end{displaymath}}
\newcommand{\beq}{\begin{equation}}
\newcommand{\eeq}{\end{equation}}
\newcommand{\bit}{\begin{itemize}}
\newcommand{\eit}{\end{itemize}}
\newcommand{\ben}{\begin{enumerate}}
\newcommand{\een}{\end{enumerate}}
\newcommand{\bfi}{\begin{figure}[htb]}
\newcommand{\bpfi}{\begin{figure}[p]}

\newcommand{\tbd}{{\bf tbd} }

\shorttitle{IR-Radio Relation: Selection Biases \& Redshift Evolution}
\shortauthors{Sargent et al.}

\begin{document}


\title{The VLA-COSMOS Perspective on the IR-Radio Relation. I. New Constraints on Selection Biases and the Non-Evolution of the IR/Radio Properties of Star Forming and AGN Galaxies at Intermediate and High Redshift}

\author{M.~T. Sargent\altaffilmark{1, $\star$},
E. Schinnerer\altaffilmark{1},
E. Murphy\altaffilmark{2},
H. Aussel\altaffilmark{3},
E. Le Floc'h\altaffilmark{4, 5},
D.~T. Frayer\altaffilmark{6},
A. Mart\'inez-Sansigre\altaffilmark{1, 7},
P. Oesch\altaffilmark{8},
M. Salvato\altaffilmark{9, 10},
V. Smol\v{c}i\'{c}\altaffilmark{10},
G. Zamorani\altaffilmark{11},
M. Brusa\altaffilmark{12},
N. Cappelluti\altaffilmark{12},
C.~M. Carollo\altaffilmark{8},
O. Ilbert\altaffilmark{13, 5},
J. Kartaltepe\altaffilmark{14},
A.~M. Koekemoer\altaffilmark{15},
S.~J. Lilly\altaffilmark{8},
D.~B. Sanders\altaffilmark{5},
N.~Z. Scoville\altaffilmark{10}}

\altaffiltext{$\star$}{E-mail: \texttt{markmr@mpia.de}}

\altaffiltext{1}{Max-Planck-Institut f\"ur Astronomie, K\"onigstuhl 17,
    D-69117 Heidelberg, Germany}
\altaffiltext{2}{{\it Spitzer} Science Center, MC 314-6, California Institute of Technology,
Pasadena, CA 91125}
\altaffiltext{3}{AIM Unit\'e Mixte de Recherche CEA CNRS Universit\'e
Paris VII UMR n158, France}
\altaffiltext{4}{CEA-Saclay, Service d'Astrophysique, F-91191 Gif-sur-Yvette,
                 France}
\altaffiltext{5}{Institute for Astronomy, 2680 Woodlawn Dr., University
of Hawaii, Honolulu, Hawaii, 96822, USA}
\altaffiltext{6}{Infrared Processing and Analysis Center, California
Institute of Technology 100-22, Pasadena, CA 91125, USA}
\altaffiltext{7}{Astrophysics, Department of Physics, University of Oxford,
                 Keble Road, Oxford, OX1 3RH, UK}
\altaffiltext{8}{Department of Physics, ETH Zurich,
                 CH-8093 Zurich, Switzerland}
\altaffiltext{9}{Max-Planck-Institut f\"ur Plasmaphysik, Boltzmanstrasse,
                 D-85741 Garching, Germany}
\altaffiltext{10}{California Institute of Technology, MC 105-24, 1200 East
                 California Boulevard, Pasadena, CA 91125, USA}
\altaffiltext{11}{INAF-Osservatorio Astronomico di Bologna, Via Ranzani 1, 
                 I-40127 Bologna, Italy}
\altaffiltext{12}{Max Planck Institut f\"ur Extraterrestrische Physik, Giessenbachstrasse 1,
                 D-85748, Garching bei M\"unchen, Germany}
\altaffiltext{13}{Laboratoire d'Astrophysique de Marseille, BP 8, Traverse du Siphon,
                 F-13376 Marseille Cedex 12, France}
\altaffiltext{14}{National Optical Astronomy Observatory,
                 950 N. Cherry Ave, Tucson, AZ 85726, USA}
\altaffiltext{15}{Space Telescope Science Institute,
                 3700 San Martin Drive, Baltimore MD 21218, USA}

\begin{abstract}
VLA 1.4\,GHz ($\sigma\sim$ 0.012\,mJy) and {\it MIPS} 24 and 70\,$\mu$m ($\sigma\sim$ 0.02 and 1.7\,mJy, respectively) observations covering the 2 deg$^2$ COSMOS field are combined with an extensive multi-wavelength data set to study the evolution of the IR-radio relation at intermediate and high redshift. With $\sim$4500 sources -- of which $\sim$30\% have spectroscopic redshifts -- the current sample is significantly larger than previous ones used for the same purpose. Both monochromatic IR/radio flux ratios ($q_{24}$ \& $q_{70}$), as well as the ratio of the total IR and the 1.4\,GHz luminosity ($q_{\rm TIR}$) are used as indicators for the IR/radio properties of star forming galaxies and AGN.\\
Using a sample jointly selected at IR and radio wavelengths in order to reduce selection biases, we provide firm support for previous findings that the IR-radio relation remains unchanged out to at least $z\sim$ 1.4. Moreover, based on data from $\sim$150 objects we also find that the local relation likely still holds at $z\in$ [2.5, 5]. At redshift $z<$ 1.4 we observe that radio-quiet AGN populate the locus of the IR-radio relation in similar numbers as star forming sources. In our analysis we employ the methods of survival analysis in order to ensure a statistically sound treatment of flux limits arising from non-detections. We determine the observed shift in average IR/radio properties of IR- and radio-selected populations and show that it can reconcile apparently discrepant measurements presented in the literature. Finally, we also investigate variations of the IR/radio ratio with IR and radio luminosity and find that it hardly varies with IR luminosity but is a decreasing function of radio luminosity.
\end{abstract}


\section{Introduction}

The global infrared (IR) and radio emission are tightly and virtually linearly correlated in a broad variety of star forming (SF) systems \citep[see][and references therein]{helou91, condon92, yun01}, thus defining what is known as the `{\it IR-radio relation}'. Studies in the nearby universe have shown that not only late-type galaxies \citep{dickeysalpeter84, helou85, wunderlich87, hummel88} ranging from normal spirals to the most vigorously star forming ultra-luminous infrared galaxies \citep[ULIRGs; e.g.][]{sandersmirabel96, bressan02} follow the relation, but also many early-type galaxies with low-level star formation \citep{wrobel88, bally89}, as well as interacting systems of mixed morphological composition \citep{domingue05}.\\
Following first indications of the correlation in ground-based observations at 10\,$\mu$m and 1.4\,GHz \citep{vanderkruit73, condon82} the ubiquity and tightness of the IR-radio relation became fully appreciated during the analysis of the combination of data from the Very Large Array (VLA) and the {\it Infrared Astronomical Satellite} ({\it IRAS}) which measured the far-IR (FIR) properties of $\sim$20,000 galaxies at $z\lesssim$ 0.15 \citep[e.g.][]{dickeysalpeter84, dejong85, helou85, yun01}. IR observations of sources at higher redshifts became available with the advent of the {\it Infrared Space Observatory} ({\it ISO}) and, in recent years, with the {\it Spitzer Space Telecope}. They have provided increasing evidence that the locally observed correlation likely holds until $z\sim$ 1 \citep{garrett02, appleton04, frayer06} and that the linearity of the correlation is maintained as far back as $z\sim$ 3 although the slope\footnote{By `slope' we mean the slope of the correlation in a plot of flux vs. flux with linearly scaled axes.} may change, especially for sub-mm galaxies (\citealp{kovacs06, vlahakis07, sajina08, murphy09b}, but see also \citealp{beelen06, ibar08}). The statistical significance of these high-$z$ studies, however, is still low because so far the number of sources detected at $z\gtrsim$ 0.5 is limited.

The very tightness of the IR-radio relation (the intrinsic dispersion in the local galaxy population is less than a factor 1.5), combined with the fact that it spans five orders of magnitude in bolometric luminosity has provided a useful tool for numerous astrophysical applications and motivated continued study of the relation. \cite{bressan02}, for example, use the measured IR/radio flux ratio to determine the evolutionary stage of systems undergoing a starburst, depending on whether they show excess IR or radio emission with respect to the average locus of star forming galaxies (SFGs). The consideration of radio-excess outliers to the IR-radio relation also has been used to select galaxies in which an active galactic nucleus (AGN) rather than star formation is the dominant source of IR and radio emission \citep[e.g.][]{donley05, park08}. At high redshift the IR-radio relation has been used to compute distance estimates of sub-mm galaxies lacking optical counterparts \citep{carilliyun99, dunne00}, as well as to estimate the contribution of SFGs to the cosmic radio background \citep{haarsmapartridge98}.\\
In the absence of AGN activity both (F)IR and radio flux measurements are in principle unbiased tracers of star formation since the radiation in these spectral regions is not attenuated by dust as opposed to the often heavily obscured emission at ultraviolet (UV) or optical to near-IR (NIR) wavelengths. The observation that a correlation between the IR and radio exists implies that the thermal IR dust emission and the radio continuum flux -- which is a frequency-dependent mixture of thermal emission and a non-thermal synchrotron component -- have a common origin. Soon after the discovery of the IR-radio relation the birth and demise of massive ($>$5\,$\solm$) stars was identified as its likely cause \citep[see][and references therein]{harwitpacini75, condon92}. Deviations from exact proportionality of the IR and radio emission as well as changes in the IR-radio relation with cosmic time thus potentially imply changes in the mechanisms steering star formation, or that IR and radio emission are not equally good proxies of star formation under all circumstances. However, several decades after the discovery of the relation many details concerning the physical processes which shape it still need to be settled.\\
A number of models to explain the global correlation between the {\it integrated} IR and radio fluxes of SFGs have been advanced: they include the calorimeter model \citep{voelk89, pohl94, lisenfeld96, thompson06, thompson07}, the `optically thin' scenario \citep{helou93} and the `linearity by conspiracy' picture of \cite{bell03}. Models focusing on the magnetohydrodynamic state of the interstellar medium (ISM) are often constructed in order to explain the correlation on both kpc scales {\it within} galaxies as well as globally \citep[e.g.][]{bettens93, niklasbeck97, groves03, murgia05}. Many of these fare well in reproducing multiple aspects of the correlation but either cannot satisfy all observational constraints or make predictions which still await confirmation.\\
On the observational side further insight into the (astro-)physical processes shaping the IR-radio relation has been gained by studying the IR-radio relation on small scales in resolved nearby galaxies \citep[e.g.][Dumas et al., in prep.]{beckgolla88, bicayhelou90, murphy08} which revealed that IR/radio properties vary inside spiral galaxies or by investigating the impact of environmental effects \citep{millerowen01, reddy04, murphy09a}. Another important finding was that the linearity of the relation is maintained with both the thermal and non-thermal radio emission \citep{priceduric92} taken separately, as well as for the warm and cold dust components \citep{pierini03}.

\noindent As a quantitative measure of the correlation \cite{helou85} introduced the logarithmic flux density ratio 
\begin{equation}
q_{\rm FIR} = {\rm log}\left(\frac{F_{\rm FIR}}{3.75\times10^{12}\,{\rm W\,m^{-2}}}\right) - {\rm log}\left(\frac{S_{\nu}(1.4\,{\rm GHz})}{\rm W\,m^{-2}\,Hz^{-1}}\right)~.
\label{eq:FIRq}
\end{equation}
Here $F_{\rm FIR}$ is the rest frame FIR flux -- traditionally computed as a linear combination of measurements in the IRAS 60 and 100\,$\mu$m bands under the assumption of a typical dust temperature of $\sim$30\,K -- and 3.75$\times$10$^{12}$\,Hz is the `central' frequency of the FIR-window (42.5-122.5\,$\mu$m). $S_{\nu}(1.4\,{\rm GHz})$ is the monochromatic rest frame 1.4\,GHz flux density.\\
Strong improvements of the sensitivity of IR observatories over the last decade, in particular in the mid-IR (MIR) with the {\it MIPS} photometer on {\it Spitzer}, have promoted the use of monochromatic flux ratios, e.g. at 24 or 70\,$\mu$m \citep[e.g.][]{appleton04, ibar08, seymour09}:
\begin{equation}
q_{\rm 24\,[70]} = {\rm log}\left(\frac{S_{\nu}(24\,[70]~{\rm {\mu}m})}{S_{\nu}(1.4\,{\rm GHz})}\right)~,
\label{eq:monchromqs}
\end{equation}
where both flux densities are specified in units of ${\rm W\,m^{-2}\,Hz^{-1}}$. This approach is convenient for evolutionary studies of the IR-radio relation at intermediate and high redshift as it avoids the computation of IR luminosities from a single IR flux measurement, usually made at 24\,$\mu$m due to the high sensitivity of the according {\it MIPS} filter. However, it still requires the computation of $K$-corrections, if the flux densities in equation (\ref{eq:monchromqs}) are to be given in the rest frame.\\
Since current radio surveys at centimeter wavelengths are generally significantly shallower than the MIR photometry several authors have carried out radio stacking experiments \citep{boyle07, beswick08, garn09a} with the aim of determining the IR/radio properties of the faint IR population. This is of particular interest in view of the potential for detecting changes in the IR-radio relation brought about by relativistic cooling of cosmic ray electrons by inverse Compton scattering off photons from the cosmic microwave background (CMB). This cooling may overwhelm -- at least at low decimeter frequencies -- the synchrotron losses in the ISM of normal galaxies at $z\gtrsim$ 0.5 \citep{carilli08} if the energy density of their magnetic fields is similar in strength to that typically measured in spiral arms \citep[$U_B \sim10^{12}$\,erg\,cm$^{-3}$;][]{beck05}.
Alas, the stacking results have produced strongly discrepant results, both among different cosmological survey fields and with respect to recent studies of sources directly detected in both the IR and radio. An additional concern of particular relevance in evolutionary studies is the bias that is introduced by constructing samples from different selection bands in flux limited surveys.

\noindent In this paper we focus on the impact of selection biases and apply the statistical technique of survival analysis to our data which permits the inclusion of constraints from flux limits in the study. This approach is a significant improvement over limiting the sample to those sources which are detected at both IR and radio wavelengths. These topics are discussed in \S\,\ref{sect:selsurv}. Our data sets contain an unprecedented amount and quality of information gathered as part of the COSMOS survey \citep{scoville07}. This is reflected by a sample which contains significantly more sources in the redshift range $z\gtrsim$ 0.5, in which the sources in previous studies began to taper out. Moreover, a large fraction (33\%) of spectroscopically determined redshifts, which are supplemented by accurate photometric redshift estimates, in combination with flux constraints at both 24 and 70\,$\mu$m for each source in the sample make for improved estimates of IR luminosities (\citealp[e.g.]{murphy09b}; Kartaltepe et al., subm.). They are an important prerequisite for placing accurate constraints on the evolution of the IR-radio at high redshift. We introduce our data and the subsequently analyzed samples in \S\,\ref{sect:data}. For those interested, a detailed review of the different catalogs and images used in the paper and the band-merging of this information into a rich multi-wavelength data is provided in Appendices \ref{appsect:data} and \ref{appsect:bandmerge}. \S\S\,\ref{sect:SFvsAGN} and \ref{sect:SEDfitting} deal with methodological considerations applying to the identification of star forming sources and AGN galaxies as well as to the derivation of IR luminosities. Following the section on biasing and our treatment of flux limits in \S\,\ref{sect:selsurv}, the bulk of our analysis is then presented in \S\,\ref{sect:results}. Our most important results are:\\
(a) the constancy of the IR-radio relation as parametrized by the flux ratios $q_{24}$, $q_{70}$ and $q_{\rm TIR}$ out to $z\sim$ 1.4, as well as for a sub-sample of high-$z$ sources at 2.5 $<z<$ 5,\\
(b) the identification of selection biases as a potential explanation for discrepant average IR/radio flux ratios measured in previous studies, and\\
(c) the observation that over the last 10 billion years the distribution of IR/radio ratios of optically selected, radio-quiet\footnote{Objects referred to as `radio-quiet AGN' in the rest of the paper are understood to be sources in which the radio emission from the AGN does not contribute significantly to the total energy emitted at 1.4\,GHz.} AGN has been very similar to that of star forming galaxies \\
We discuss and summarize these findings in \S\S\,\ref{sect:discussion} and \ref{sect:conclusions}, respectively. 

\noindent Throughout this article we adopt the WMAP-5 cosmology defined by $\Omega_m=$ 0.258, $\Omega_{\Lambda}+\Omega_m=$ 1 and a present-day Hubble parameter of 71.9 km\,s$^{-1}$\,Mpc$^{-1}$ \citep{dunkley09}. Magnitudes are given in the AB-system of \cite{oke74} unless  the opposite is explicitly stated and we henceforth drop the subscript `AB' in the text.

\section{Data Sets and Samples}
\label{sect:data}

In order to study biases arising from the selection of sources at either IR or radio wavelengths we have constructed both a radio- and an IR-selected sample of COSMOS galaxies. All observations and associated data sets, as well as the band-merging procedures used to identify counterparts from radio to X-ray wavelengths are described in detail in Appendices \ref{appsect:radiodata}-\ref{appsect:redshifts} and Appendix \ref{appsect:bandmerge}, respectively. Here we only briefly present the primary data sets (\S\,\ref{sect:IRradiodata}) and review the most important properties of our {\it radio-selected} and {\it IR-selected} samples (\S\,\ref{sect:sampreview}). This section also contains a summary of the ancillary multi-wavelength photometry and the redshift information which is available for our sources (\S\,\ref{sect:ancilldata}).

\subsection{Radio and IR Data}
\label{sect:IRradiodata}

The VLA-COSMOS Project \citep{vlacos1, vlacos2} has imaged the COSMOS field at 1.4\,GHz with the VLA to a mean sensitivity of $\lesssim$0.01 ($\sim$0.04)\,mJy/beam at the centre (edge) of the field. As the basis of our subsequent analysis we use the VLA-COSMOS `Joint' Catalog (Schinnerer et al. 2009, subm.) which contains $\sim$2900 sources detected with $S/N\geq$ 5. Nearly 50\% of these are resolved at a resolution (FWHM of synthesized beam) of 2.5$''\times$2.5$''$. Flux measurements were carried out with AIPS \citep[Astronomical Image Processing System;][]{greisen03} and have been corrected for bandwidth smearing in the case of the unresolved radio sources. Their errors are generally $\sim$17\% of the flux value.

\noindent IR data at 24 and 70\,$\mu$m were taken by the S-COSMOS Survey \citep{sanders07} using {\it MIPS} on {\it Spitzer}.\\
At 24\,$\mu$m the FWHM of the PSF is 5.8$''$ and the average $1\,\sigma$ sensitivity $\sim$0.018\,mJy over a large fraction of the imaged area. Flux measurements \citep{lefloch09} in the deep and crowded {\it MIPS} 24\,$\mu$m image were performed with the PSF-fitting algorithm DAOPHOT \citep{stetson87} which can simultaneously fit and hence de-blend multiple sources. In the following we use a flux limited catalog of COSMOS 24\,$\mu$m sources which is restricted to the range $S_{\nu}(24\,\mu{\rm m})\geq$ 0.06\,mJy. Typical flux uncertainties at 24\,$\mu$m are $\sim$8\%.\\
MIPS 70\,$\mu$m observations of the COSMOS field were carried out in parallel with the 24\,$\mu$m imaging. They have an average 1\,$\sigma$ point source noise of 1.7 mJy and a resolution of 18.6$''$. Our 70\,$\mu$m source list includes detections down to $S/N=$ 3 and was compiled as described in \cite{frayer09}. Fluxes were derived using the APEX \citep[Astronomical Point source EXtraction][]{makovoz05b} peak fitting algorithm; their average uncertainty is $\sim$16\%.

\begin{figure}
\centering
\includegraphics[scale=0.65, angle=-90]{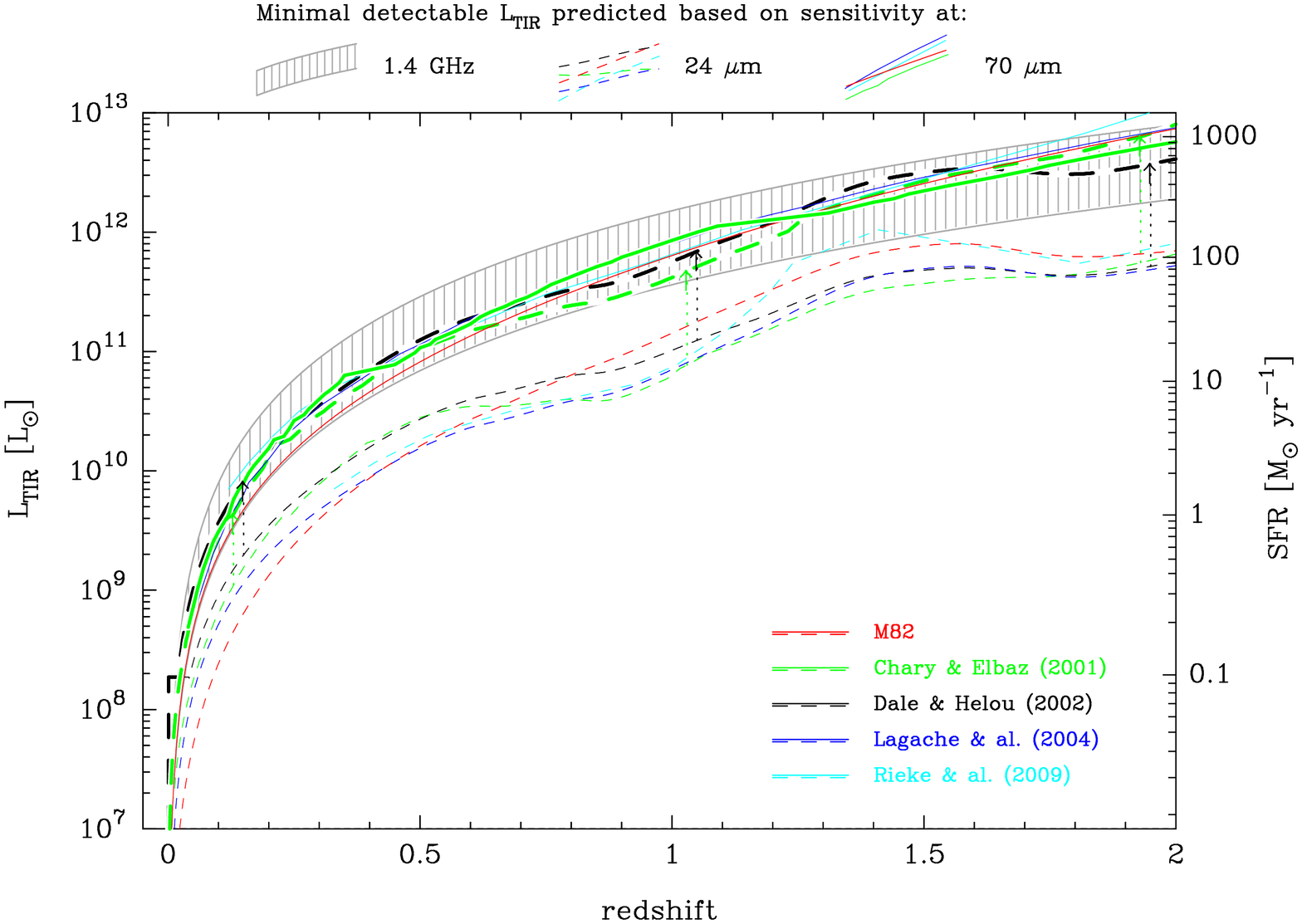}
\caption{Range of total infrared luminosities $L_{\rm TIR}$ (or star formation rates following \cite{bell03}, see right-hand vertical scale) sampled between redshift 0 and 2, given the 3$\sigma$ detection limits at 24\,$\mu$m, 70\,$\mu$m and 1.4\,GHz. The 1.4\,GHz detection limit has been converted to $L_{\rm TIR}$ assuming the average local TIR/radio ratio $\langle q_{\rm TIR}\rangle=$ 2.64 (cf. \cite{bell03, yun01} -- the upper and lower envelope of the arc filled with vertical light grey lines denote $\pm1\,\sigma$ outliers to the mean of the relation). The solid tracks following a similar locus as the 1.4\,GHz detection limits represent the smallest value of $L_{\rm TIR}$ expected to be detectable based on the sensitivity at 70\,$\mu$m. Different tracks correspond to different IR SEDs and/or libraries (see legend in lower right corner). The thicker (solid) lines emphasize the predictions from the libraries of \cite{dalehelou02} and \cite{charyelbaz01} which are used for the IR SED fitting (see \S\,\ref{sect:SEDfitting}). Predictions based on 3\,$\sigma$ flux limits at 24\,$\mu$m are plotted as thin dashed tracks. An observed flux limit of 0.3\,mJy at 24\,$\mu$m leads to a similar sampling of the IR luminosity function as is possible at 70\,$\mu$m and 1.4\,GHz. This is illustrated by the coarser dashed tracks which are joined to the fine ones with the dotted lines. Note that tracks do not always run across the entire redshift range due to the finite choice of IR luminosities represented in the various SED libaries.\label{fig:fluxlims}}
\end{figure}

\noindent In Fig. \ref{fig:fluxlims} we show the minimum total IR (TIR; 8-1000\,$\mu$m) luminosity that is detectable as a function of redshift given the sensitivity of the 1.4\,GHz (converted to an IR measurement assuming the local IR-radio relation) and 24/70\,$\mu$m data. Different IR SED template libraries (see colour scheme in the lower right corner) lead to somewhat different predictions but it is clear that the 1.4\,GHz and 70\,$\mu$m surveys have matching depths while the 24\,$\mu$m observations are about seven times deeper. A similar sampling of the IR luminosity function is achieved in all three bands if a 24\,$\mu$m flux limit of approx. 0.3\,mJy is assumed. We therefore limit our 24\,$\mu$m catalog to the range $S_{\nu}(24\,\mu{\rm m})\geq$ 0.3\,mJy when we construct our IR-selected sample but allow fainter counterparts of 1.4\,GHz sources to be included in the radio-selected sample. The use of these flux limited samples has the immediate consequence that we only detect the brightest ULIRGs at $z\gtrsim$ 1.5 while the average luminosity of our sources is much lower at, e.g. $z\sim$ 0.5, where most sources belong to the LIRG class.

\subsection{Ancillary COSMOS Data}
\label{sect:ancilldata}

Optical data and photometric redshifts are taken from the COSMOS photometry catalog of \cite{ilbert09a} which lists more than 600,000 COSMOS galaxies with $i^+\lesssim$ 26 detected in a region roughly contiguous with the area covered by the VLA-COSMOS survey. The wavelength range covered by these observations (30 broad, medium and narrow band filters) extends all the way from the UV at 1550 \AA\ to the MIR at 8 $\mu$m. \cite{capak07, capak08} provide a complete description of these observations.\\
Spectroscopic data has been gathered for more than 20,000 sources in the COSMOS field, e.g. by the zCOSMOS survey \citep{lilly07} and SDSS \citep{york00}, or in Magellan/IMACS and Keck/Deimos follow-up observations dedicated to specific (classes of) sources (\citealp[e.g.][]{trump07}; \citealp{trump09}; Kartaltepe et al., in prep.; Salvato et al., in prep). If a reliable spectroscopic redshift is available it is favoured over the photometric redshift estimate. The choice of the best possible distance measurement for our radio and IR sources is described in detail in \S\,\ref{appsect:redshifts}.

\noindent The {\it XMM-Newton} COSMOS Survey \citep{hasinger07, cappelluti07, cappelluti09} has detected a total of 1887 bright ($\gtrsim$2$\times10^{-15}$ erg\,cm$^{-2}$\,s$^{-1}$ in the 0.5-10\,keV band) X-ray sources over 90\% (1.92 deg$^2$) of the COSMOS field. A large fraction of these are associated with AGN and hence provide a means of identifying AGN-powered radio and IR sources in our sample which is complementary to our primary classification scheme introduced in \S\,\ref{sect:SFvsAGN}. For our subsequent analysis we rely on a list of XMM sources \citep{cappelluti09} with unique and secure optical counterparts (see Brusa et al., in prep.) and SED fits to the UV to MIR photometry performed by \cite{salvato09}. 

\subsection{Description of the Samples}
\label{sect:sampreview}

Due to the differing characteristics (resolution, astrometric accuracy) of the radio and IR data the determination of counterparts at other wavelengths differed somewhat for the radio- and the IR-selected sample. Which candidate counterparts are incorporated in the final sample and which are rejected is determined by the goals of this study: in our case it is more important to select objects with the cleanest possible radio and IR flux measurements rather than having a statistically complete sample. The details of the band-merging between the IR and the radio catalogs and the subsequent exclusion of ambiguous counterparts are presented in detail in Appendix \ref{appsect:bandmerge}. Here we summarize the most important properties of the radio- and IR-selected samples.

\subsubsection{The Radio-Selected Sample}
\label{sect:radiosamp}
\begin{figure}
\epsscale{.9}
\plotone{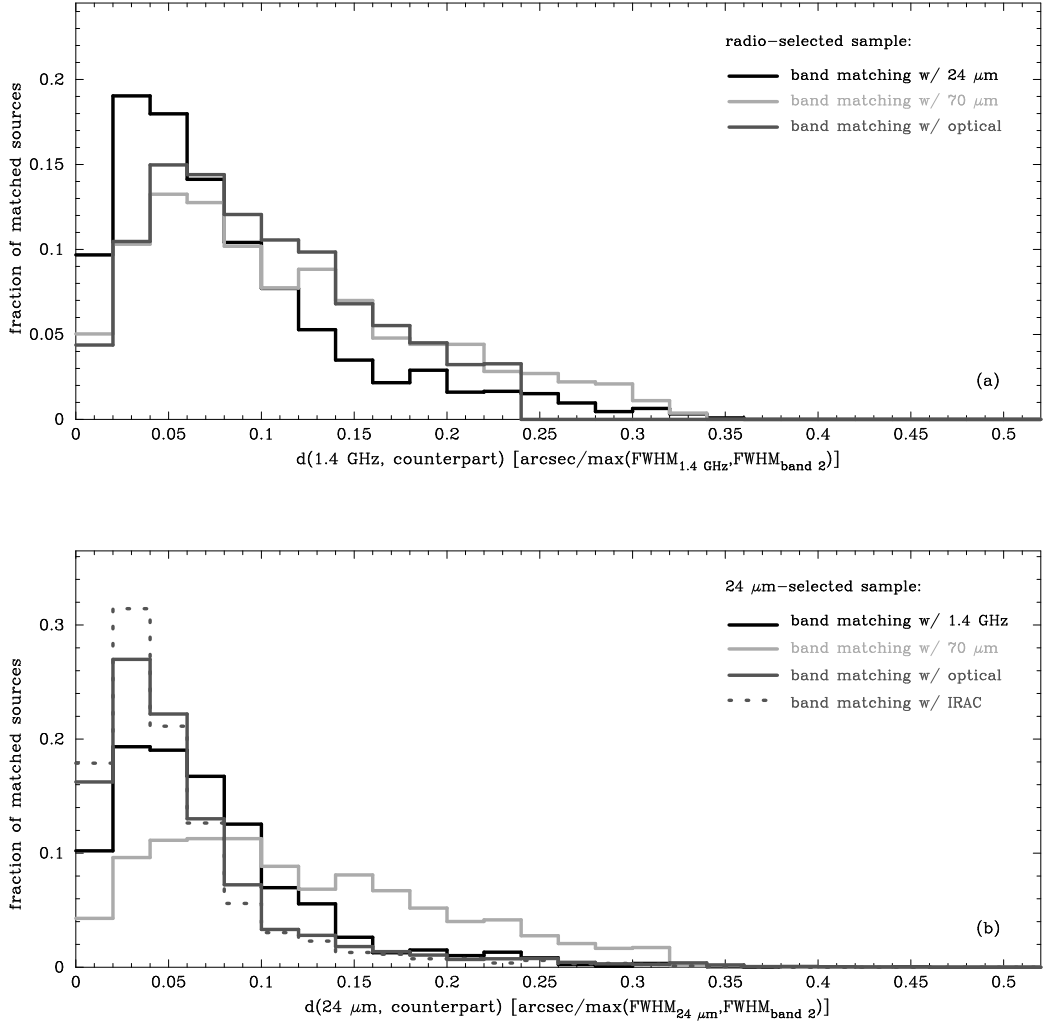} 
\caption{{\it (a)} - Radio-selected sample; separation between radio source position and nearest catalog counterpart at 24 (black line), 70\,$\mu$m (light grey line) and at optical wavelengths (dark grey). The distance is expressed in fractions of the FWHM of the larger PSF of the two bands involved in the match. (The resolution is 0.6$''$, 2.5$''$, 5.8$''$ and 18.6$''$ in the optical and at 1.4\,GHz, 24\,$\mu$m and 70\,$\mu$m, respectively. The IRAC PSF has a FWHM which ranges from 1.6$''$ to 2$''$ between the 3.6\,$\mu$m and the 8\,$\mu$m channel. For the histogram in the lower panel of the figure we assume an average FWHM of 1.8$''$.)\newline
{\it (b)} - IR-selected sample; separation (defined as in (a)) between 24\,$\mu$m source position and the corresponding counterpart at 1.4\,GHz (black line), 70\,$\mu$m (light grey line) and at optical (solid dark grey line) and NIR (IRAC) wavelengths (dotted dark grey histogram).\label{fig:ctp_dists}}
\end{figure}

\begin{figure}
\epsscale{.9}
\plotone{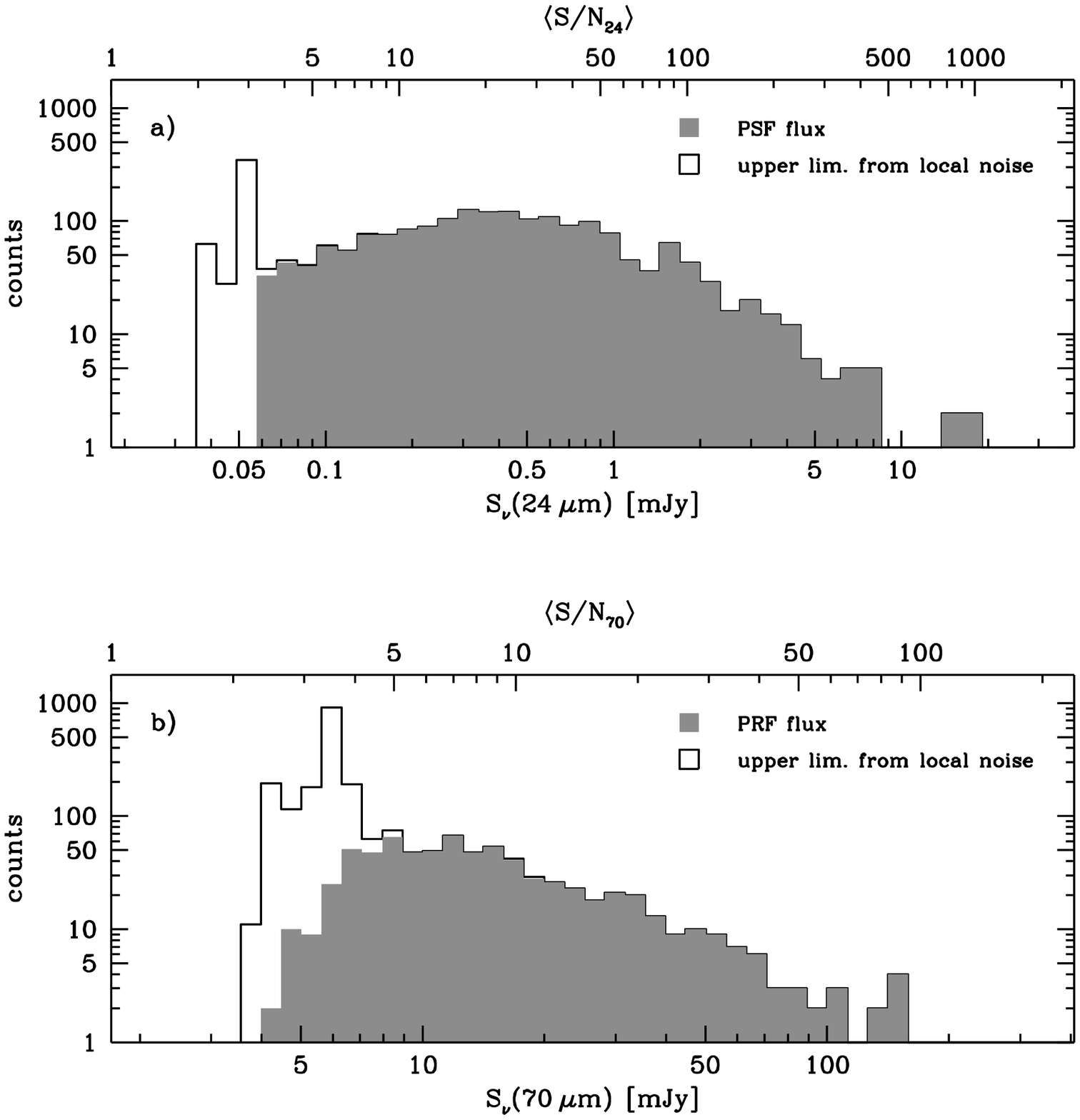} 
\caption{{\it (a)} - Histogram showing the available flux information for the 24\,$\mu$m counterparts to radio-selected COSMOS sources in a given bin of 24\,$\mu$m flux density (grey area - PSF-flux measurement with $S/N>$ 3; white area - 3\,$\sigma$ upper limit from local noise).\newline
{\it (b)} - Same information as displayed in upper panel but for flux constraints on 70\,$\mu$m counterparts of radio-selected sources (grey area - PRF-flux measurement with $S/N>$ 3; white area - 3\,$\sigma$ upper limit from local noise).\newline
The 24\,$\mu$m flux distribution shows a very sharp transition from PSF-fitted detections to upper limits. This is due to the fact that the 24\,$\mu$m catalog is a flux-limited catalog while the 70\,$\mu$m catalog is selected according to $S/N$.\label{fig:radioctp_flxchar}}
\end{figure}

\begin{figure}
\centering
\includegraphics[scale=0.7]{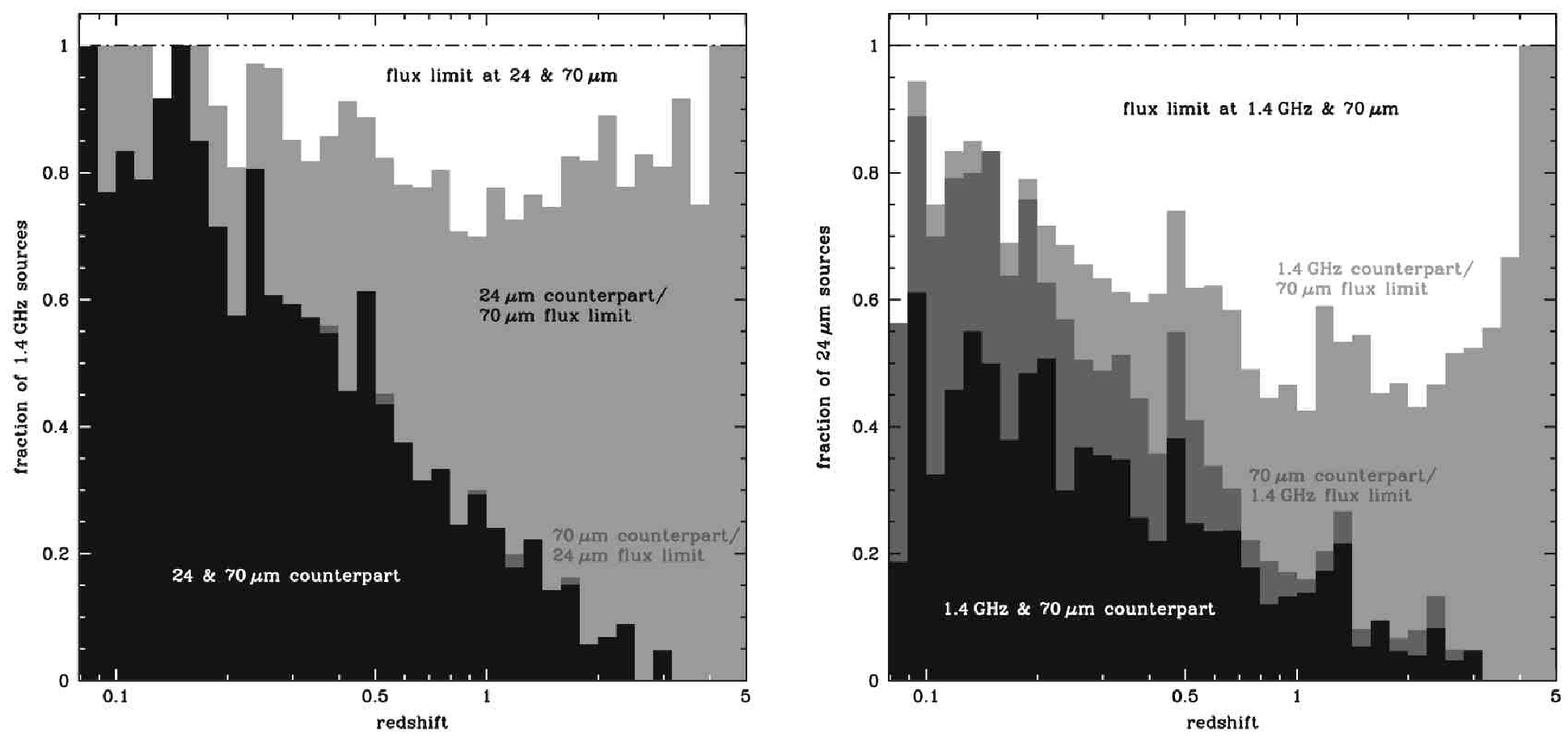}
\caption{Bar diagrams showing the quality of the available flux information at different redshifts for sources in the radio-selected ({\it left}) and IR-selected sample ({\it right}).\newline {\it Left} -- Dependence on redshift of the fraction of radio-selected sources (referred to the total number of objects in a given redshift slice) with a positive identification in both the 24\,$\mu$m and 70\,$\mu$m catalog (black histogram), as well as the fractional contribution of sources with flux limits in either one of the two bands (light grey -- detected 24\,$\mu$m counterpart, upper flux limit at 70\,$\mu$m; dark grey -- upper flux limit at 24 $\mu$m, detection at 70\,$\mu$m) or in both (white area).\newline
{\it Right} -- As for the first panel but for the IR-selected sample (restricted to sources with $S_{\nu}{\rm (24\,\mu m)\geq 0.3\,mJy}$). Black -- counterpart detected at both 70\,$\mu$m and 1.4\,GHz; dark grey -- detected counterpart at 70\,$\mu$m, upper flux limit at 1.4\,GHz; light grey -- detected counterpart at 1.4\,GHz, upper flux limit at 70\,$\mu$m; white -- upper flux limits at both 70\,$\mu$m and 1.4\,GHz.\label{fig:flxchar_vs_z}}
\end{figure}

Based on the positions of $\geq$5\,$\sigma$ 1.4\,GHz detections in the VLA-COSMOS Deep Project image we searched for IR counterparts in the S-COSMOS 24 and 70\,$\mu$m catalogs which have $S/N\gtrsim$ 3. Counterparts were determined by direct positional matching of radio and IR coordinates with search radii corresponding to approx. FWHM/3 of the IR PSFs of the respective {\it MIPS} bands. If no counterpart was found, a 3\,$\sigma$ point-source detection limit was determined based on the corresponding uncertainty images. Radio sources with ambiguous IR counterparts -- i.e. in the presence of more than one potential counterpart or if the counterpart had not been uniquely assigned to a single radio source -- have been excluded from the analysis of the paper. The match with the COSMOS multi-wavelength and spectroscopy catalogs provides distance estimates for 73\% of the radio-selected sample as well as photometry from the UV to the MIR which is used to separate galaxies dominated by star formation or AGN emission (see \S\,\ref{sect:SFvsAGN}). In the upper panel of Fig. \ref{fig:ctp_dists} we show histograms of the separation between radio source positions and the location of the optical and IR counterparts. Note that the distance is normalized by the width of the broader PSF of the two involved bands. Fig. \ref{fig:radioctp_flxchar} shows the 24 and 70\,$\mu$m flux distribution of the radio sources, including information on whether the flux constraint is a well-defined measurement or an upper flux limit.\\
Fig. \ref{fig:flxchar_vs_z} (left-hand panel) explicitly shows how the fraction of sources that have a directly detected counterpart in either or both of the {\it MIPS} bands or only upper flux limits changes as a function of redshift (38\% of the redshifts are spectroscopically,  64\% photometrically determined).

\subsubsection{The IR-Selected Sample}
\label{sect:IRsamp}

\begin{figure}
\epsscale{.9}
\plotone{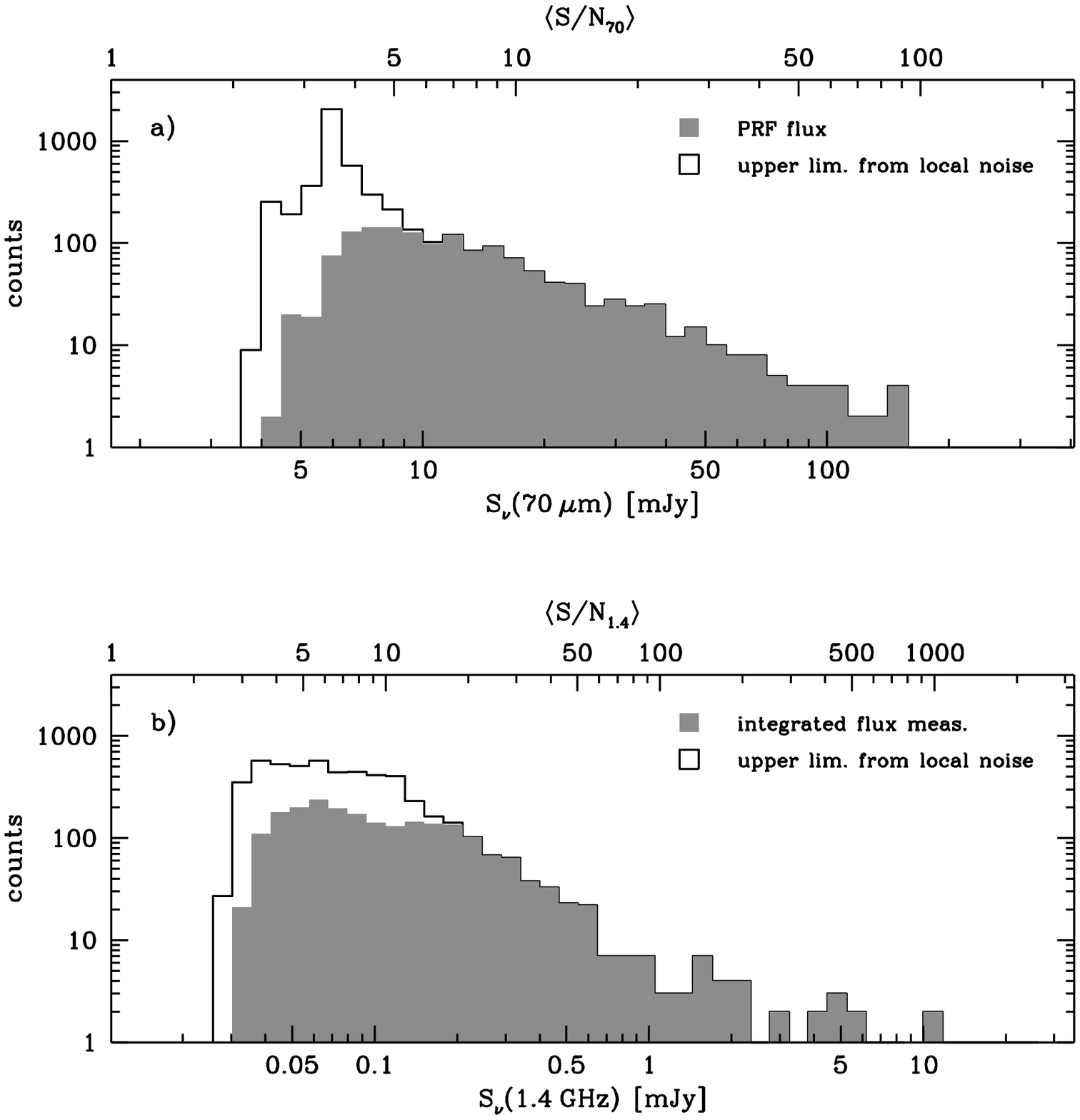} 
\caption{{\it (a)} - Histogram displaying the contribution of well-defined flux measurements and upper flux limits for the 70\,$\mu$m counterparts of S-COSMOS sources (with $S_{\nu}{\rm (24\,\mu m)\geq 0.3\,mJy}$) selected at 24\,$\mu$m as a function of their 70\,$\mu$m flux density (grey area - PRF-flux measurement with $S/N>$ 3; white area - 3\,$\sigma$ upper limit local noise).\newline
{\it (b)} - Same information as in the upper panel of the figure but for flux constraints at 1.4\,GHz (grey area - flux measurement with $S/N>$ 3; white area - 3\,$\sigma$ upper limit from local noise). \label{fig:IRctp_flxchar}}
\end{figure}

The IR-selected sample is based on sources listed in the S-COSMOS 24\,$\mu$m catalog that satisfy the criterion $S_{\nu}(24\,\mu{\rm m})\geq$ 0.3\,mJy. This criterion ensures that the IR-selected sample is well matched to the 70\,$\mu$m and 1.4\,GHz data as far as the sampling of the IR luminosity function is concerned. To reduce the likelihood of false identifications due to the positional uncertainty of the 24\,$\mu$m sources we searched for IRAC counterparts, the positions of which were used as a prior in the subsequent band-merging with the other wavelengths. If no IRAC counterpart was available we also admitted unambiguous matches with optical sources. 70\,$\mu$m counterparts to the 24\,$\mu$m sources with $S/N\geq$ 3 were determined and validated following exactly the same approach as in the radio-selected sample. For those 24\,$\mu$m sources which did not already have a known radio counterpart (determined in the construction of the radio-selected sample) we checked whether they are associated with a counterpart having $S/N>$ 3. All new detections satisfying this criterion were then added to the list of radio counterparts with $S/N\geq$ 5 that were already known from the construction of the radio-selected sample. 24\,$\mu$m sources that are undetected at 70\,$\mu$m and/or 1.4\,GHz are assigned 3\,$\sigma$ upper flux bounds.\\
The distributions of separations between all 24\,$\mu$m sources and their counterparts in the optical, far-IR and 1.4\,GHz maps are given in Fig. \ref{fig:ctp_dists}.b. The contribution of flux limits and well-defined flux measurements as a function of flux and $S/N$ at 70\,$\mu$m and 1.4\,GHz is illustrated in Fig. \ref{fig:IRctp_flxchar}. Finally, in the right-hand panel of Fig. \ref{fig:flxchar_vs_z} we show at different redshifts which fraction of the IR-selected sample has direct detections or upper flux density limits at 70$\mu$m and/or 1.4\,GHz. Spectroscopic or photometric redshift measurements are available for 80\% of the objects in the IR-selected sample. The remaining sources are either not bright enough for spectroscopy or have flux information in too few bands to derive a photometric redshift based on SED fitting.

\subsubsection{The Jointly Selected Sample}
\label{sect:jointsamp}

The jointly selected sample is the union of the radio- and IR-selected samples presented in \S\S\,\ref{sect:radiosamp} and \ref{sect:IRsamp}. As such it contains 6863 sources: 1560 sources that are only detected at 1.4\,GHz, 3960 sources that are only detected at 24\,$\mu$m and, finally, 1341 sources which are selected at both wavelengths. In Table \ref{tab:zstats} we summarize the available redshift information for the jointly radio- and IR-selected sample, as well as separately for the radio- and IR-selected samples.

\begin{figure}
\centering
\includegraphics[scale=0.65]{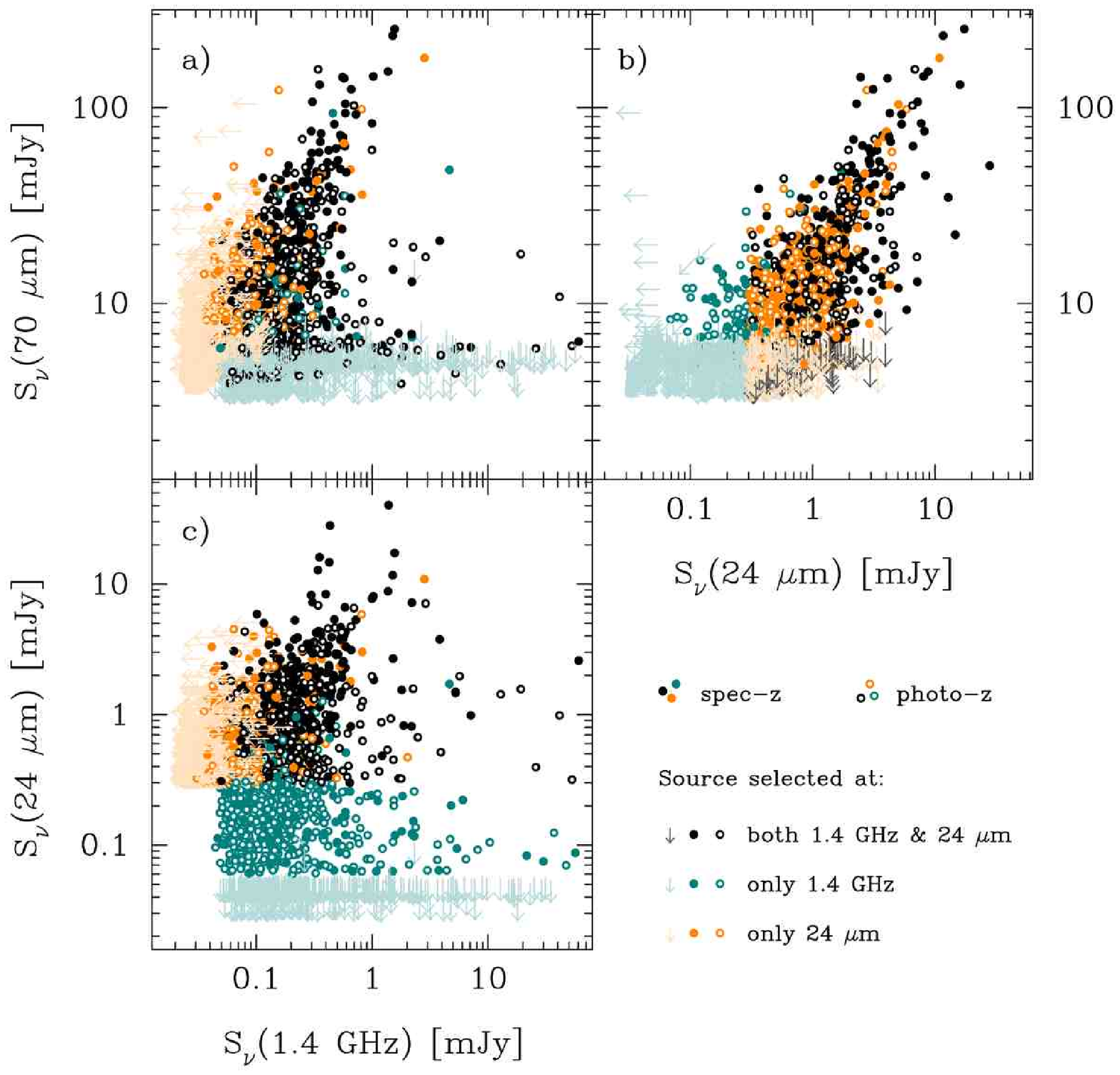}
\caption{Location of sources selected either only at 24\,$\mu$m (orange symbols) or 1.4\,GHz (green) or present in both catalogs (black) in plots comparing fluxes that will be used for the computation of their IR/radio properties: {\it (a)} -- 70\,$\mu$m flux constraints $S_{\nu}(70\,\mu{\rm m})$ as a function of 1.4\,GHz flux density $S_{\nu}(1.4\,{\rm GHz})$; {\it (b)} -- $S_{\nu}(70\,\mu{\rm m})$ vs. $S_{\nu}(24\,\mu{\rm m})$; {\it (c)} -- $S_{\nu}(24\,\mu{\rm m})$ vs. $S_{\nu}(1.4\,{\rm GHz})$.\newline
The IR-selected sample comprises sources plotted in orange and black. The union of green and black symbols defines the radio-selected sample (see \S\,\ref{sect:sampreview}). In panel {\it c)} the straight cut-offs confining the distribution of the green points on the left and that of the orange points on the bottom reflect the lower flux limits of the radio- and IR-selected sample, respectively. \label{fig:flxvsflx}}
\end{figure}

\noindent In Fig. \ref{fig:flxvsflx} the IR and radio fluxes of our sources are compared. The colour coding of the data points distinguishes three kinds of sources: in black those which have entered both the 1.4\,GHz catalog as well as the 24\,$\mu$m catalog (restricted to sources with flux density larger than 0.3\,mJy), in green 1.4\,GHz sources without counterpart in the 24\,$\mu$m catalog and in orange those 24\,$\mu$m-detected sources which do not have a counterpart in the VLA-COSMOS Joint catalog. The sources from these three different categories have been added to the plot in random order to 
prevent that the symbols of the initially plotted category are systematically hidden by the successively overplotted data in common regions of flux space. Fig. \ref{fig:flxvsflx}.c confronts the fluxes in the two selection bands; the empty rectangle in the lower left corner of this panel reflects the selection criteria at 1.4\,GHz and 24\,$\mu$m. Since the 24\,$\mu$m catalog is flux limited, essentially all upper 24\,$\mu$m flux limits lie at or below the critical flux threshold; upper 1.4\,GHz flux limits for undetected 24\,$\mu$m sources on the other hand are also encountered at higher 1.4\,GHz flux values than the sharp cut-off at $S_{\nu}(1.4\,{\rm GHz})\approx$ 0.05\,mJy because the radio-catalog was constructed using a $S/N$ criterion. Note that the region where both the 1.4\,GHz and 24\,$\mu$m flux density clearly exceed the respective selection thresholds contains some sources which are not included in both the catalog of 24\,$\mu$m and that of 1.4\,GHz detections (cf. orange and green symbols in the area where $S_{\nu}(1.4\,{\rm GHz})>$ 0.1\,mJy and $S_{\nu}(24\,\mu{\rm m})>$ 0.3\,mJy). Two reasons can be responsible for this: {\it (a)} minor incompleteness of the catalogs, or {\it (b)} spatial variations in the background noise which, at a given flux, lead to certain sources not being detected at the significance level required for inclusion in the original source list.

\section{Identification of Star Forming Galaxies}
\label{sect:SFvsAGN}
\begin{figure}
\epsscale{.9}
\plotone{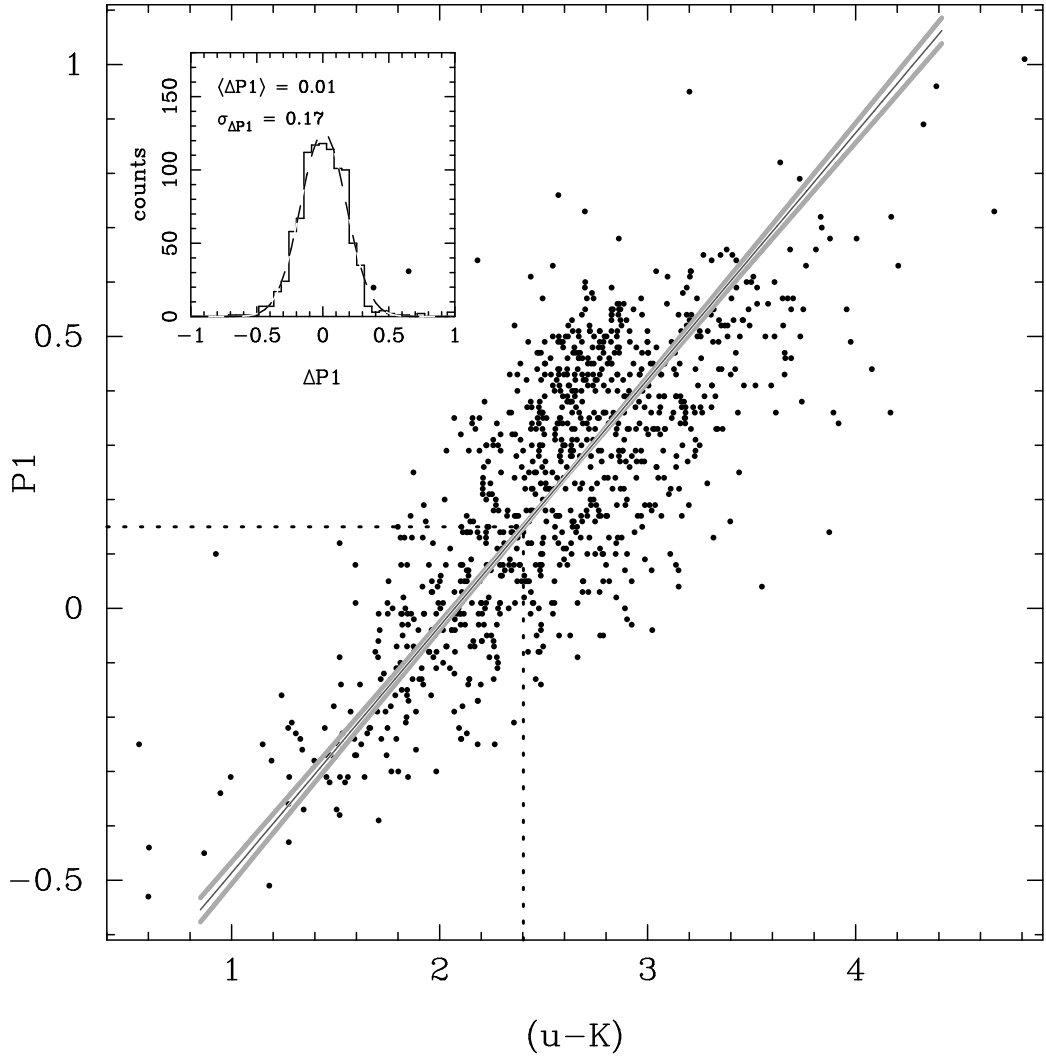} 
\caption{Correlation between the rest frame colours P1 and ($u-K$) for radio-selected sources from the VLA-COSMOS survey. The bisector fit -- see equation (\ref{eq:P1umK_corr}) -- to the relation is shown with a black line; grey lines illustrate the uncertainty in the fit. The separation between SF and AGN galaxies of \cite{smolcic08} at P1= 0.15 corresponds to a cut at ($u-K$) = 2.42 (cf. dotted line). Note that due to our treatment of composite SF/AGN sources we adopt a slightly different colour threshold for the selection of SF systems (see following figure). Upper left corner: histogram of offsets $\Delta P1$ (measured parallel to the vertical axis for each data point) from the black trend line. The dispersion of the relation is $\sigma_{\Delta P1} =$ 0.17.\label{fig:P1equivcol}}
\end{figure}

Both star formation and AGN activity cause the host galaxy to (re-)emit at (mid-)IR and radio wavelengths. To study the IR/radio properties of these two distinct populations separately, information from different regions of the electromagnetic spectrum is thus required. \cite{smolcic08} devised a method which, in a statistical sense, is capable of selecting star forming and AGN galaxies with a simple cut in rest frame optical colour. It relies on the tight correlation \citep{smolcic06} between the rest frame colours of emission line galaxies and their position in the BPT diagram \citep{baldwin81} and was developed and calibrated with radio sources at $z\lesssim$ 1.3 using the principal component colour\footnote{P1 and its homologue P2 are linear combinations of the narrow band (modified) Str\"omgren filter magnitudes ($uz$, $vz$, $bz$, $yz$; \cite{odell02}) in the wavelength range 3500-5800\,\AA; see \cite{smolcic08} for the definitions and additional details.} henceforth referred to as `P1'. It can, however, be easily adapted to other rest frame colours because galaxy SEDs from the near-UV to the NIR represent a one-parameter family \citep{obric06, smolcic06}. Here we use the combination of the filters $u$ and $K$ to select AGN and SFGs. This choice is motivated by the desire to apply the classification to both the radio- and IR-selected sample; the likely presence of dust-obscured star forming systems in the IR-selected sample requires the inclusion of a red band, to prevent, as best possible, dust-reddened star forming sources from being mistaken for red, early-type AGN host galaxies.\\
In Fig. \ref{fig:P1equivcol} we show the correlation of P1 \citep[computed according to][]{smolcic08} and ($u-K$) for $\sim$950 VLA-COSMOS sources, for which both P1 and ($u-K$) were available. rest frame ($u-K$) colours were computed with ZEBRA \citep[{\it Zurich Extragalactic Bayesian Redshift Analyzer};][]{feldmann06} which was used to find the best-fitting SED template to the COSMOS photometry in the medium and broad band filters $u^*$, $B$, $V$, $g^+$, $r^+$, $i^+$, $z^+$, $J$ and $K_s$, as well as in the four IRAC channels given the known redshift (cf. \S\,\ref{appsect:redshifts}). Note that the magnitudes $u$ and $K$ used here are computed in Johnson-Kron-Cousins filters rather than the COSMOS filters. An ordinary least squares (OLS) bisector fit \citep{isobe90} accounts for the fact that both colours are subject to uncertainty and returned a best-fit correlation given by
\begin{equation}
P1 = -0.94(\pm0.0006) + 0.45(\pm0.01)\times(u-K)~,
\label{eq:P1umK_corr}
\end{equation}
which is indicated in grey in Fig. \ref{fig:P1equivcol}. The criterion $P1\leq$ 0.15 of \cite{smolcic08} for the separation of SF ($P1\leq$ 0.15) and AGN sources ($P1>$ 0.15) thus corresponds to ($u-K$) = 2.42. Note that due to our treatment of composite SF/AGN sources we adopt a slightly different colour threshold for the selection of SFGs (see following paragraph and Fig. \ref{fig:uK_SFprob}).

\noindent From Fig. 24 of \cite{smolcic08} (reproduced in the upper left corner of Fig. \ref{fig:uK_SFprob}) it is obvious that the tails of the distribution of star forming and AGN systems in P1 colour space extend into the selection regions for AGN and star forming sources, respectively. Moreover, about 10\% of the sample on which the classification scheme was developed are `composite' systems and found on either side of the divide\footnote{The classification of sources in the reference sample of \cite{smolcic08} into AGN, SF, and composite galaxies is based on their position in the optical spectroscopic diagnostic (BPT) diagram \citep{baldwin81}.}. When a source is classified as SF or AGN based purely on its rest frame optical colour there thus is a non-negligible probability of assigning it to the false category. For some purposes, e.g. when estimating which fraction of AGN systems have similar IR/radio properties as star formers, it is thus useful to adopt a probabilistic approach. Given the distributions $N_{\rm SF}$, $N_{\rm AGN}$ and $N_{\rm compo.}$ (see Fig. \ref{fig:uK_SFprob}.a) a possible definition for an {\it effective} probability ${\rm Pr_{eff.}\,(SF)}$ of correctly classifying a source as star forming at a given rest frame optical colour is
\begin{equation}
{\rm Pr_{eff.}\,(SF)} \approx {\rm Pr\,(SF)} + \frac{N_{\rm SF}}{N_{\rm SF} + N_{\rm AGN}}\times{\rm Pr\,(compo.)}\,,
\label{eq:SFprob}
\end{equation}
where
\begin{eqnarray*}
{\rm Pr\,(SF)} &=& \frac{N_{\rm SF}}{N_{\rm SF} + N_{\rm AGN} + N_{\rm compo.}}\\
{\rm Pr\,(compo.)} &=& \frac{N_{\rm compo.}}{N_{\rm SF} + N_{\rm AGN} + N_{\rm compo.}}~.
\end{eqnarray*}
In setting up (\ref{eq:SFprob}) we have assigned composite systems to the SF and AGN population according to the relative abundance $N_{\rm SF}/N_{\rm AGN}$ of SF and AGN sources at the particular colour. ${\rm Pr_{eff.}\,(SF)}$ is {\it \`a priori} given as a probability as a function of P1 through the distributions $N_{\rm SF}$, $N_{\rm AGN}$ and $N_{\rm compo.}$ presented in \cite{smolcic08}. However, it may be directly converted to the desired dependency on ($u-K$) by convolving the expression in (\ref{eq:SFprob}) with the distribution of P1 at fixed ($u-K$) colour (see inset of Fig. \ref{fig:P1equivcol}), which reflects the range of probabilities ${\rm Pr_{eff.}\,(SF\,|\,P1)}$ that contribute to ${\rm Pr_{eff.}\,(SF\,|\,}(u-K))$. In the upper right panel of Fig. \ref{fig:uK_SFprob} we show the distributions ${\rm Pr_{eff.}\,(SF/AGN\,|\,P1)}$ obtained according to equation (\ref{eq:SFprob}) and smoothed with a three-point running average (black curve -- star forming sources; dark grey curve -- AGN systems). Its convolution with a standard normal curve leads to the probability distribution ${\rm Pr_{eff.}\,(SF/AGN\,|\,}(u-K))$ shown in the lower panel of Fig. \ref{fig:uK_SFprob} which uses the same colour scheme as in panel {\it (b)}. The uncertainty in the best-fit correlation between P1 and ($u-K$) has been translated into an error in the probability function which is shown as a light grey area to either side of the black line giving ${\rm Pr_{eff.}\,(SF\,|\,}(u-K))$ in panel {\it (c)}. Due to the small uncertainties in the OLS bisector line parameters of equation (\ref{eq:P1umK_corr}) the dispersion $\sigma_{\Delta P1}$ is the most important factor that determines the differences in the shape of ${\rm Pr_{eff.}\,(SF/AGN\,|\,P1)}$ and ${\rm Pr_{eff.}\,(SF/AGN\,|\,}(u-K))$.\\
If one assigns composite objects to the SF and AGN population according to equation (\ref{eq:SFprob}) the point of equal probability of correctly classifying objects as SF or AGN, respectively, is reached at ($u-K$) = 2.36. This value is only slightly different from the direct translation (see previous paragraph) of the original definition in \cite{smolcic08}. In the remainder of the paper we will use the ($u-K$) = 2.36 threshold to separate SFGs from sources with emission that is dominated by AGN activity\footnote{When writing about and plotting probabilities we will henceforth use Pr\,(SF) as a shorthand for ${\rm Pr_{eff.}\,(SF\,|\,}(u-K))$.}.

\noindent Apart from the tails in the colour distribution of AGN and SF systems which cross the colour threshold, three additional effects could reduce the accuracy of the classification scheme.\\
First of all, a general evolution of the SF and AGN population to bluer colours at high redshift would lead to increasing contamination by AGN of the high-z population of SFGs if the colour cut is not adapted. \cite{smolcic08} have shown that an unchanging threshold is adequate until at least $z\sim$ 1.3. In Fig. \ref{fig:probcol_distr} we plot the distribution of ($u-K$) colours of our sources and follow the evolution of the median colour for both the IR- (dark grey histogram) and the radio-selected sample (light grey histogram; sources common to both samples lie within the hatched area). We find no evidence for a  strong evolution of average colours in either of the two samples out to $z\sim$ 2, and out to $z\sim$ 3 only by a small amount. Hence we apply the selection criterion uniformly to all sources, regardless of their redshift, except for the objects at the highest redshift where the medians have begun to change appreciably (see lower right panel of Fig. \ref{fig:probcol_distr}).\\
Secondly, nonperiodic flux variations of active galaxies will affect the choice of the best-fitting SED if photometric measurements are not simultaneously carried out over the whole spectrum. Since the rest frame optical colours are determined using SED templates this can cause misclassifications of AGN or SFGs with colours close to the threshold ($u-K$) = 2.36. A variability analysis (M. Salvato, private communication) of our 1.4\,GHz sources revealed that maximally 20\% of these display strong variability \citep[defined as ${\rm VAR}>$ 0.25; cf. equation (1) in][]{salvato09}. The true fraction of affected sources is likely to be smaller because inaccuracies in the photometry can artificially raise the variability parameter.\\
Finally, we cannot exclude that some unobscured Type 1 AGN with a blue ($u-K$) colour will be assigned to the SF category in our classification scheme. It is also possible that a number of dust-reddened starburst galaxies end up being classified as AGN, even though we used a red filter to define our rest frame colour on which we base the separation into SFGs and AGN. In the calibration sample of \cite{smolcic08} this kind of contamination amounted to less than $\sim$10\% (see their Appendix B2).

\section{IR SED Template Fitting}
\label{sect:SEDfitting}
Data from lensed high-$z$ galaxies \citep{siana08, gonzalez09} and from recent deep FIR surveys have shown that the SEDs of local star forming galaxies reproduce the SEDs of high redshift galaxies well out to $z\lesssim$ 1.5 \citep[e.g.][]{elbaz02, magnelli09, murphy09b}. However, it has also been reported that the SEDs of some IR-selected galaxies at high redshift can differ from local templates both at MIR \citep{rigby08} and FIR \citep{symeonidis08} wavelengths, conceivably due to intrinsic scatter in the physical properties of these sources which deviate from the median trend that the empirical galaxy templates represent.\\
Following the procedure described in \citet{murphy09b}, we derive infrared luminosities ($L_{\rm TIR}$) by fitting the 24 and 70\,$\mu$m data points to the \citet{charyelbaz01} SED templates and integrating between 8 and 1000~$\mu$m. This wavelength range is in principle also sampled by S-COSMOS observations at 8 and 160\,$\mu$m \citep{sanders07, frayer09} but we restrict ourselves to the two aforementioned bands because ({\it i}) at $z>$ 0.6 a 8\,$\mu$m measurement would include stellar light (which starts to dominate at the SED at rest frame 5\,$\mu$m), while we are fitting pure dust templates; and ({\it ii}) the shallower coverage and broad PSF of the 160\,$\mu$m observations complicate the identification of unambiguous counterparts. Our choice of the \citet{charyelbaz01} templates is motivated by the fact that they have been found to exhibit 24/70\,$\mu$m flux density ratios that are more representative \citep{magnelli09} of those measured for galaxies at $z \sim1$ compared to the \citet{dalehelou02} or \citet{lagache03} templates. \\ 
For the cases where a source is detected firmly at 24 and 70~$\mu$m, the best-fit SEDs are determined by a $\chi^2$ minimization procedure whereby the SED templates are allowed to scale such that they are being fitted for luminosity and temperature separately. Consequently, the amplitude and shape of the SEDs scale independently to best match the observations. The input photometry is weighted by the $S/N$ ratio of the detection if it is a well-defined measurement, and the normalization constant is determined by a weighted sum of observed-to-template flux density ratios for all input data used in the fitting. 
In the cases where only an upper limit is available at 70~$\mu$m, the latter is not incorporated into the $\chi^2$ minimization but used to reject fits which have flux densities above the associated measured limit.\\
Errors on the best-fitting value of $L_{\rm TIR}$ are determined by a standard Monte Carlo approach using the photometric uncertainties of the input flux densities which reflect both calibration errors ($\sim$2\% at 24 \citep{engelbracht07} and $\sim$5\% at 70\,$\mu$m \citep{gordon07}) and the uncertainties in the PSF-fitting (generally of order $S_{\rm PSF}/(S/N)$, where $S_{\rm PSF}$ is the flux returned by the PSF fit).\\
If a source is only detected at 24~$\mu$m, we also fit the photometry using the SED templates of \citet{dalehelou02} and define the best estimate of the IR luminosity as the average $L_{\rm TIR}$ from the two separate fits.

\section{Selection Effects and Statistical Treatment of Flux Limits}
\label{sect:selsurv}

\subsection{Shifts Between the Average IR/Radio Ratios of Flux Limited Samples}
\label{sect:selbias}

The selection effects that are the topic of this section arise in flux limited samples when flux information from one of the selection bands is directly used in the computation of the quantity being studied. In the present case the critical quantity is the logarithmic IR/radio flux ratio $q$, but analogous effects need to be considered in the context of studies of the distribution of spectral indices at different radio frequencies \citep[e.g.][]{kellermann64, condon84}, of X-ray to optical continuum slopes of AGN \citep{francis93} or of the $M_{\bullet} - \sigma$ and $M_{\bullet} - L$ relationships \citep{lauer07}.\\
In Fig. \ref{fig:qselsketch} we illustrate the origin of the selection effect: consider the left hand panel in which the IR-to-radio SEDs of three sources with different observed bolometric flux are distributed along the vertical axis. Each of these three SEDs splits into three branches at the peak of the SED, thereby schematically reflecting the range of observed IR/radio ratios (from top to bottom: 3\,$\sigma$ radio-excess outlier -- dashed line; average source -- solid line; and 3\,$\sigma$ IR-excess outlier -- dotted line). If we impose the indicated selection threshold at 1.4 GHz (red line) the resulting sample will contain ({\it i}) all sources of the brightest flux class, regardless of their IR/radio ratio; ({\it ii}) the source with an average IR/radio ratio and the radio-excess source from objects of the intermediate flux class and; ({\it iii}) in the faintest flux bin only the radio-excess sources. Since the fainter sources are more abundant (as parametrized by the slope of the differential source counts $\nicefrac{dN}{dS} \propto S^{-\beta}$, with $\beta>$ 0) this results in a surplus of radio-excess sources and consequently a {\it low} average IR/radio ratio in a radio-selected sample. The right hand side of Fig. \ref{fig:qselsketch} shows that an IR-selected sample is biased in the opposite direction, i.e. towards {\it high} IR/radio ratios.

\noindent The analytical expression for the difference between the average IR/radio ratio of IR- and radio-selected samples is \citep{kellermann64, condon84, francis93, lauer07}:
\begin{equation}
\Delta q_{\rm bias} = {\rm ln}(10)\,(\beta-1)\,\sigma_q^2~.
\label{eq:qbias}
\end{equation}
It thus depends on $\beta$, the power law index of the source counts, and on $\sigma_q$, which is the dispersion of the IR/radio relation. Note that this offset will occur {\it regardless of the relative depth} of the two involved bands. An estimate of the `intrinsic' (i.e. unbiased) IR/radio ratio can be obtained by constructing the sample using an unrelated selection criterion like optical luminosity, mass or morphological type \citep{lauer07}.

\noindent Since the recent work on the evolution of the IR-radio relation at intermediate and high redshift was often based on flux limited surveys, we would expect most of the findings to be affected by this selection bias to a certain extent. In Table \ref{tab:prevwork} we have collected the selection criteria and average values of $q$ (final column) that were published in the literature during the last decade.\\
We see that broadly speaking the various IR/radio diagnostics have values $q_{24}\approx$ 1, $q_{70}\approx$ 2.1, $q_{\rm FIR}\approx$ 2.3 and $q_{\rm TIR}\approx$ 2.6. These different values are not the result of selection effects but reflect if the IR filter covers a wavelength range that is close to the IR SED peak or a part of the SED with lower energy content. In the following paragraph we will discuss the plausible influence of selection effects on the various measurements of $q_{24}$ and $q_{\rm FIR}$, in particular.\\
Due to the high sensitivity of the {\it MIPS} 24\,$\mu$m band many of the papers listed in Table \ref{tab:prevwork} have studied the IR/radio ratio $q_{24}$. The radio-selected samples of \cite{appleton04} and \cite{ibar08} find that $q_{24}\in$ [0.94, 1], depending on the choice of the IR template used for the $K$-correction. The local IR-selected sample of \cite{rieke09} on the other hand has a mean $q_{24}$ of $\sim$1.25 and shows some signs of variations with IR luminosity. The offset between the means of the radio-selected samples and the IR-selected data set is $\sim$0.3 dex, in good agreement with the predicted $\Delta q_{\rm bias}=$ 0.31 of equation (\ref{eq:qbias}) if we set $\sigma_q\approx$ 0.3 in accordance with observations \citep[e.g.][]{yun01, bell03, appleton04, ibar08} and under the simplified assumption of Euclidean source counts ($\beta=2.5$). In the case of the FIR/radio flux ratio $q_{\rm FIR}$ we can compare the two radio-selected (sub)samples of \cite{garrett02} and \cite{sajina08} that have $q_{\rm FIR}\approx$ 2 with a jointly radio- and sub-mm selected mean of 2.07 from \cite{kovacs06} and mean values $q_{\rm FIR}\in$ [2.2, 2.4] for IR-selected \citep{younger09, sajina08} or essentially volume limited samples in \cite{bell03} and \cite{yun01}. As with $q_{24}$ there is thus evidence of a $\sim$0.3 dex shift in $q_{\rm TIR}$ between radio-selected and other samples. As far as we know no measurement of $q_{70}$ in an IR-selected sample exists but the compilation in Table \ref{tab:prevwork} shows that reassuringly all determinations of $q_{70}$ based on radio-selected samples \citep{appleton04, frayer06, seymour09} are quite similar.\\
The radio stacking experiments of \cite{boyle07}, \cite{beswick08} and \cite{garn09a} do not fit the picture which is probably due to the different nature of the analysis. Nevertheless, it cannot be excluded that part of the variations in the other studies are due to field-to-field variance or different assumptions about IR SEDs and the radio spectral slope. To this end we will test in \S\,\ref{sect:results} whether or not the offset between our IR- and radio-selected samples -- that have been consistently constructed from the same parent data sets --  conforms to our expectation. If so, it would be strong support for selection effects alone being able to reconcile the seemingly discrepant measurements of average IR/radio properties in the literature.

\subsection{Derivation of Distribution Functions with Survival Analysis}
\label{sect:briefsurv}

Discarding the information from undetected counterparts introduces a second source of bias in addition to the selection effects mentioned in \S\,\ref{sect:selbias}. It arises from the unrepresentative sampling of the true distribution function of IR/radio ratios by sources which are directly detected in all involved bands. We would like to emphasize that the shift $\Delta q_{\rm bias}$ in equation (\ref{eq:qbias}) is the difference between the mean $q$ of IR- and radio-selected samples with {\it correctly sampled} distribution functions. $\Delta q_{\rm bias}$ cannot be compensated by accounting for upper or lower limits on $q$ due to undetected IR or radio counterparts in the two different samples; as discussed in the previous section, the mean IR/radio ratio measured in an IR- and radio-selected sample only brackets the value one would measure with an unbiased data set which we can best approximate by a sample jointly selected at IR and radio wavelengths (see \S\,\ref{sect:jointsamp}).\\

\noindent The ratio of two flux constraints that could be either a well-defined measurement or an upper limit will render an upper or lower bound, a well-defined value or be indeterminate (if both numerator and denominator are limits). Since we use the pooled information from a radio- and IR-selected sample in this study the latter case never occurs. We do expect, however, to encounter upper limits on IR/radio ratios from radio-selected sources that are not detected in the IR or lower limits if the radio counterpart of an IR-selected source was too faint to be detected (cf. \S\S\,\ref{appsect:radiodata} \& \ref{appsect:IRdata}).\\
Let $q_i$ ($i=$ 1, ... , $n$) be the actual value of the flux ratio for each of the $n$ sources in a suitably defined sample (e.g. the population in a certain slice of redshift). As a consequence of the noise characteristics in the radio and IR images $q_i$ can only be measured if it lies in the interval $[q^-_i, q^+_i]$, where $q^-_i$ and $q^+_i$ are upper and lower limits on the flux ratio, respectively. These limits may be different for each source. Our knowledge about the distribution of IR/radio ratios after carrying out all our measurements can thus be summarized with two vectors of variables, $\bm{\mathcal{Q}}$ and $\bm{\delta}$:
\begin{eqnarray}
\mathcal{Q}_i &=& \textrm{ max(min($q_i$, $q^+_i$), $q^-_i$)}\nonumber \\
\\
\delta_i &=& \begin{cases}
\quad -1 & \quad \mbox{if }q_i<q^-_i\\
\quad  0 & \quad \mbox{if }q_i \in [q^-_i, q^+_i]\\
\quad  1 & \quad \mbox{if }q_i>q^+_i
\end{cases} \nonumber
\label{eq:doubcensodef}
\end{eqnarray}
In {\it survival} or {\it life time analysis} the action of imposing measurement limits is referred to as `censoring'. A variable $\mathcal{Q}_i$ is said to be {\it left censored} if $\mathcal{Q}_i<q^-_i$ and {\it right censored} if $\mathcal{Q}_i>q^+_i$. If both kinds of censoring occur in a data set it is called {\it doubly censored}, otherwise one talks of {\it single censoring}. During the remainder of the paper we will use the terms `limit' and 'censored measurement' interchangeably.

\noindent In Appendix \ref{appsect:survival} we sketch the steps that are involved in going from the information ($\bm{\mathcal{Q}}$, $\bm{\delta}$) to the distribution function of the IR/radio ratios. Inferring the true distribution of the $q_i$ of a sample is essential for the calculation of its average IR/radio properties. In \S\,\ref{sect:results} we will construct distribution functions for data sets that are both singly and doubly censored. Recipes for dealing with the former case are plentiful in texts on survival analysis (see e.g. \cite{feigelson85} for applications to astronomy) such that we only include some brief remarks in \S\,\ref{appsect:singlecense}. Since the more general case of double censoring is not as widely used in astronomical applications, the most important formulae and useful computational guidelines are provided in Appendix \ref{appsect:doubcensetheory}.\\
The methods described in Appendix \ref{appsect:survival} have been implemented using Perl/PDL\footnote{The Perl Data Language (PDL) has been developed by K. Glazebrook, J. Brinchmann, J. Cerney, C. DeForest, D. Hunt, T. Jenness, T. Luka, R. Schwebel, and C. Soeller and can be obtained from http://pdl.perl.org} scripts written by M.T.S. . Their correct functionality was verified with examples in the literature. In particular, we checked that our implementation of the algorithm for the calculation of the doubly censored distribution function \citep{schmitt85} -- when applied to the special case of singly censored data -- gave the same results as the scripts based on the Kaplan-Meier product limit estimator (\cite{kaplanmeier58}; see also Appendix \ref{appsect:singlecense}).

\section{Results}
\label{sect:results}

The main focus of this section is the search for changes with redshift of the average IR/radio ratio in the SF population. We track evolutionary trends in the range $z<$ 1.4 for both monochromatic and TIR/radio flux ratios in \S\S\,\ref{sect:monochromqs} and \ref{sect:TIRqs}, and separately consider a sample of high redshift ($z\gtrsim$ 2.5) sources in \S\,\ref{sect:highz}. \S\,\ref{sect:agnfract} is dedicated to the IR/radio properties of AGN hosts and in \S\,\ref{sect:qvslum} we study variations of IR/radio ratios with luminosity.\\
Previous studies have carried out similar analyses using a variety of IR/radio diagnostics. These include {\it MIPS}-based monochromatic flux ratios $q_{24}$ and $q_{70}$ (e.g. \cite{appleton04, ibar08, seymour09}; see equation (\ref{eq:monchromqs}) for the definition of $q_{24\,[70]}$) which we discuss in \S\,\ref{sect:monochromqs}. Other studies have used the FIR (42.5-122.5\,$\mu$m) to radio flux ratio $q_{\rm FIR}$ \citep[see equation (\ref{eq:FIRq}); e.g.][]{garrett02, kovacs06, sajina08}, or the ratio of total IR luminosity ($L_{\rm TIR}$) to radio luminosity \citep[e.g.][]{murphy09b}:
\begin{equation}
q_{\rm TIR} = {\rm log}\left(\frac{L_{\rm TIR}}{3.75\times10^{12}\,{\rm W}}\right) - {\rm log}\left(\frac{L_{\rm 1.4\,GHz}}{\rm W\,Hz^{-1}}\right)~.
\end{equation}
Total infrared luminosities $L_{\rm TIR}$ (in units of [W]) are calculated by integrating the SED between 8 and 1000\,$\mu$m. The rest frame 1.4\,GHz luminosity (expressed in [W\,Hz$^{-1}$]) is
\begin{equation}
L_{\rm 1.4\,GHz} = \frac{4\pi D_L(z)^2}{(1+z)^{1-\alpha}}\,S_{\nu}({\rm 1.4\,GHz})~,
\label{eq:L20}
\end{equation}
where $S_{\nu}({\rm 1.4\,GHz})$ is the integrated radio flux density of the source and $D_L(z)$ the luminosity distance. The $K$-correction $K_{\rm 1.4\,GHz}(z) = (1+z)^{-(1-\alpha)}$ depends on the spectral index $\alpha$ of the synchrotron power law $S_{\nu} \propto \nu^{-\alpha}$. For the rest of the analysis we will assume that $\alpha=$ 0.8 \citep{condon92}. We will return to the TIR/radio flux ratios in \S\,\ref{sect:TIRqs}.

\subsection{Monochromatic IR/Radio Properties of Star Forming and AGN Galaxies}
\label{sect:monochromqs}

\subsubsection{Observed Flux Ratios}
\label{sect:obsmonochrom}

The observed 24\,$\mu$m/1.4\,GHz flux ratio $q_{\rm 24,\, obs}$ is plotted against redshift in Fig. \ref{fig:uncorrq24} for SFGs ({\it top}) and AGN ({\it bottom}). Sources are assigned to the two categories depending on whether Pr\,(SF) is larger or smaller than 50\% (cf. \S\,\ref{sect:SFvsAGN}).\\
While there clearly are many radio-loud sources in our AGN sample, Fig. \ref{fig:uncorrq24}.b shows that a majority of the objects assigned to the AGN category displays very similar IR/radio ratios as the SFGs. We will discuss this observation in more detail in \S\,\ref{sect:agnfract}. At the same time, the sample of SFGs also includes a number of radio-excess sources. They usually have photometric redshift estimates and mostly lie at 1 $<z<$ 3. This roughly corresponds to the redshift range in which photometrically determined redshifts are subject to the largest uncertainty because the 4000\,{\AA} break is only sampled by broad and widely spaced photometric bands. As a consequence, absorption features and emission lines from AGN and SF systems often interfere with each other in the same filter. Even though we did attempt to remove all unreliable redshifts -- as described in \S\,\ref{appsect:redshifts} -- it thus seems probable that at least some of these cases are due to wrong distance estimates and hence to the selection of an inappropriate optical SED. Since this results in a faulty ($u-K$) colour, the source in question could then have been assigned to the SF rather than the AGN category. Another possibility is that the peak of AGN activity at $z\sim$ 2 \citep[e.g.][]{wolf03,richards06} also influences the SF sample due to the statistical nature of the identification of SFGs and because especially AGN in composite systems could have been classified as star forming. Lastly, we tried to assess if unobscured Type 1 AGN represent a significant fraction of the nominally star forming radio excess sources in the pertinent redshift range. Based on the confidence class (see \S\,\ref{appsect:specz}) of those objects for which follow-up spectroscopy was available, we estimate that only $\sim$5\% are quasars classified as SF due to their blue colour.\\
The mean value of $q_{\rm 24.\, obs}$ decreases as a function of redshift. We will show later on (see Fig. \ref{fig:q24AGNfrac}) that this decrease agrees with the variations local LIRGs (detectable only out to $z\sim$ 1; see vertical dotted lines in panel ({\it a})) and ULIRGs would display if redshifted.

\noindent In Fig. \ref{fig:uncorrq70} we plot the observed 70\,$\mu$m/1.4\,GHz flux ratio $q_{\rm 70.\, obs}$ of our sources as a function of redshift. All symbols and colours are exactly as in the previous figure. Note that censored measurements due to flux limits at 70$\mu$m are more frequent than was the case for $q_{\rm 24.\, obs}$ because the 70\,$\mu$m observations are much shallower. The observed flux ratio before $K$-correction shows the same decline at higher redshifts as was seen for the observed 24\,$\mu$m/1.4\,GHz flux ratios.

\subsubsection{Evolution of $q_{24}$ and $q_{70}$ with Redshift}
\label{sect:monochromevo}

As described in \S\,\ref{sect:SEDfitting} all sources classified as star forming were fit with IR SEDs in order to derive IR luminosities. As a by-product of the template fitting we can immediately obtain rest frame (i.e. $K$-corrected) 24 and 70\,$\mu$m flux densities by convolving the best-fitting SED with the filter response functions of {\it MIPS}. In the following we define the $K$-corrected 24 and 70\,$\mu$m fluxes as the average of the values obtained from the libraries of \cite{charyelbaz01} and \cite{dalehelou02} and use them to construct monochromatic rest frame flux density ratios $q_{24,\,0}$ and $q_{70,\,0}$. The associated 1.4 GHz flux densities have been $K$-corrected according to equation (\ref{eq:L20}).\\

\noindent To quantify the evolution of $q_{24,\,0}$ in the joint IR- and radio-selected sample we:
\begin{enumerate}
\item Bin the data such that each redshift slice contains an identical number of objects ($\sim$250). The number of bins is kept limited to guarantee that the distribution function of $q_{24,\,0}$ is sampled sufficiently.
\item Run the iterative algorithm outlined in \S\,\ref{appsect:doubcensetheory} to find the cumulative distribution function of $q_{24,\,0}$ at each redshift. The median immediately follows from this computation as does the scatter in the population which we obtain by fitting a Gaussian distribution with known mean (equal to the previously determined median) to the distribution function. The choice of the Gaussian is motivated by the shape of the local IR-radio relation \citep{helou85, yun01, bell03}.
\item Determine the evolution of the average IR/radio ratio by fitting a linear trend line to the medians. Only measurements at $z\leq$ 1.4 are considered for this since the scatter at higher redshifts is found to increase abruptly, thus making the determination of the median uncertain.
\end{enumerate}
Steps 1 and 2 are are also carried out individually for the sample of IR- and radio-selected galaxies. The results are shown in Fig. \ref{fig:SFcorrq24_medevo}. Since the cumulative distribution function is normalized it lies in the range between zero and unity and can thus be regarded as the probability of obtaining a measurement of $q_{24,\,0}$ which is less than -- in the case of the radio-selected sample (light grey curve) -- or in excess of -- for the IR-selected sample (dark grey curve) -- the ordinate. The distribution function of the doubly censored union of the IR- and radio-selected samples is also parametrized such that it runs from 0 to 1 with increasing $q_{24,\,0}$. It is plotted in black together with a dashed red line which shows the corresponding best-fitting Gaussian distribution. The intersection of the black curve with the 50\% probability line (dotted horizontal line) defines the median value of $q_{24,\,0}$.\\
Fig. \ref{fig:SFcorrq24_medevo} demonstrates that the median of the radio-selected population lies systematically below that of the IR-selected objects. The shift is approximately 0.35 dex at low redshift and grows to about 0.7 dex beyond $z\sim$ 1. The increase is probably caused by the intrinsically higher scatter in the IR-radio relation at high luminosities \citep{yun01, bressan02}, possibly in combination with the reduced reliability of photometric redshifts and/or some falsely classified AGN that begin to affect the sample starting at $z\sim$ 1.3. A shift of $\sim$0.35 dex as observed at $z<$ 1 where the accuracy of our measurements is highest agrees fairly well with the prediction of equation (\ref{eq:qbias}) and is hence a likely explanation for differences between previously reported average IR/radio properties of both local and high-$z$ galaxies \citep[e.g.][]{appleton04, ibar08, rieke09}.

\noindent In the lower panel of Fig. \ref{fig:SFcorrq24_medevo} we plot the medians $\langle q_{24,\,0}\rangle$ of the jointly IR- and radio-selected SFGs (black dots) at different redshift on top of the $K$-corrected values $q_{24,\,0}$ (colours and symbols are identical to those in Fig. \ref{fig:uncorrq24}.a). The error bars mark the 95\% confidence interval associated with the median. Table \ref{tab:q24info} lists the median and scatter of $q_{24,\,0}$ which were determined with survival analysis in each of the redshift bins of Fig. \ref{fig:SFcorrq24_medevo}. In addition to the measurements carried out on the jointly IR- and radio-selected sample the table also provides the according values for the IR- and radio-selected samples individually.\\
Using the 2\,$\sigma$ errors on the medians as weights we fit them with a model of linear redshift evolution. The best-fitting trend of $\langle q_{24,\,0}\rangle$ vs. $z$ is shown in black in the lower window of Fig. \ref{fig:SFcorrq24_medevo} (see Table \ref{tab:medqevo} for the parametrization of the line). Because the fit was carried out with respect to linear redshift space while the plot has a logarithmically scaled redshift axis it is curved. Within the errors the slope $\nicefrac{d\langle q_{24,\,0}\rangle}{dz} = -0.015\pm0.136$ is consistent with no evolution of the $K$-corrected 24\,$\mu$m/1.4\,GHz ratio at $z\leq$ 1.4 (the maximal distance out to which the precision of the photometric redshifts is high). The $y$-axis intercept of the trend line at $z=$ 0 is in agreement with the recent analysis of \cite{rieke09} who find that $q_{24,\,0}=$ 1.22 with a scatter of 0.24. In our sample we find $\langle q_{24,\,0}\rangle_{z=0} = $ 1.28$\pm$0.10 (where the error states the formal 1\,$\sigma$ uncertainty from the linear fit). For a comparison between the average IR/radio properties of radio-selected samples we can refer to the studies of \cite{appleton04} and \cite{ibar08}; they report an average $q_{24,\,0}=$ 0.94\,-\,1, depending on the IR-SED adopted to $K$-correct to the rest frame. These values agree well with the range of medians $\langle q_{24,\,0}\rangle \in$ [0.8, 1] measured for radio-selected COSMOS data at intermediate redshift (cf. left-most column of Table \ref{tab:q24info}).

\noindent Our convention for choosing SFGs states that Pr\,(SF) must be at least 50\%. In Fig. \ref{fig:SFcorrq24varevol} we assess how changing this threshold affects the redshift evolution that is inferred from the data. The variation of the parameters of the best-fitting evolutionary trend line is shown in the upper- and lower-most window of Fig. \ref{fig:SFcorrq24varevol} ($y$-axis intercept and slope, respectively). A black symbol in the middle of the displayed data range marks the results that were shown in Fig. \ref{fig:SFcorrq24_medevo}. They are fully consistent with the evolution found if a more conservative threshold -- e.g. at Pr\,(SF) = 66\% -- for the selection had been chosen. It is interesting that the inclusion of a significant fraction of sources with a probability of up to 80\% of being AGN does not alter the results either. This is a strong indication that our sample of optically selected AGN contains many objects with IR/radio properties that closely resemble those of SF systems. Similar observations were made by, e.g., \cite{sopp91} and \cite{roy98}, who studied local samples of radio-quiet quasars and/or Seyfert 1 sources lacking a compact nucleus. The middle row of Fig. \ref{fig:SFcorrq24varevol} shows that while the average values of $q_{24,\,0}$ are similar for many SFGs and AGN the latter are subject to a larger scatter as was previously found by, e.g., \cite{condon82, obric06, mauchsadler07}.\\
As an additional test of the robustness of our findings we checked if the evolutionary trend in SF samples selected through ($u-K$) or P1 differs. For the radio-selected sample where both colours were available we found equivalent results regardless of the chosen approach.

\noindent Figures \ref{fig:SFcorrq70_medevo} and \ref{fig:SFcorrq70varevol} (the results of which are summarized in Table \ref{tab:q70info}) repeat the analysis of Figs. \ref{fig:SFcorrq24_medevo} and \ref{fig:SFcorrq24varevol} for the $K$-corrected 70\,$\mu$m/1.4\,GHz flux ratio. Note that in comparison with Fig. \ref{fig:uncorrq70} the number of censored measurements is much smaller when we consider rest frame IR/radio ratios $q_{70,\,0}$ rather than observed flux ratios $q_{\rm 70,\,obs}$. The reason is that we do not require a 70\,$\mu$m detection for the IR template fitting but also fit objects which have a limit at 70\,$\mu$m and a direct detection at 24\,$\mu$m.\\
The plot of $q_{70,\,0}$ vs. redshift (lower panel of Fig. \ref{fig:SFcorrq70_medevo}) as well as Fig. \ref{fig:SFcorrq70varevol}, which illustrates the stability of the findings with respect to changes in the selection criterion for SFGs, show that $q_{70,\,0}$ behaves in a similar way as was found for $q_{24,\,0}$. The extrapolated average 70\,$\mu$m/1.4\,GHz flux ratio at $z=$ 0, $\langle q_{70,\,0}\rangle_{z=0}$ equals 2.31$\pm$0.09 and if AGN are included the scatter in the relation increases in analogy to what was found for $q_{24,\,0}$. As for $q_{24}$, the evolutionary slope $\nicefrac{d\langle q_{70,\,0}\rangle}{dz} = -0.123\pm0.135$ (slope and normalization of the evolutionary trend line are logged in Table \ref{tab:medqevo}) is consistent with zero.\\
 The average IR/radio properties of the radio-selected samples of \cite{appleton04} and \cite{frayer06}\footnote{Although based on a catalog of xFLS 70\,$\mu$m sources the sample of \cite{frayer06} becomes essentially a radio-selected sample at the stage when sources without a counterpart in the 1.4\,GHz radio catalog of the FLS field \citep{condon03} are removed from the sample.} are 2.16$\pm$0.17 and 2.10$\pm$0.16, respectively. Although the agreement with our findings is not quite as good as in the case of $q_{24}$, they are are nevertheless consistent (within both the formal error and the scatter) with the range of medians $\langle q_{70,\,0}\rangle \in$ [1.7, 2.1] at intermediate redshift in our radio-selected sample (see Table \ref{tab:q70info}). To our knowledge there so far has been no comparable study which uses an IR-selected sample to compute an average $q_{70}$.

\subsection{Evolution of TIR/Radio Flux Ratios with Redshift}
\label{sect:TIRqs}
In the local universe the correlation of IR and radio flux is tightest if integrated (F)IR luminosities rather than monochromatic flux ratios are considered. To complement the analysis of \S\,\ref{sect:monochromqs} we thus show in this section the correlation of TIR (8-1000\,$\mu$m) and 1.4\,GHz luminosity as parametrized by the TIR/radio ratio $q_{\rm TIR}$ for our VLA- and S-COSMOS data.

\noindent The computation of the distribution functions for the parameter $q_{\rm TIR}$ is carried out following the same steps described in \S\,\ref{sect:monochromevo}. The results are shown for a number of redshift bins in Fig. \ref{fig:SFqTIR_medevo} where we also compare the median derived for the jointly IR- and radio-selected SFGs with the local value of $\langle q_{\rm TIR}\rangle_{z=0} = 2.64\pm$0.02 (\cite{bell03}; vertical dashed line). Our average values $\langle q_{\rm TIR}\rangle$ in the range $z<$ 1.4 (see Table \ref{tab:qTIRinfo}) lie to either side and always remain well within the dispersion of the local measurement of \cite{bell03}.\\
The evolution of $q_{\rm TIR}$ is shown in the lower panel of Fig. \ref{fig:SFqTIR_medevo} using the same presentation of the data as for the monochromatic IR/radio flux ratios. Since the latter were derived based on the IR templates which are used here to calculate the integrated IR luminosity, we expect by construction that the evolutionary trend is in good qualitative agreement with the findings of \S\,\ref{sect:monochromqs}. For the same reason we cannot expect to observe a reduced scatter in the values of $q_{\rm TIR}$ with respect to those of the monochromatic flux ratios as the spread in the properties of the best-fitting IR SEDs must manifest itself in $q_{24,\,0}$ and $q_{70,\,0}$ as well.\\
The line parameters for the evolution of $\langle q_{\rm TIR}\rangle$ are given together with those of $\langle q_{24,\,0}\rangle$ and $\langle q_{70,\,0}\rangle$ in Table \ref{tab:medqevo}: in contrast to $q_{24,\,0}$ and $q_{70,\,0}$ the best-fitting evolutionary trend for $\langle q_{\rm TIR}\rangle$ suggests a decrease of the average TIR/radio ratio by 0.35 dex out to $z\sim$ 1.4. However, this slope is detected at the 2\,$\sigma$ significance level and predicts a median $\langle q_{\rm TIR}\rangle$ at $z\sim$ 1.4 that still lies within the dispersion measured in our lowest redshift bin. It thus seems unlikely that the evolutionary signal is real, especially in view of the results of \S\,\ref{sect:highz} where we measure an average $\langle q_{\rm TIR}\rangle_{z>2.5}$ that is in excellent agreement with the local value for a subset of highly redshifted galaxies in the COSMOS field. An examination of the evolutionary slopes for $\langle q_{24,\,0}\rangle$, $\langle q_{70,\,0}\rangle$ and $\langle q_{\rm TIR}\rangle$ in Table \ref{tab:medqevo} shows that they become more negative along this sequence. This could be related to a number of radio-excess sources with 1 $\lesssim z \lesssim$ 3 which are part of our optically selected SF sample (visible as a diffuse cloud of upper limits and detections below the main locus of symbols in all our plots of $q$ vs. $z$; see also our comment in \S\,\ref{sect:obsmonochrom}) and that tend to lower the average IR/radio in this redshift range. If these objects were falsely classified composite sources or AGN the increased emission at 24\,$\mu$m from their hot dust might be able to compensate the radio-excess, thus leading to zero evolution in $q_{24,\,0}$ as observed. $q_{70,\,0}$ and $q_{\rm TIR}$ on the other hand sample mainly IR light from star formation and hence are lowered in the presence of excess radio emission. This scenario can also explain why the evolutionary slope of $\langle q_{24,\,0}\rangle$ is insensitive to the selection criterion for SFGs (cf. Fig. \ref{fig:SFcorrq24_medevo}) while it varies in the same sense as described above in the case of $\langle q_{70,\,0}\rangle$ and $\langle q_{\rm TIR}\rangle$.

\subsection{AGN with Similar IR/Radio Properties as Star Forming Galaxies}
\label{sect:agnfract}

The analysis of the previous sections revealed (cf. Figs. \ref{fig:SFcorrq24varevol}, \ref{fig:SFcorrq70varevol} and \ref{fig:SFqTIRvarevol}; also Figs. \ref{fig:uncorrq24} and \ref{fig:uncorrq70}) that the IR/radio properties of SFGs are shared by many of the AGN-bearing systems in our sample. In this section we will study this in more detail. We first test (\S\,\ref{sect:XMMonSF}) if it remains valid for a subsample of sources which are detected in X-rays and have been found to host an AGN using a different approach than the classification scheme introduced in \S\,\ref{sect:SFvsAGN}. In \S\,\ref{sect:cohabfract} we then compute (in different redshift bins at $z\lesssim$ 1-1.4) the relative frequency of AGN and SF sources as a function of the IR/radio ratio.

\subsubsection{IR/Radio Properties of X-ray Detections}
\label{sect:XMMonSF}

At the sensitivity of the {\it XMM-Newton} observations of the COSMOS field a large fraction of the detected sources is expected to be powered by AGN. This is confirmed by \cite{salvato09} who have shown that $\sim$70\% of the {\it XMM-Newton} sources have UV to NIR SEDs which contain an AGN component. In Fig. \ref{fig:XMMqs} we compare the observed 24\,$\mu$m and 70\,$\mu$m to radio flux density ratios $q_{\rm 24,\,obs}$ and $q_{\rm 70,\,obs}$ of X-ray detected AGN hosts at different redshifts with the predicted IR/radio properties of model SFGs (coloured tracks\footnote{The tracks are constructed by taking the ratio of the $K$-corrections between ({\it i}) the flux density at the rest frame ($\lambda$) and redshifted effective wavelength ($\lambda/(1+z)$) of the {\it MIPS} filter and, analogously, ({\it ii}) that applied to the 1.4\,GHz band. The $K$-correction $K(z)$ is defined as the ratio of the rest frame luminosity $L_{\nu}(\lambda)$ and the luminosity at wavelength $\lambda/(1+z)$ which is sampled by the observer's measurement of the flux $S_{\nu}(\lambda)$:
\begin{equation}
L_{\nu}(\lambda) = L_{\nu}(\nicefrac{\lambda}{1+z})\times K(z) = 4\pi D_L(z)^2\,S_{\nu}(\lambda)\times K(z)~.
\end{equation}
\noindent Here $D_L(z)$ is the luminosity distance. The $K$-corrections used in the conversion of observed IR flux measurements at 24 and 70\,$\mu$m to rest frame quantities depends on the shape of the IR SED of the galaxies. For the radio flux it has the form given after equation (\ref{eq:L20}).}; see footnote and the text of \S\,\ref{sect:cohabfract} for additional details). Note that according to the analysis of \cite{salvato09} the AGN contribution to the UV-NIR SED exceeds 50\% for most of these sources. From Fig. \ref{fig:XMMqs} it is obvious that a majority of the {\it XMM-Newton} sources have IR/radio ratios that are perfectly consistent with those expected for starbursts. They are genuine examples of active galaxies in which the AGN, although significantly contributing to the SED at optical and X-ray wavelengths, does not cause significant excess radio emission. Fig. \ref{fig:XMMqs} therefore is strong evidence that the findings of \S\S\,\ref{sect:monochromqs} and \ref{sect:TIRqs} cannot be ascribed to an inadequacy of the method we adopted to distinguish between AGN and SFGs.

\subsubsection{The Relative Abundance of AGN and SFGs on the Star Forming Locus}
\label{sect:cohabfract}

We first define the `main' locus of SFGs in a plot of {\it observed} IR/radio ratio vs. redshift. Working with observed flux densities is necessary because we want to avoid imposing a template fit with the IR SED of a SFG on an AGN-bearing source even if it quite probably shares similar IR/radio properties. At each redshift the star forming locus is centred on the average value -- $\langle q_{\rm 24\,[70],\,template}\rangle$ -- of $q_{\rm 24,\,obs}$ as predicted by the model SEDs of sources in the observable range of IR luminosities. We then consider a region between +2\,$\sigma_q$ and -2\,$\sigma_q$ around $\langle q_{\rm 24\,[70],\,template}\rangle$ in which we chart the relative frequency ${\rm f_{AGN}}(q, z)$ of AGN. The analysis is restricted to this $\pm2\,\sigma$ band because beyond it the sparse sampling of the distribution function of the IR/radio ratios leads to unwanted fluctuations of ${\rm f_{AGN}}(q, z)$. The value of\,$\sigma_q$ is a representative average of the scatter in our data for the SF population at $z\lesssim$ 1, i.e. 0.35 dex.\\
In the upper panel of Fig. \ref{fig:q24AGNfrac} (Fig. \ref{fig:q70AGNfrac} shows the same information for $q_{70}$) we show the expected variations in $q_{24,\,obs}$ of high-$z$ galaxies assuming that their SEDs at IR and radio wavelengths are similar to those of local SFGs. SEDs from three different template libraries -- as well as that of the starburst M82 -- are shown for different $L_{\rm TIR}$. The tracks are normalized at $z=$ 0 using the best-fit evolutionary trend line displayed in Fig. \ref{fig:SFcorrq24_medevo}. In the background we re-plot (cf. Fig. \ref{fig:uncorrq24}.a) the observed 24\,$\mu$m/1.4\,GHz ratios of our sample of SF galaxies in order to show how they nicely follow the tracks of the local SEDs. The solid black lines delineate the $\pm 2\sigma$ band centred on $\langle q_{\rm 24,\,template}\rangle$. The jumps at $z\sim$ 0.5 and 1 occur because the averaging of the IR templates is performed with a discrete and restricted set of IR templates. Since we merely use these boundaries to define the parameter space for the subsequent analysis the discontinuities are inconsequential.\\
The expression for the relative AGN abundance which accounts for censored measurements and the use of discrete probability bins is (see derivation in Appendix \ref{appsect:cohabderiv})
\begin{equation}
{\rm f_{AGN}}(q, z) = \frac{\sum_{i=1}^{n}\,N_{i,\, {\rm eff.}}\times \langle {\rm 1-Pr\,(SF)}\rangle_i}{\sum_{i=1}^{n}\,N_{i,\, {\rm eff.}}\times \langle {\rm Pr\,(SF)}\rangle_i}~.
\label{eq:effagnfrac}
\end{equation}
Here the summation with respect to $i$ extends over a finite number $n$ of probability bins. In the case of $q_{24}$ we grouped sources into bins of width $\Delta$Pr\,(SF) = 0.1 in order to have a sufficient number of measurements, and thus to ensure a well-behaved estimate of the distribution function $f(q)$ (computed according to equations (\ref{eq:schmittprob}) and (\ref{eq:schmittcoupled})) in each probability bin. 

\noindent In the lower panel of Fig. \ref{fig:q24AGNfrac} we present the function ${\rm f^{24\,\mu m}_{AGN}}(q, z)$ in four redshift bins covering $z\lesssim$ 1.4. The zero-point of the $x$-axis has been renormalized to the average $q_{24}$ of the IR templates at the centre of the redshift slice. A value of 2 (0.5) on the $y$-axis implies that at a given value of $q_{24}$ the relative abundance of AGN and SF systems is 2:1 (1:2). Within the errors ${\rm f^{24\,\mu m}_{AGN}}(q, z)$ is consistent with being unity across the whole width of the star forming locus at all redshifts. Deviations from the generally smooth variations of ${\rm f^{24\,\mu m}_{AGN}}(q, z)$ can occur on the edge of the assessed range of $q_{24}$ due to fluctuations caused by poor statistics. There is weak evidence for a gradual decrease of the AGN fraction from about $\nicefrac{2}{3}$ to roughly $\nicefrac{1}{3}$ as one goes from the region which hosts sources with radio-excess to that populated by sources with excess IR emission. This trend is barely significant but interestingly enough it tilts in the opposite direction as would be expected if, e.g., AGN activity were to manifest itself by exciting increased hot dust emission in the MIR. (Note that in general the radio emission could also be altered by the presence of an AGN, thus making the observed slope less easily interpretable. However, the fact that the distribution of ${\rm f^{70\,\mu m}_{AGN}}$ -- which has a radio contribution that is identical to that in ${\rm f^{24\,\mu m}_{AGN}}$ -- is essentially flat, suggests that the radio emission is not strongly affected by the AGN.)\\
The calculation of ${\rm f^{70\,\mu m}_{AGN}}(q, z)$ involved slightly wider probability bins of width $\nicefrac{1}{8}$ (to ensure convergence of the distribution function) and was limited to $z\lesssim$ 1.1 due to the ubiquity of 70\,$\mu$m non-detections at higher redshift (cf. upper panel of Fig. \ref{fig:q70AGNfrac}). ${\rm f^{70\,\mu m}_{AGN}}(q, z)$ appears to be a constant function of $q_{\rm 24,\,obs}$ with no traces of being tilted as detected with marginal significance for ${\rm f^{24\,\mu m}_{AGN}}(q, z)$, except maybe in the redshift bin $z\in$ [0.72, 1.06]. Overall, we can thus deduce that our optically selected AGN and SFGs occupy the SF locus in very similar proportions. A possible explanation for this is that both the IR and radio emission are predominantly powered by star formation rather than AGN activity. It is also conceivable, however, that other (combinations of) astrophysical processes conspire to place AGN hosts close to the IR-radio relation \citep[e.g.][]{sanders89, colina95}.

\subsection{Variations of IR/Radio Ratios with Luminosity}
\label{sect:qvslum}
In a recent work on local IR galaxies \cite{rieke09} have found evidence of variations in the $K$-corrected average 24\,$\mu$m/1.4\,GHz flux ratio $q_{24,\,0}$ with IR luminosity $L_{\rm TIR}$. According to their analysis $q_{24,\,0}$ is a constant function of luminosity at $L_{\rm TIR}\leq 10^{11}\,L_{\sun}$ and then begins to rise with increasing luminosity. Using the NVSS- and IRAS-detected SDSS galaxies, Mori{\'c} et al. (in prep.) see an opposite trend of decreasing FIR/radio ratio when they examine $q_{\rm FIR}$ vs. $L_{\rm 1.4\,GHz}$ for various types of active galaxies (both star forming and AGN-bearing).

\noindent We investigate whether or not the star forming sources in our sample show any evidence of variations of $q_{\rm TIR}$ with IR or radio luminosity. Since our $K$-corrected monochromatic IR/radio ratios are based on the best-fitting TIR SEDs, all luminosity-dependent trends they display will be qualitatively identical to those measured for $q_{\rm TIR}$. Comparisons with previous studies are therefore possible even if these used a different IR/radio parameter.\\
Note that the fact that we are plotting $q_{\rm TIR}$ against luminosity implies that upper and lower limits cannot always be unambiguously placed along the ordinate. An example are the radio-selected sources in the upper panel of Fig. \ref{fig:SFqTIR_vs_lum} of which we merely know that they must lie to the lower left of their limits. They are indicated by an arrow pointing diagonally downward. The calculation of the median $q_{TIR}$ in a given bin of luminosity should correct for measurements that in truth belong to a fainter luminosity range. To account for this we construct broad luminosity bins ($\Delta {\rm log}(L) \approx$ 1) and assume that most of the censored measurements would come to lie in the next lower luminosity bin (anything fainter would imply that they are more than 3\,$\sigma$ outliers to the IR-radio relation). We can then `average' away the effect of falsely assigned measurements by ({\it i}) computing the median $\langle q_{\rm TIR}\rangle$ in two sets of luminosity bins which are offset by half a bin width and then ({\it ii}) averaging the two estimates of the median thus obtained and reporting the new value half way between the centres of the two involved bins along the luminosity axis. The medians themselves are calculated by applying survival analysis to the jointly IR- and radio-selected data as previously done in \S\S\,\ref{sect:monochromqs}-\ref{sect:agnfract}.

\noindent The results of this procedure are shown in the larger two windows in Fig. \ref{fig:SFqTIR_vs_lum}. Using the COSMOS data we see no evidence of an increase in the IR/radio ratio at $L_{\rm TIR} \sim 10^{11}\,L_{\sun}$ as suggested by \cite{rieke09}. We do detect a higher value of $\langle q_{\rm TIR}\rangle$ in the brightest IR luminosity bin but this increase happens around $L_{\rm TIR} \sim 10^{13}\,L_{\sun}$, similar to the results of \cite{younger09}. It should be mentioned, however, that the methodology used by \cite{rieke09} to derive $q_{24,\,0}$ differs significantly from the one used here in that it involves -- for example -- luminosity-dependent (and template-based) conversions of {\it IRAS} 25\,$\mu$m flux densities to 24\,$\mu$m {\it MIPS} equivalent values.\\
While no universal trend for variations of $q$ with IR luminosity are detected in our sample we do find that $q_{\rm TIR}$ is a decreasing function of radio luminosity (see lower-most window in Fig. \ref{fig:SFqTIR_vs_lum}). The trend is consistent and increases rapidly at $L_{\rm 1.4\,GHz} \sim 10^{24}$\,W/Hz. This could potentially be the effect of contaminating AGN at high radio luminosities in our optically selected sample of SFGs. However, the fact that Mori{\'c} et al. (in prep.) see a similar trend in local SF, composite and AGN-bearing systems which have been classified based on the standard optical line emission ratios \citep{kauffmann03,kewley06} suggests that the trend is genuine.\\
The two narrower windows in Fig. \ref{fig:SFqTIR_vs_lum} show the variations of the dispersion of $q_{\rm TIR}$ with IR ({\it top}) and radio luminosity ({\it bottom}). In the low-redshift samples of \cite{yun01} and \cite{bressan02} an increase in scatter with infrared luminosity is detected. In the present data a similar -- albeit very weak -- tendency is seen; the reduced accuracy of the $L_{\rm TIR}$ measurements of the high-$z$ galaxies likely masks most of the trend if present. The plot of $\sigma_{q_{\rm TIR}}$ vs. $L_{\rm 1.4\,GHz}$, on the other hand shows a clear increase in the scatter which starts to manifest itself at the same radio luminosity at which the strong decline of $\langle q_{\rm TIR}\rangle$ sets in.

\subsection{The IR-Radio Relation at $z>$ 2.5}
\label{sect:highz}
While in the previous sections we usually tacitly plotted data points from high-$z$ sources the fitting of evolutionary trends in \S\S\,\ref{sect:monochromqs} and \ref{sect:TIRqs} was restricted to galaxies at $z<$ 1.4. This corresponds to the redshift at which the 4000\,{\AA} break leaves the reddest Subaru band with deep coverage \citep{taniguchi07}, the $z$-band. After $z\sim$ 1.4 the break is constrained by the NIR data of the $J$, $H$ and $K_s$ bands (McCracken et al. 2009, subm.; Capak et al., in prep.). These exposures of the COSMOS field, however, are two magnitudes shallower and have gaps between filters, leading to large uncertainties in the photometric redshift estimates. Beginning from about $z\sim$ 2.5 the Ly$\alpha$ (1215\,\AA) break enters the wavelength range covered by the ground based photometry \citep{capak07, taniguchi07}. As a consequence the accuracy of the photometric redshift improves to again $\sigma(\nicefrac{\Delta z}{(1+z)})\simeq$ 0.03.\\
In an assessment of ongoing spectroscopic follow-up observations of high-$z$ sources in the COSMOS field, Capak et al. (in prep.) find that photometric redshift estimates of genuine high-$z$ sources may be scattered to low redshift due to confusion between the Ly$\alpha$ and 4000\,{\AA} break. Most of the confusion is due to regions of the Ly$\alpha$ forest which are not as opaque as expected and/or light from nearby foreground galaxies contaminating the apertures. Conversely, there is little evidence for any upward scattering of galaxies at low and intermediate redshift to $z\gtrsim$ 2.5. This implies that sources with photometric redshift estimates $>$2.5 represent, with high likelihood, a clean -- albeit not complete -- sample of high-$z$ objects.

\noindent Our sample contains more than $\sim$140 sources at redshift $z>$ 2.5, of which approx. 60\% have direct detections at 1.4\,GHz and in at least one {\it MIPS} filter. As far as we are aware, this is the largest sample of high-$z$ sources so far, for which it is possible to study the IR-radio relation based on direct detections rather than flux limits. We must point out, however, that only 2\% of the high-$z$ sources have a direct detection at 24 {\it and} 70\,$\mu$m while the SEDs of the remaining 98\% are only constrained by a direct detection at 24\,$\mu$m and an upper flux limit at 70\,$\mu$m. Accordingly, the calculated values of $L_{\rm TIR}$ luminosities must be regarded as fairly rough estimates of the true IR luminosity of these sources as they are primarily based on measurements made at a rest frame wavelength of $\sim$6\,$\mu$m. \cite{murphy09b} caution that the IR luminosities of high luminosity and high redshift sources ($\nicefrac{L_{\rm TIR}}{L_{\sun}} > 10^{12.5}$; $z>$ 1.4) are generally overestimated by a factor of $\sim$4 even after subtraction of a flux contribution from AGN. However, in view of the COSMOS study of Kartaltepe et al. (subm.) -- who, in the same range of IR luminosities, do not see this trend and instead report that IR luminosities based solely on 24\,$\mu$m data tend to be underestimated in general -- we refrained from applying any corrections to our data.\\
Bearing in mind these uncertainties we plot the TIR/radio ratios of our high-$z$ sources in Fig. \ref{fig:SFqTIR_medevohighz} (left panel). For illustrative purposes the measurements of $q_{\rm TIR}$ are coloured according to their probability Pr\,(SF). We caution, however, that this classification is based on the fiducial ($u-K$) cut used throughout the paper so far and that the evidence presented in Fig. \ref{fig:probcol_distr} indicates that this threshold is no longer appropriate at $z\gtrsim$ 3. In view of this we do not distinguish between star forming systems and AGN for the high-$z$ sources but use this global sample to derive the average IR/radio flux ratio. The right hand side shows the distribution function of $q_{\rm TIR}$ which is broad ($\sigma=0.59\pm$0.05) and has a median of 2.71$^{+0.09}_{-0.14}$. This value is in good agreement with the local measurement of \cite{bell03} (dashed line) and is almost identical to the average value of 2.76$^{+0.02}_{-0.08}$ we find for the COSMOS data in our lowest redshift bin in \S\,\ref{sect:TIRqs}. The average IR/radio properties of our high redshift sample -- the most distant sources of which are detected when the universe was only $\sim$1.5 Gyrs old --  are thus very similar to those observed in the local universe. It is important to remember, however, that at $z\geq$ 2.5 the COSMOS data contains mostly extremely IR-luminous HyLIRGS ($L_{\rm TIR}> 10^{13}\,L_{\sun}$) which are a very different kind of object than those encountered at $z<$ 0.5 where the majority of our sources have $10^{11} < \nicefrac{L_{\rm TIR}}{L_{\sun}} < 10^{12}$ (cf. Fig. \ref{fig:fluxlims}).

\section{Discussion}
\label{sect:discussion}

Various parameterizations of the IR-radio relation exist. The flux ratios $q_{\rm TIR/FIR}$ and $q_{70}$ predominantly reflect the IR and radio emission of the ISM which is caused by two stages in the life cycle of massive stars; ({\it i}) the main sequence phase during which UV light is converted into FIR emission by dust grains and ({\it ii}) supernovae explosions inducing synchrotron emission when their shock waves accelerate cosmic ray electrons in the galactic magnetic field. The parameter $q_{24}$, on the other hand, is more sensitive to hot dust emission triggered by AGN activity. Several recent papers \citep[e.g.][]{garrett02, appleton04, frayer06, ibar08, younger09, murphy09b} using 1.4\,GHz data provide consistent evidence that the local IR-radio relation holds out to high redshift. An identical conclusion has been reached using radio flux density measurements at 610\,MHz rather than 1.4\,GHz \citep{garn09b}.\\
A majority of the samples previously used to study the IR-radio relation in distant galaxies contain only several dozen to roughly a hundred galaxies at $z\lesssim$ 0.5. The COSMOS sample used in the present analysis increases the number of sources at redshifts $z\gtrsim$ 0.5 by at least a factor of 5. Also, it is probably the first data set in which the number of directly measured IR/radio ratios at high redshift is larger than the number of censored values.\\
We have studied the IR/radio properties of both SFGs and AGN. Various complementary indicators of AGN activity are present in the COSMOS data base. Rather than identifying AGN based on a combination of multiple parameters we have chosen to work with a single statistical criterion based on the work by \cite{smolcic08} which classifies radio- and IR-detected sources as SF or AGN based on their rest frame optical colour. This statistical approach allows us to treat our sources in a probabilistic way which is especially advantageous for the assessment of systematics inherent in the selection of SFGs and AGN.

\subsection{The IR-Radio Relation at Intermediate and High Redshift}

At $z<$ 1.4 where both photometric and spectroscopic redshifts have a high accuracy we found no compelling evidence of an evolving IR-radio relation. The mean IR/radio flux ratios in this redshift range are $\overline{q}_{24,\,0}=1.26\pm0.13$, $\overline{q}_{70,\,0}=2.23\pm0.13$ and $\overline{q}_{\rm TIR}=2.57\pm0.13$, where the first two are computed using $K$-corrected IR and radio flux densities. With the exception of the highest redshift bin, the median IR/radio ratios in the different redshift bins covering the range $z\in$ [0, 1.4[ are offset from the averages $\overline{q}$ by at most half the dispersion in the local IR-radio relation. An alternative to testing the constancy of rest frame IR/radio flux ratios derived using IR SEDs of local SFGs, is to form IR/radio ratios with observed flux densities (cf. \ref{sect:obsmonochrom}). Variations with redshift of $q_{\rm 24,\,obs}$ and $q_{\rm 70,\,obs}$ in the SF population can then be compared with the changes that would be expected for redshifted local galaxies \citep[e.g.][]{ibar08, seymour09} in order to assess if the IR/radio properties of the latter are compatible with those of high-$z$ systems. Our analysis has shown that the decline with redshift of observed and predicted IR/radio ratios are indeed in good agreement.\\
Since measurements of $q_{24}$ and $q_{70}$ have been carried out with flux-limited data sets which likely are subject to some selection band-related bias (see \S\S\,\ref{sect:selbias} and \ref{sect:biasconclu}) only our value of $q_{\rm TIR}$ lends itself to a straightforward comparison with local measurements. \cite{bell03} find $q_{\rm TIR}= 2.64\pm0.02$ for a sample of local SFGs and show that this figure is in excellent agreement with the {\it IRAS}-based FIR measurement $q_{\rm FIR} = 2.34\pm0.01$ of \cite{yun01}. The COSMOS measurements of the mean IR/radio properties of high redshift galaxies thus are fully consistent with the local average. Furthermore, a subsample of highly redshifted ($z>$ 2.5) COSMOS galaxies has a median 2.71$^{+0.09}_{-0.14}$ which also agrees well with both the COSMOS data at low redshift and independent local measurements.\\

\noindent Due to Malmquist bias we detect systems with very differing star formation rates (SFR) in the low and high redshift universe where only extreme starbursts of the HyLIRG class (SFR $\gtrsim 10^3 M_{\odot}$\,yr$^{-1}$) are visible. That starburst galaxies which we observe when the universe was just 10\% of its current age follow the same IR-radio correlation as local galaxies runs counter to expectation which would predict higher-than-average IR/radio ratios for such systems. \cite{lisenfeld96} have shown that a strong (and prompt) enhancement of the magnetic field strength is required lest a deficit of synchrotron emission develop due to high inverse Compton losses of the cosmic ray electrons in the strong radiation field generated by the starburst.\\
On the other hand, it could be that precisely this inability to detect sources with SFRs of `normal' (disk) galaxies at higher redshift is hiding changes in the IR/radio ratios of the SF population. In particular, such changes are expected as soon as the energy density of the cosmic microwave background (CMB) exceeds that of the galactic magnetic fields, whereupon inverse Compton losses off CMB photons begin to dominate synchrotron emission. Based on the typical magnetic field strengths in spiral arms (a few $\mu$G), \cite{carilli08} estimate that this could be the case for regular disks as early as $z\sim$ 0.5.\\
Regardless of the remaining uncertainties surrounding the maintenance of the local IR-radio relation out to high redshift the observational fact {\it per se} is an important confirmation of the central assumption used in studies that have computed the SF history of the universe \citep{haarsma00, seymour08, dunne09, smolcic09a} using deep radio surveys, namely that of an unchanging proportionality between SFR and radio luminosity. The constancy of the IR-radio relation not only implies that IR and radio measurements are equally good tracers of star formation out to high redshift. It also suggests that the physical processes of massive star formation when the universe was only 1-2 Gyrs old used to be strikingly similar to those at play in the local universe.\\
Currently available observations are not yet capable of revealing the exact workings of the astrophysical mechanisms that cause galaxies at intermediate and high redshift to lie on the IR-radio relation. The improvements in sensitivity and spatial resolution with {\it Herschel} and the extended Very Large Array (EVLA) will be a crucial step forward in this respect. A better sampling of the FIR emission from distant (U)LIRGs will reveal if their SEDs indeed match those of their local analogues as this study and previous work generally assume. Similarly, a systematic study of the range of radio spectral indices in SFGs is important to quantify the inaccuracies that are introduced by assuming a single generic value. Clarifying both these issues is essential if, e.g., trends for a luminosity dependence of $\overline{q}$ \citep[see][and also our Fig. \ref{fig:SFqTIR_vs_lum}]{rieke09} are to be put on a solid basis.

\noindent Apart from SFGs, the IR-radio relation is also observed by many AGN-bearing and/or composite systems \citep[e.g][]{sanders89, marx94, sopp91, roy98, murphy09b, seymour09}, albeit with a larger dispersion \citep{condon82, obric06, mauchsadler07}. In the COSMOS sample we also observe that AGN and SFGs often have very similar values of $q_{24}$ and $q_{70}$. This is true not only for optically-selected AGN but also for X-ray detected sources, implying that the phenomenon is not merely an artifact of our statistical colour criterion for the separation between AGN and SFGs. It should be emphasized that the finding is {\it not} an artificial consequence of template fitting; the abundance fraction of AGN and SFGs has been derived using observed IR/radio ratios and hence does not involve any assumptions about the value of the radio spectral index or the shape of the IR SED.\\
We find that our optically-selected AGN and SFGs populate the locus of the correlation in nearly equal proportions out to at least $z\sim$ 1. It is important to bear this in mind when IR/radio ratios are used -- possibly in combination with other indicators -- to distinguish between AGN and SF sources \citep[e.g.][]{donley05, park08, seymour08}: radio-excess outliers can only be used to single out radio-loud AGN rather than a complete AGN sample and, conversely, selecting only objects that follow the correlation will result in a mixture of radio-quiet AGN and SFGs rather than a pure SF sample. The fact that an AGN is present does not exclude coeval star formation in the host galaxy \citep[e.g.][]{silverman09}. However, whether or not it is responsible for the similar in IR/radio properties of SFGs and AGN is still debated, with both supporting and contesting evidence being advanced \citep[e.g. in the case of PG QSOs: see][]{sanders89, barthel06}.

\subsection{Biases Revisited}
\label{sect:biasconclu}
Even though the current data set has been selected both in the IR and radio, the fact that the average IR/radio ratios of the jointly selected sample are usually close to those of the IR-selected sample suggests that the average values $\langle q\rangle$ we quote are not the `intrinsic' value one would hope to find in an entirely unbiased sample. By separately studying an IR- and radio-selected sample of SFGs, however, it is at least possible to bracket the unbiased average IR/radio properties. As expected the jointly selected sample always lies within this region (with the exception of one case -- see Table \ref{tab:qTIRinfo} -- where the amount by which the median of the jointly selected sample exceeds that of the IR-selected sample is still smaller than the width of the bins used for the calculation of the distribution functions of $q$ in the doubly censored data set).

\noindent The shift between the average IR/radio properties of an IR- and a radio-selected sample are in principle predictable based on the dispersion of the IR-radio relation and the slope of the differential source counts (cf. equation (\ref{eq:qbias})). While comparable to the locally measured dispersion at intermediate redshift, the scatter of the relation in our COSMOS sample is significant at $z>$ 1 where it is twice as large as at low redshift. Part of the increase could be due to an intrinsically higher scatter $\sigma_q$ at high luminosities \citep{yun01, bressan02} but probably is also a consequence of the limitations that are inherent in the calculation of IR luminosities when the IR SED is constrained by few points\footnote{Note, however, that in this respect the current data still represents an improvement over previous survey-scale samples.}. Under the simplified assumption of Euclidian source counts ($\beta =$ 2.5) equation (\ref{eq:qbias}) predicts the offset $\Delta q_{\rm bias}\approx$ 0.35 dex, found in our lowest redshift bins. It also makes a fair prediction of a shift of $\sim$0.7 dex between the IR- and radio-selected sample at $z\sim$ 1 if one accounts for the larger scatter and the finding that at faint fluxes IR \citep{chary04, papovich04} and radio \citep{richards00, fomalont06, bondi08} source counts are sub-Euclidean\footnote{The larger measured scatter $\sigma_q$ tends to increase $\Delta q_{\rm bias}$ while the sub-Euclidean counts (with $\beta\sim$ 1.5; see references in text) have the opposite effect (with respect to a Euclidean slope). Consider the different measurements of $\langle q_{\rm TIR}\rangle$ at $z\in$ [1.14, 1.51[ (see third row from bottom in Table \ref{tab:qTIRinfo}) as an illustration that the combination of the two factors leads to a prediction of $\Delta q_{\rm bias}$ which agrees excellently with the data. Using equation (\ref{eq:qbias}) with $\beta=$ 1.5 and $\sigma_q=$ 0.78 one finds $\Delta q_{\rm bias}=$ 0.70. The observed shift between the median $\langle q_{\rm TIR}\rangle$ in the IR- and radio-selected sample, on the other hand, is $\sim$0.67.}.\\
Figs. \ref{fig:SFcorrq24_medevo}, \ref{fig:SFcorrq70_medevo} and \ref{fig:SFqTIR_medevo} show that apart from biasing the average IR/radio ratio, selection effects can also produce spurious evolution. Based on the radio-selected sample alone we would infer a decrease of the mean $\langle q\rangle$ out to $z\sim$ 1 (see also numerical values in Tables \ref{tab:q24info}, \ref{tab:q70info} and \ref{tab:qTIRinfo}). As the only recent study, \cite{seymour09} measure an average value of $q_{70}$ for their radio-selected sample that is reduced by 0.25 dex at $z\sim$ 1 with respect to low redshift. This is close to the evolution $\nicefrac{d\langle q_{70}\rangle}{dz}\approx$ 0.2-0.3 dex which we see for the radio-selected COSMOS sample. Based on the argument outlined above a simple explanation for this could be selection effects. However, as \cite{seymour09} adopt the different approach of  stacking radio sources that are not detected at 70\,$\mu$m rather than including them in the analysis in the form of flux limits as we have done, other explanations cannot be ruled out.

\noindent As stated in \S\,\ref{sect:selbias} there is ample evidence from the results summarized in Table \ref{tab:prevwork} that the offset predicted by equation (\ref{eq:qbias}) not only occurs in our data but also can reconcile most apparently discrepant measurements of mean IR/radio ratios in the literature. The one exception to this generally encouraging agreement are the highly inconsistent radio stacking results of \cite{boyle07}, \cite{beswick08} and \cite{garn09a} who have all studied the mean $q_{24}$ as a function of IR flux. \cite{garn09a} in particular pointed out that the field-to-field variation of the mean IR/radio ratio can be considerable. The prospects are good that the issue will soon be resolved with the aid of significantly deeper EVLA observations at the $\sim$$\mu$Jy level that will even directly detect the radio emission of the faintest 24\,$\mu$m sources.

\section{Summary and Concluding Remarks}
\label{sect:conclusions}

On the preceding pages we have discussed the IR/radio properties of star forming galaxies (SFGs) and active galactic nuclei (AGN) in the redshift range 0 $<z<$ 5. Our analysis has benefitted from the extensive multi-wavelength coverage of the COSMOS field: each of our sources has a flux constraint at 24\,$\mu$m, 70\,$\mu$m and 1.4\,GHz from {\it Spitzer/MIPS} and VLA observations, multiwavelength photometry in $\sim$30 bands from the UV to the NIR, and in some cases X-ray detections. Roughly one in three of our sources has a spectroscopic, the rest an accurate photometric redshift measurement.\\
Our primary focus was the evolution of the IR-radio relation with cosmic time. With a total of approx. 4500 sources our sample is the largest one which has so far been used to study the IR-radio relation at intermediate and high redshift. This is particularly true at redshift $z>$ 2.5 where we have detected nearly 150 sources of which, for the first time, more objects have direct IR and radio detections rather than upper flux limits in one of the two bands. To our knowledge this is also the first time the evolution of both monochromatic and integrated IR/radio ratios has been consistently derived using objects from the same field. The computation of IR luminosities using complete SED template libraries represents a further improvement over previous work which has often relied on single starburst templates, e.g. that of M82.

\noindent The average IR/radio ratio is subject to selection biases in that it depends on the band in which a population is selected. We have shown that the average IR/radio properties of IR- and radio-selected samples of galaxies differ by an amount which is in agreement with theoretical expectations and that studies relying on objects selected in only one of the two bands run the risk of inferring spurious evolutionary trends. For this reason, we base our analysis on a sample jointly selected at IR {\it and} radio wavelengths in order to eliminate biases as best possible. Furthermore, we make frequent use of the methods of survival analysis which permit us to include all information carried by flux limits from sources that have failed to be detected at either IR or radio wavelengths.

\noindent To summarize our findings:
\begin{enumerate}
\renewcommand{\labelenumi}{(\roman{enumi})}
\item The median IR/radio ratios of SFGs show little variation in the redshift range $z<$ 1.4 and the IR-radio relation remains similarly tight out to $z\sim$ 1 as it is in the local universe (see \S\S\,\ref{sect:monochromqs} and \ref{sect:TIRqs}). Above $z\sim$ 1 the dispersion in the COSMOS population increases, probably due to the reduced accuracy of our measurements and/or an intrinsically larger dispersion of IR/radio ratios at high luminosities.
\item For a sample of sources with high confidence redshift estimates in the range $z>$ 2.5, the average IR/radio ratio is still the same as that found in the local universe (see \S\,\ref{sect:highz}). 
\item At both 24 and 70\,$\mu$m many of our optically-selected AGN have similar IR/radio ratios as SFGs (see \S\,\ref{sect:agnfract}). The relative abundance of AGN and SFGs in our sample is about 1:1. The same applies to X-ray detected AGN, of which a large fraction has IR/radio ratios which lie in the range measured for star forming systems.
\item The median IR/radio ratios of SFGs consistently decrease as a function of radio luminosity. On the other hand, they remain constant over three orders of magnitude in IR luminosity. Only at the highest IR luminosities ($L_{\rm TIR}\gtrsim 10^{13}L_{\odot}$) has a tendency for an increase been detected (see \S\,\ref{sect:qvslum}).
\item Apparently discrepant measurements of the average IR/radio ratio reported in the literature can be reconciled if one properly accounts for the selection band of the respective samples (see \S\,\ref{sect:biasconclu}).
\end{enumerate}
\renewcommand{\labelenumi}{\arabic{enumi}.}

\noindent In this work we have applied a statistically more sophisticated treatment of IR/radio ratios than has previously been adopted. It accounts for both detection limits and selection biases inherent in the data, and our results provide firm support for previous reports that the IR-radio relation remains unchanged out to high redshift ($z\gtrsim$ 4). The observed constancy is striking evidence that the interplay between the life cycle of massive stars and the interstellar matter (ISM) has followed a very similar pattern for more than 10 billion years. IR and radio measurements apparently represent equally good tracers of star formation over much of the history of the universe. This has been a critical assumption underlying measurements of the cosmic star formation history with deep radio surveys, and fundamental to the estimation of redshifts for optically undetected sub-mm galaxies.\\
The upcoming generation of IR and radio observatories like {\it Herschel} and the Extended Very Large Array (EVLA) are expected to be able to perform measurements of sufficient sensitivity and spatial resolution to provide clues on the mechanism shaping the IR-radio relation in distant star forming galaxies with moderate star forming rates. This will be a major advance over the mere observation of the phenomenon as is presently the case.

\clearpage
\appendix

\section{Data Description}
\label{appsect:data}

In the following, additional information on our data sets is provided. Since the determination of source fluxes and positions is central to the accuracy of our IR/radio ratios and the band-merging between the observations at different wavelengths, we describe the derivation of these quantities in particular detail.

\subsection{VLA-COSMOS Radio Observations}
\label{appsect:radiodata}

Following the completion of the VLA-COSMOS Large Project \citep{vlacos2} which had mapped the entire 2 deg$^2$ COSMOS field at 1.4\,GHz with the VLA in antenna configurations A and C for 275 hours, the central 0.84 square degrees were re-imaged during an additional 62 hours in configuration A in the spring of 2006. The resulting VLA-COSMOS `Deep Project' mosaic has a resolution (FWHM of synthesized beam) of 2.5$''\times$ 2.5$''$ and a pixel scale of 0.35$''$/px. The mean $rms$ sensitivity is $\lesssim$0.01\,mJy/beam at the centre of the field and more than 60\% (80\%) of the field has an $rms$ level better than 0.02 (0.03)\,mJy (see Fig. 3 in Schinnerer et al. 2009, subm.). The data reduction was carried out with AIPS \citep[Astronomical Image Processing System;][]{greisen03} and followed standard procedures described in \cite{vlacos2} and (2009, subm.).

\noindent The AIPS source/component finding task \texttt{SAD} was used to detect radio sources in the Deep Project mosaic in multiple iterations with cut-off levels of successively lower $S/N$. The resulting source list was then combined with the VLA-Large Project catalog \citep{vlacos2} to create a list of 1.4\,GHz sources (henceforth referred to as the `{\it Joint catalog}') that are at least of 5\,$\sigma$ above the local background. Of the $\sim$2900 sources in the Joint catalog 51\% are found to be unresolved. In these cases the integrated flux density is set equal to the peak flux density of the sources which was measured with the AIPS task \texttt{MAXFIT}. In order to correct for bandwidth smearing a position-dependent correction based on a model sensitivity map of the Deep Project mosaic has been applied to the peak flux values. Integrated flux densities \citep[which are not affected by bandwidth smearing; e.g.][]{bondi08} for resolved radio sources were determined by integrating over the size of the best-fitting elliptical Gaussian component with the task \texttt{JMFIT}. For another 131 (4.6\%) of the sources in the catalog which were best fit by the sum of more than one Gaussian flux component, the task \texttt{TVSTAT} was used to measure the flux within a manually defined contour around the source. The distribution of 1.4\,GHz flux density values in the Joint catalog is shown in Fig. 15 of Schinnerer et al. (2009) where a detailed description of the construction of the catalog is given. The flux measurements carried out on the VLA-COSMOS radio maps have been compared to those of about 300 sources also observed at 1.4\,GHz in the context of the NVSS and FIRST surveys \citep{condon98, white97}. The agreement between the different data sets is reasonable except for a number of NVSS sources where the VLA-COSMOS observations have probably resolved out a large extended flux component \citep[cf.][]{vlacos2}.\\
For the multiple component sources which often have a complicated radio morphology due to outflows or continuum emission from star formation activity we set the most likely source position by visual inspection of HST ACS images \citep{koekemoer07}. In all other cases the source position is located at the peak of the radio emission as determined by \texttt{MAXFIT}. This definition was also adopted for a small fraction of multiple component sources that could not be reliably associated with an optical source. The typical accuracy of the radio positions is 0.13$''$ as shown by \cite{vlacos2}.

\noindent While the Joint catalog serves as the basis of our radio-selected sample (introduced in \S\,\ref{sect:radiosamp}) we will also use the {\it MIPS} 24\,$\mu$m catalog described in \S\,\ref{appsect:24umdata} to construct a sample of infrared-selected sources. In addition to counterparts with $S/N\geq$ 5 from the Joint catalog we also allow for sources having $S/N>$ 3 in the presence of IR detections. To find additional sources in this $S/N$ range we ran \texttt{MAXFIT} at the position of IR detections without a counterpart in the Joint catalog. The \texttt{MAXFIT} box size is chosen in accordance with the uncertainty in the position of the IR source (see Appendix \ref{appsect:bandmerge}). Since the vast majority of sources at these low detection levels are likely unresolved (radio sources with flux density $\lesssim$0.1\,mJy typically have sub-arcsecond sizes, see \cite{bondi08} and references therein) their integrated flux is equated to the peak flux density.\\
If the extracted peak flux density has a significance level of less than $3\,\sigma$ or if \texttt{MAXFIT} fails to converge, upper limits on the total flux from the local sky background are used as flux constraints for a potential source at that position. The noise in the Deep Project image was estimated with the AIPS task \texttt{RMSD} in a box of dimensions 105$''\times$105$''$ that was moved across the survey area in steps of 2.45$''$ in R.A. and Dec.. We adopt an upper flux limit of 3 times the local $rms$ noise level for the rest of the analysis for undetected radio sources. The numeric value of this upper bound is obtained by reading the noise image at the pertinent position.

\subsection{S-COSMOS Infrared Observations}
\label{appsect:IRdata}

\subsubsection{MIPS 24\,$\mu$m Data}
\label{appsect:24umdata}
During {\it MIPS} Cycle-2 and Cycle-3 the S-COSMOS project \citep{sanders07} imaged the whole COSMOS field at 24\,$\mu$m in medium and slow scan mode, respectively. The data from both {\it MIPS} cycles was coadded and combined with the MOPEX package (\citealp[MOsaicker and Point source EXtractor; ][]{makovoz05a}) after elimination of transient sources such as asteroid. The resulting mosaic has a pixel scale of 1.2$''$/px, while the FWHM of the {\it MIPS} 24\,$\mu$m PSF is 5.8$''$. About 90\% of the area was mapped with a median integration time of $\sim$3400 seconds resulting in a $1\,\sigma$ sensitivity of $\sim$0.018\,mJy. Over the remaining $\sim$10\% of the field the average integration time was roughly 7000 seconds, leading to an equivalent depth of $1\,\sigma\sim13~\mu$Jy. A noise map for the S-COSMOS 24\,$\mu$m observations was generated using the associated coverage map. The details of the {\it MIPS} 24\,$\mu$m data reduction and source extraction procedures are spelled out in \cite{lefloch09}. In the following we summarize the points that are most relevant to our work with respect to the 24\,$\mu$m source catalog.

\noindent Source detection in the 24\,$\mu$m map was carried out with SExtractor \citep{bertin96} which returned positions that served as input to the PSF-fitting algorithm DAOPHOT \citep{stetson87}. DAOPHOT performs simultaneous PSF fitting to multiple sources which is an important advantage for the deep and crowded {\it MIPS} 24\,$\mu$m images where objects may be blended. By inserting and re-extracting artificial sources in exactly the same manner as true astronomical sources, \cite{lefloch09} found that the approach of PSF fitting is reliable down to a flux density of $\sim$60$~\mu$Jy which in most regions of the mosaic corresponds to a $S/N$ of about 3. At this detection threshold there are $\sim$50,000 sources in the area covered by the 1.4\,GHz observations. Given the resolution of $\sim$6$''$ of the {\it MIPS} images most of these are not resolved. As described in \cite{lefloch09}, the fidelity of the S-COSMOS 24\,$\mu$m flux calibration was ascertained by checking that the $K_s-[24]$ colours for stars in the COSMOS field listed in the catalog of the 2MASS survey \citep{jarrett00} did not deviate from the expected relation. Note also that total flux measurements at 24\,$\mu$m account for all the flux in the extended wings of the PSF. The astrometry of the {\it MIPS} data is usually accurate to within a few tenths of an arcsecond which was confirmed by the cross-correlation with the 2MASS catalog, which revealed only a small systematic offset of 0.3$''$ in declination that was subsequently corrected in the catalog of 24\,$\mu$m sources.

\noindent Since the 24\,$\mu$m catalog extends to a lower detection threshold of $\sim$3\,$\sigma$ we do not attempt to extract extra sources in the vicinity of radio detections. If no 24\,$\mu$m counterpart to a radio source is found during the band-merging (see Appendix \ref{appsect:bandmerge}) we directly assign an upper $3\,\sigma$ flux limit based on the value of the noise map at the according position. The conversion from surface brightness noise (as specified by the noise image) and point source detection noise (as required for a flux limit) was derived by scaling the median value of all pixels with an exposure time of $\sim$3400 seconds in the noise image to the average $1\,\sigma$ sensitivity level of $\sim$0.018\,mJy. The adopted value of the conversion factor is 2.3\,mJy\,(MJy/sr)$^{-1}$.

\subsubsection{{\it MIPS} 70\,$\mu$m Data}
\label{appsect:70umdata}
MIPS 70\,$\mu$m observations of the COSMOS field were carried out in parallel with the 24\,$\mu$m imaging using the scan mapping mode \citep{frayer09}. The FWHM of the data is 18.6$''$ and the pixel scale 4$''$/px.  As in the case of the 24\,$\mu$m survey a limited area has a high coverage by repeated observations with a total exposure time of $\sim$2800 seconds. The median effective exposure time over the remaining 90\% of the survey area is 1350 seconds leading to an average point source noise ($1\,\sigma$) of 1.7 mJy. The 70\,$\mu$m observations were reduced with the SSC pipeline tools GeRT and MOPEX according to standard procedures for {\it MIPS}-Ge survey data. Special attention was given to the creation of an accurate noise image which represents the uncertainty owing to both small scale (scatter in repeated observations of each sky pixel) and large spatial scale noise properties (i.e. pixel-to-pixel correlated noise and confusion noise). All steps leading to the final data products are described in \cite{frayer09}.

\noindent Source detection and extraction was performed within the MOPEX package \citep{makovoz05b} using the APEX (Astronomical Point source EXtraction) peak algorithm, augmented by additional specialized scripts. The noise image was employed to detect and then fit peaks with $S/N>$ 3 using the point source response function (PRF). The final source list has been cleaned of spurious detections along the first Airy ring of the bright 70\,$\mu$m sources. We checked all objects down to the flux level at which it would take a $>$2.5\,$\sigma$ noise fluctuation in order for a spurious source to be flux-boosted to $S/N \geq$ 3 in the first Airy ring (the amplitude of which is about 2\% of the peak).\\
The final list of 70\,$\mu$m sources contains almost 3000 entries in the region of the S-COSMOS survey that overlaps with the VLA-COSMOS area. Flux measurements in the catalog have been corrected for the additional flux outside the PRF image (i.e. beyond the first Airy ring) using Spitzer Tiny Tim models \citep{krist02} and placed on a constant $\nu S_{\nu}$ scale. This colour-correction is accurate within 2\% for a wide range of galaxy SEDs across the filter bandpass. The calibration of the 70\,$\mu$m data was confirmed to be consistent with the official MIPS calibration \citep{frayer09} which is accurate to 7\% \citep{gordon07}. Finally, a small positional offset in declination was found in the comparison of radio and 70\,$\mu$m source positions. However, at $\sim$0.2$''$ it is significantly smaller than the scatter measured for the positions of individual sources which amounts to about 2$''$.

\noindent To provide upper $3\,\sigma$ flux limits at 70\,$\mu$m whenever the band-merging with the radio or 24\,$\mu$m fails to identify a 70\,$\mu$m counterpart we convert the local surface brightness noise estimate given in the noise image to a point source noise. After accounting for the flux in the extended wings of the PSF the corresponding conversion factor is 14.9 mJy (MJy/sr)$^{-1}$.

\subsection{Multi-Wavelength Photometry}
\label{appsect:photomcat}

The COSMOS photometry catalog is an $i$-band selected catalog with PSF-matched photometry (FWHM = 0.6$''$) from 30 broad, medium and narrow band filters extracted at the positions of the Subaru $i^+$ band detections. The wavelength range covered by these observations extends all the way from the UV at 1550 \AA\ to the MIR at 8 $\mu$m. \cite{capak07, capak08} provide a complete description of the observations and data reduction leading to the compilation of the multi-wavelength data set. Here we use the photometry catalog compiled by \cite{ilbert09a} which lists more than 600,000 COSMOS galaxies with $i^+\lesssim$ 26 detected in a region roughly contiguous with the area covered by the VLA-COSMOS observations.

\subsection{X-Ray Observations}
\label{appsect:xmm}

The COSMOS field has been observed with {\it XMM-Newton} for a total of $\sim$1.5$\times10^6$\,seconds ($\sim$400 hours), resulting in a homogeneous depth of $\sim$5$\times10^4$\,sec \citep{hasinger07, cappelluti07, cappelluti09} over 1.92 deg$^2$. The associated catalog includes 1887 point-like sources chosen to have a high probability of being a reliable detection\footnote{The maximum likelihood threshold ensures that a catalogued XMM source has a probability of at most $\sim$5$\times10^{-5}$ of being a spurious detection.} in at least one of the soft (0.5-2\,keV), hard (2-10\,keV) or ultra-hard (5-10\,keV) bands down to limiting fluxes of $\sim$5$\times10^{-16}$, $\sim$3$\times10^{-15}$ and $\sim$5$\times10^{-15}$ erg\,cm$^{-2}$\,s$^{-1}$, respectively (see \cite{cappelluti09} for additional details). Moreover, the central part of the field has been observed with {\it Chandra} \citep{elvis09}, providing precise positions for about half of the XMM sources. Brusa et al. (in prep.) have determined optical counterparts to more than 90\% of the XMM catalog sources. In the present work we use only those with unique/secure optical counterparts and a successful fit to the UV to MIR SED \citep{salvato09}. The optical position has been used to associate our radio and IR sources to XMM detections using the same search radii as adopted for the band-merging of our radio and IR sources with the optical photometry (see Appendix \ref{appsect:bandmerge}).

\subsection{Redshift Information}
\label{appsect:redshifts}

The catalog of \cite{ilbert09a} contains photometric redshift estimates for all tabulated objects with the exception of (i) sources lying in a `masked' region of the optical imaging due to a bright star in their vicinity, or (ii) sources that are detected at X-ray wavelengths, in which case a special set of AGN-templates was employed to derive a photometric redshift \citep{salvato09}.\\
Spectroscopic data has already been gathered for more than 20,000 sources in the COSMOS field (\citealp[e.g.][]{lilly07}; \citealp{trump07}; \citealp{trump09}; \citealp{lilly09}; Kartaltepe et al., in prep.; Salvato et al., in prep). Whenever possible, we give precedence to redshift information from spectroscopy. In all other cases we assign a photometric redshift to the sources in our radio- and IR-selected samples. The details of the band-merging of the radio and IR catalogs with the optical data are presented in Appendix \ref{appsect:bandmerge} while the next two paragraphs summarize the properties and screening of the available redshift information itself.

\subsubsection{Spectroscopic Redshifts}
\label{appsect:specz}
At the time of writing the data base of reduced spectroscopic observations in the COSMOS field included more than 14,000 sources. Most of these (75\%) were observed by VIMOS in the context of the zCOSMOS survey, observations with DEIMOS and IMACS(3) account for another $\sim$12\% each and a small fraction of sources has LRIS, SDSS or FORS1 spectroscopy. In some cases spectra of the same source have been obtained by more than one of these instruments. If these measurements do not agree within experimental error we disregard the one with the lower quality flag. The quality of the redshift measurement for our spectroscopically observed optical counterparts is judged by the spectroscopic confidence classes\footnote{\texttt{http://archive.eso.org/cms/eso-data/data-packages/zcosmos-data-release-dr2/index\_html\#release\_notes}} of the zCOSMOS survey. In particular, we accepted only those objects with confidence flag 3 and 4 (regardless of the decimal place, see below), as well as 2.5, 2.4, 8.5, 9.3, 9.5 and 1.5\footnote{Also Type 1 quasars (i.e. broad line objects), to the flag values of which 10 has been added, and objects serendipitously covered by the slit of another spectroscopic target (the confidence flag of such objects are marked by a prepended `2') were admitted into the sample as their statistical reliability is equal to the classes listed in the body of the text.}. The decimal place indicates the level of agreement between the spectroscopic and photometric redshift measurements. Our choice of acceptable flag values ensures that the reliability of all measured spectroscopic redshifts is better than 99\% \citep{lilly09}.\\
The classification with decimal places was not available for some of the sources observed by follow-up campaigns other than zCOSMOS. In such cases we updated the spectroscopic confidence flag as required by comparison with the photo-z estimates of \cite{ilbert09a} and \cite{salvato09}. For spectroscopic targets with unsatisfactory confidence flags ($<$3\% of the spectroscopic targets) we checked whether a reliable photometric redshift estimate was available according to the selection criteria described in the next section and were thus able to recover redshift information for all but 4\% of these. In Table \ref{tab:zstats} we summarize the availability of redshift information for our sources.

\subsubsection{Photometric Redshifts}
\label{appsect:photoz}
Sources successfully matched to an entry in the COSMOS photometry catalog generally have a precise photometric redshift estimate derived with the code {\it Le Phare} \citep{ilbert09a}. The average number of bands available for the computation of photometric redshifts for the matched radio and 24\,$\mu$m sources was 29. The accuracy of the photometric redshifts was calibrated with the help of more than 4000 high-confidence redshifts from zCOSMOS \citep{lilly07, lilly09} sources with $i\leq$ 22.5 and verified using faint 24\,$\mu$m sources (Kartaltepe et al., in prep.) in the range $i>$ 22.5. At $z<$ 1.25 \cite{ilbert09a} found that the dispersion $\sigma(\nicefrac{\Delta z}{(1+z)})$ of photo-z measurements is 0.007, 0.013 and 0.051 in the Subaru $i^+_{\textrm{AB}}$-band magnitude ranges of $i^+_{\textrm{AB}} \leq$ 22.5, 22.5 $< i^+_{\textrm{AB}}\leq$ 24 and $i^+_{\textrm{AB}} >$ 24, respectively. Beyond redshift $z\sim$ 1.25 the statistical accuracy of the photometric redshifts abruptly decreases by a factor of $\sim$3 with respect to the range $z<$ 1.25 because the Balmer break is redshifted into the NIR filters which have less sensitive photometry and non-contiguous wavelength coverage. We use these values of the photo-z dispersion to remove sources in our sample with uncertain photometric redshifts; the criterion which each source must satisfy to remain in the sample is that its photo-z error be smaller than $2\,\sigma$ of the dispersion at a given magnitude and redshift. We also eliminate sources with a $\chi_{\textrm{phz}}^2$ above a threshold corresponding to two standard deviations in an ideal $\chi^2$-distribution (with the number of degrees of freedom equal to the difference between the number of filters used for deriving the redshift and the number of free parameters, namely three, i.e., the redshift, the template type and the template normalization). In practice very few sources ($<$0.5\% in the combined radio- and IR selected samples) are rejected due to the latter criterion. A somewhat larger number has been excluded due to a broad redshift probability distribution but they nevertheless represent only $\sim$1\% and $\sim$2\% of the radio- and IR-selected sample, respectively. The vast majority of sources without spectroscopically constrained redshift have a photometric redshift estimate within the statistically expected accuracy (46.4\% and 55.5\% in the radio- and IR-selected samples, respectively).\\
The template library used by \cite{ilbert09a} consists of star-forming and passive galaxy SEDs but includes none that reflect the features expected in spectra of sources with a dominant contribution from an AGN. \cite{salvato09} have computed photometric redshifts for all $XMM$-detected sources in the COSMOS photometry catalog using templates with varying AGN contributions. Their redshift measurements account for source variability and achieve an accuracy of $\sigma(\nicefrac{\Delta z}{(1+z)})<$ 0.015 at  $i^+_{\textrm{AB}} <$ 24.5 for both type 2 and type 1 AGN and QSOs out to $z\sim$ 4.5. As no confidence intervals were available for the best-fitting AGN photo-z estimates and since the statistical dispersion is far smaller than the width of the redshift bins studied below we have kept all sources with redshifts derived from the AGN-template library in the sample.

\section{Band-Merging}
\label{appsect:bandmerge}

\subsection{Definition of the Reference Position}
\begin{itemize}
\item {\it IR-selected sample} -- The $\sim$6$''$ FWHM of the 24\,$\mu$m PSF and {\it Spitzer} pointing uncertainties result in a non-negligible uncertainty on the source centroids of 24\,$\mu$m sources. In order to have a more precise reference for the positional matching with other catalogs we searched for IRAC counterparts to each 24\,$\mu$m source (the FWHM of the IRAC PSF is about $\nicefrac{1}{3}$ of that at 24\,$\mu$m). Since the IRAC imaging of the COSMOS field is shallower than that performed at optical wavelengths but still sufficiently deep to detect a counterpart for most 24\,$\mu$m sources this approach simultaneously reduces the likelihood of assigning the wrong optical counterpart in the subsequent band-merging with the optical photometry catalog (see last paragraph of this section).\\
We correlated the 24\,$\mu$m catalog with the IRAC catalog \citep{sanders07, ilbert09b} using a matching radius of 2$''$ and found 45,827 unambiguous counterparts (92\% of all sources in the 24\,$\mu$m catalog). In 572 cases more than one IRAC source was found within the search radius. Unless the closest of the potential counterparts was at least twice as close to the 24\,$\mu$m position as the other candidates these sources (1\% of the catalog) were excluded from the sample. In the case of 3668 24\,$\mu$m sources for which no IRAC counterpart could be determined we searched the photometry catalog for optical counterparts within 0.6$''$, the FWHM of the ground-based imaging. If this match was ambiguous the source was disregarded. Unique optical counterparts were found for another 1386 sources, however, and the final percentage of 24\,$\mu$m sources with reference coordinates thus increased to 94\%.
\item {\it Radio-selected sample} -- The 1.4\,GHz source positions are accurate to within about 0.1$''$, which is sufficiently accurate that they can be used directly as the reference position for the identification of optical and IR counterparts.
\end{itemize}

\subsection{Search for 24\,$\mu$m Counterparts (Radio-Selected Sample)}
Starting with the radio source position we identified 2233 objects (77\% of the radio catalog) that have a single 24\,$\mu$m counterpart within 2$''$ while 647 (22\%) radio sources remain unmatched. Based on their radio properties the 24\,$\mu$m non-detections are primarily \tbd. For $<$1\% of the radio-selected sample the match was ambiguous. This subset includes two pairs of radio sources which have been assigned the same 24\,$\mu$m counterpart and 17 for which more than one IR-detection was present within the search radius (mostly due to a complicated 24\,$\mu$m-morphology or the occasional over-deblended source).

\subsection{Search for Radio Counterparts (IR-Selected Sample)}
Since the identification of 24\,$\mu$m counterparts to 1.4\,GHz sources in the immediately preceding paragraph was based on the precise radio positions these pairs of objects can be taken over directly into the IR-selected sample\footnote{If the according 24\,$\mu$m source has no trustworthy IRAC or optical counterpart, the 24\,$\mu$m-radio association in question is not included in the sample. This affects only a very small fraction of the IR-selected sample since the fraction of sources for which no reference coordinate could be determined is $<$1\%.}. Note that subsequently we search for additional radio detections with $S/N_{1.4\,{\rm GHz}}\geq$ 3 in the vicinity of all 24\,$\mu$m sources without a radio counterpart in the radio catalog described in \S\,\ref{appsect:radiodata}. The box in which we check for these additional faint radio sources has dimensions 1.75$''\times$1.75$''$ (i.e. 5$\times$5 px$^2$ at the pixel scale of the Deep Project image) which corresponds to the FWHM of the PSF of the first two IRAC bands. Radio fluxes for an additional 639 24\,$\mu$m sources could be measured in this way.

\subsection{Search for 70\,$\mu$m Counterparts}
We allow for a maximal separation of 6$''$ between reference coordinates and an accepted 70\,$\mu$m source. Due to the broad FWHM (18.6$''$) of the {\it MIPS} 70\,$\mu$m PSF there is a relatively large probability that the emission from more than one 24\,$\mu$m or 1.4\,GHz source is blended into a single 70\,$\mu$m source with no distinguishable secondary peaks by which the individual components could be separated. This occurred for $\sim$7\% of the 1.4\,GHz sources with a potential counterpart at 70\,$\mu$m. In the IR-selected sample 21\% of the matched sources had a potential 70\,$\mu$m counterpart that lay within less than 6$''$ of at least one more 24\,$\mu$m source. Unless these ambiguous 70\,$\mu$m associations are at least twice as close to the nearest 24\,$\mu$m or radio source as to the next best candidate they have been excluded for the analysis of the paper. Ultimately, this was the fate of 70\% (50\%) of the radio (24\,$\mu$m) sources with blended 70\,$\mu$m photometry. In the interest of achieving a high accuracy of the measured IR/radio properties we prefer this approach to, e.g., attempting to fit the 70\,$\mu$m flux distribution with two components with peak positions fixed to the reference position.\\ 
To summarize, in we have found $\sim$1500 reliable 70\,$\mu$m counterparts to 24\,$\mu$m sources. A further 60 sources in the IR-selected catalog had more than one counterpart within 6$''$ and were removed together with the 24\,$\mu$m objects that are associated with a blended 70\,$\mu$m detection. The majority of the 24\,$\mu$m sources (67\%) has no directly detected counterpart. In the radio-selected sample we were able to identify $\sim$820 unambiguous 70\,$\mu$m counterparts while excluding $\sim$60 sources because they either were associated with a blended 70\,$\mu$m source or had an ambiguous match. For some 70\% of the radio sources no 70\,$\mu$m counterpart could be found.

\subsection{Search for Optical Counterparts and Spectra}
A search radius of 0.6$''$ is used for the cross-correlation of the radio-selected sample with the COSMOS photometry catalog and the spectroscopy catalog. This figure corresponds to the FWHM of the PSF-matched ground-based photometry. Whenever more than one optical counterpart is found within this distance the according radio source is removed from the catalog. This choice was made because any photometric redshift derived for the source would necessarily be unreliable as both sources contribute to the measured flux. The same applies to spectroscopic observations which have a slit width of $\sim$1$''$ and are thus treated analogously because sources within 0.5$''$ could potentially lead to ambiguous spectral features.\\
To find optical counterparts and spectroscopic information for sources in the IR-selected catalog we use a search radius of 1$''$ which accounts for the somewhat larger uncertainty in the predominantly IRAC-based reference positions. Also, we ease the rejection criterion for ambiguous matches somewhat in that we accept optical or spectroscopic counterparts which are more than twice as close as the other candidate(s), provided the next best candidate lies beyond 0.6$''$.

\section{Derivation of Distribution Functions with Survival Analysis}
\label{appsect:survival}

\subsection{Doubly Censored Data}
\label{appsect:doubcensetheory}

To derive the underlying distribution $f(q)$ of IR/radio ratios $q$ for a sample in which measurements of $q$ can be limited both from above and below\footnote{The formalism summarized here assumes a random censoring model in which measurements are censored independently of their location in the distribution.} due to non-detections either in the IR or radio band we use the method of \cite{schmitt93}. Their approach is based on maximum likelihood estimation and requires no assumptions about the form of the true distribution of $q$ (it is thus said to be non-parametric). If represented in the form of a cumulative distribution, $F(q)$, the estimator is equivalent to the probability of measuring a value of $q$ smaller than a given ratio (which we will denote by a capital Q):
\begin{equation*}
F(q) \equiv P(q \leq {\rm Q})~.
\end{equation*}
In the following we summarize the steps involved in the construction of $F(q)$.

\noindent To begin with, the $n$ data points $\mathcal{Q}_i$ are arranged in $M$ bins\footnote{Since the bin width can always be decreased to the point that each bin contains only one (untied) detection this assumption is not restrictive.} with ascending bin centres ${\rm Q}_j> {\rm Q}_{j+1}$, $j=$ 1, ..., $M$. Let $d_j$ be the number of detections in the $j$-th bin which spans the range $[{\rm Q}_j-d{\rm Q}/2, {\rm Q}_j+d{\rm Q}/2[$. For the $j$-th bin we also define $u_j$, the number of upper limits smaller than ${\rm Q}_j$ and, in analogy, $l_j$ as the number of lower limits exceeding ${\rm Q}_j$. Note that in view of the definition of $u_j$ and $l_j$ it is useful to choose bin boundaries offset by $-d{\rm Q}/2$ and $+d{\rm Q}/2$ for the upper and lower limits, respectively \citep[cf.][]{avni80}. This eases the computation of the vectors $\bm{d}$, $\bm{u}$ and $\bm{l}$ which can be easily constructed as slightly shifted histograms of the detections as well as of the upper and lower limits.\footnote{An important modification of the vectors $\bm{d}$ and $\bm{u}$ is necessary if the first uncensored value, $q_X$, occurs at a higher value of ${\rm Q}_j$ than the smallest upper limit in the data set. By construction, the algorithm of \cite{schmitt93} is not sensitive to upper limits that lie below the smallest uncensored data point. As a consequence the normalized cumulative distribution function defined in equation (\ref{eq:schmittcumul}) rises above 0 for the first time in the bin containing the smallest direct detection. The correct asymptotic value of the distribution function for $q\rightarrow{\rm min}(\mathcal{Q}_i)$, however, should not be 0 in this case but a number larger than or equal to $\#\{i, \mathcal{Q}_i < q_X\}/n$. Here $\#\{i, \mathcal{Q}_i < q_X\}$ is the number of upper limits smaller than $q_X$ and $n$ is defined immediately following equation (\ref{eq:schmittcoupled}). The correct behaviour of the cumulative distribution function can be achieved by setting $d_0=\#\{i, \mathcal{Q}_i < q_X\}$ and $u_j=$ 0 for all bins $j$ with bin centres ${\rm Q}_j<q_X$.\\ Note that even when there are lower limits that exceed the largest uncensored value, no analogous fixes are needed to correct the doubly censored distribution function in the limit $q\rightarrow{\rm max}(\mathcal{Q}_i)$.}

\noindent The value $f_j$ of the distribution function $f(q)$ in the $j$-th bin is constrained by the probability
\begin{equation}
P\sim \prod_{j=1}^{M} f_j^{d_j}~\left(\sum_{k=1}^{j-1} f_k\right)^{u_j}~\left(\sum_{k=j+1}^{M} f_k\right)^{l_j}
\label{eq:schmittprob}
\end{equation}
of obtaining the triplet ($d_j$, $u_j$, $l_j$). Intuitively equation (\ref{eq:schmittprob}) can be understood as the product of the {\it \`a priori} probabilities of a measurement in the $j$-th bin being a detection (first term), an upper limit which -- if detectable -- would in truth be located in a lower bin (second term) or an upper limit in actually located in a higher bin (third term). \cite{schmitt85} and \cite{campbell81} have shown that by introducing the likelihood and by using the constraint that $\sum_{j=1}^M f_j =$ 1 the probability in equation (\ref{eq:schmittprob}) can be used to derive a system of $M$ coupled fixed point equations for each of the $f_j$:
\begin{equation}
f_j = \frac{d_j}{n} + \sum_{k=j+1}^M \frac{u_k}{n}\left(\frac{f_j}{\sum_{m=1}^{k-1} f_m}\right) +  \sum_{k=1}^{j-1} \frac{l_k}{n}\left(\frac{f_j}{\sum_{m=k+1}^M f_m}\right)~.
\label{eq:schmittcoupled}
\end{equation}
Here $n=\sum_{j=1}^M (d_j+u_j+l_j)$ is the total number of measurements in the sample. Equations (\ref{eq:schmittcoupled}) can be solved by iteration starting with $f_j=d_j/n$ and $u_j=l_j=$ 0 and the cumulative distribution constructed according to
\begin{equation}
F_j = \sum_{k=1}^j f_k~, \qquad\qquad\qquad j=1, ..., M
\label{eq:schmittcumul}
\end{equation} 
once convergence has been achieved. This was typically the case after approximately 30-70 iterations.

\noindent Various approaches to calculate the uncertainty associated with the doubly censored cumulative distribution function $F(q)$ have been proposed \citep[see, e.g.][]{zhusun07}. Here we use the analytical formalism of \cite{turnbull74} which is based on the Fisher information matrix $\bm{J}$ with elements\footnote{If $l_M=$ 0 (and hence $F_M=$ 1) the error on $F_M$ is zero by definition. In this case is useful for computational reasons to reduce the dimensionality of the problem by one by setting $l_{M-1} = l_{M-1} + n_M$ and thereafter only considering the ($M-1$)-dimensional vectors $\bm{d}$, $\bm{u}$, $\bm{l}$ and $\bm{F}$.}
\begin{align}
J_{jj} &= \frac{d_j}{(F_j-F_{j-1})^2} + \frac{d_{j+1}}{(F_{j+1}-F_j)^2} + \frac{l_j}{(1-F_j)^2}-\frac{u_j}{F_j^2}~, & \multirow{2}{*}{$j = 1, ..., M-1$} \nonumber\\
J_{j,j+1} &= J_{j+1,j} = -\frac{d_{j+1}}{(F_{j+1}-F_j)^2}~, \nonumber\\
\\[-3ex]
J_{MM} &= \frac{d_M}{(F_M-F_{M-1})^2} + \frac{l_M}{(1-F_M)^2}-\frac{u_M}{F_M^2}~,\nonumber\\*[2ex]
J_{ij} &= 0~, \qquad &{\rm for}\quad \left|i-j\right| > 1. \nonumber
\label{eq:covarelem}
\end{align}
$\bm{J}$ is a symmetric Jacobi matrix in which only the diagonal and off-diagonal elements are non-zero. In the limit of large samples its inverse, $\bm{V}$, is an asymptotically unbiased estimate of the covariance matrix of the $F_j$ ($j=1, ..., M$) such that one has
\begin{equation}
\sigma_{F_j}^2 = {\rm var}(F_j) = V_{ii}^2~.
\label{eq:doubcensevar}
\end{equation}
Following \cite{feigelson85} and references therein it is then possible to define 2\,$\sigma$ confidence intervals of the form
\begin{equation}
F_j^{\rm exp(\pm 1.96\,\sigma_{F_j})}~,
\label{eq:doubcenseerr}
\end{equation}
which always lie in the permissible range [0, 1].\\

\noindent By applying the algorithm of equation (\ref{eq:schmittcoupled}) to singly censored data sets it was possible to check whether its output agrees with that obtained using the formalism of Appendix \ref{appsect:singlecense}. We found that the variance in equation (\ref{eq:doubcensevar}) needed to be doubled in order to achieve consistency of the confidence intervals in equation (\ref{eq:doubcenseerr}) with those computed with the algorithm specifically designed to deal with singly censored data. For the sake of consistency the uncertainties on the doubly censored distribution functions stated throughout the paper have thus been computed using twice the nominal value of equation (\ref{eq:doubcensevar}).

\noindent It can be useful to convey information on the shape of a (cumulative) distribution in form of percentiles. Given $F_j$ and the associated confidence intervals, the $p$-th percentile is defined as that $q$ for which $F(q)=\nicefrac{p}{100}$. Similarly, the points where the curves defined by the upper and lower confidence intervals -- given by equation (\ref{eq:doubcenseerr}) -- of $F(q)$ are equal to $\nicefrac{p}{100}$ are used as a measure of the uncertainty on the sought percentile.

\subsection{Singly Censored Data}
\label{appsect:singlecense}

A formalism which is identical to or simplified with respect to that of the previous section can be used when dealing with single censoring. If, for instance, we consider a radio-selected sample its distribution of $q$ is censored from the left and can be represented as
\begin{eqnarray}
\mathcal{Q}_i &=& \textrm{ max($q^-_i$, $q_i$)}\nonumber \\
\\
\delta_i &=& \begin{cases}
\quad -1 & \quad \mbox{if }q_i<q^-_i\\
\quad  0 & \quad \mbox{if }q_i \in [q^-_i, q^+_i]
\end{cases} \nonumber
\label{eq:singcensodef}
\end{eqnarray}
in analogy to equations (\ref{eq:doubcensodef}). The treatment of singly censored data has been discussed in detail in \cite{feigelson85} and \cite{schmitt85} with an emphasis on astronomical applications. The former paper also describes how to transform left censored into right censored data, the treatment of which has been documented more frequently in the literature for historical reasons (see references in the two aforementioned articles). Here we simply would like to point out that equations (\ref{eq:schmittcoupled}), which constrain the distribution function of doubly censored data, remain valid in the case of single censoring. However, the fact that either all of the $u_i$ or $l_i$ as defined in \S\,\ref{appsect:doubcensetheory} equal zero allows the coupled system of equations to be solved analytically in terms of the cumulative distribution function $F_j$ through the Kaplan-Meier product limit estimator \citep{kaplanmeier58}. To do so we follow exactly the prescriptions of \cite{feigelson85}.

\section{Derivation of the relative AGN frequency ${\rm f_{AGN}}(q, z)$}
\label{appsect:cohabderiv}

\begin{figure}
\epsscale{.8}
\plotone{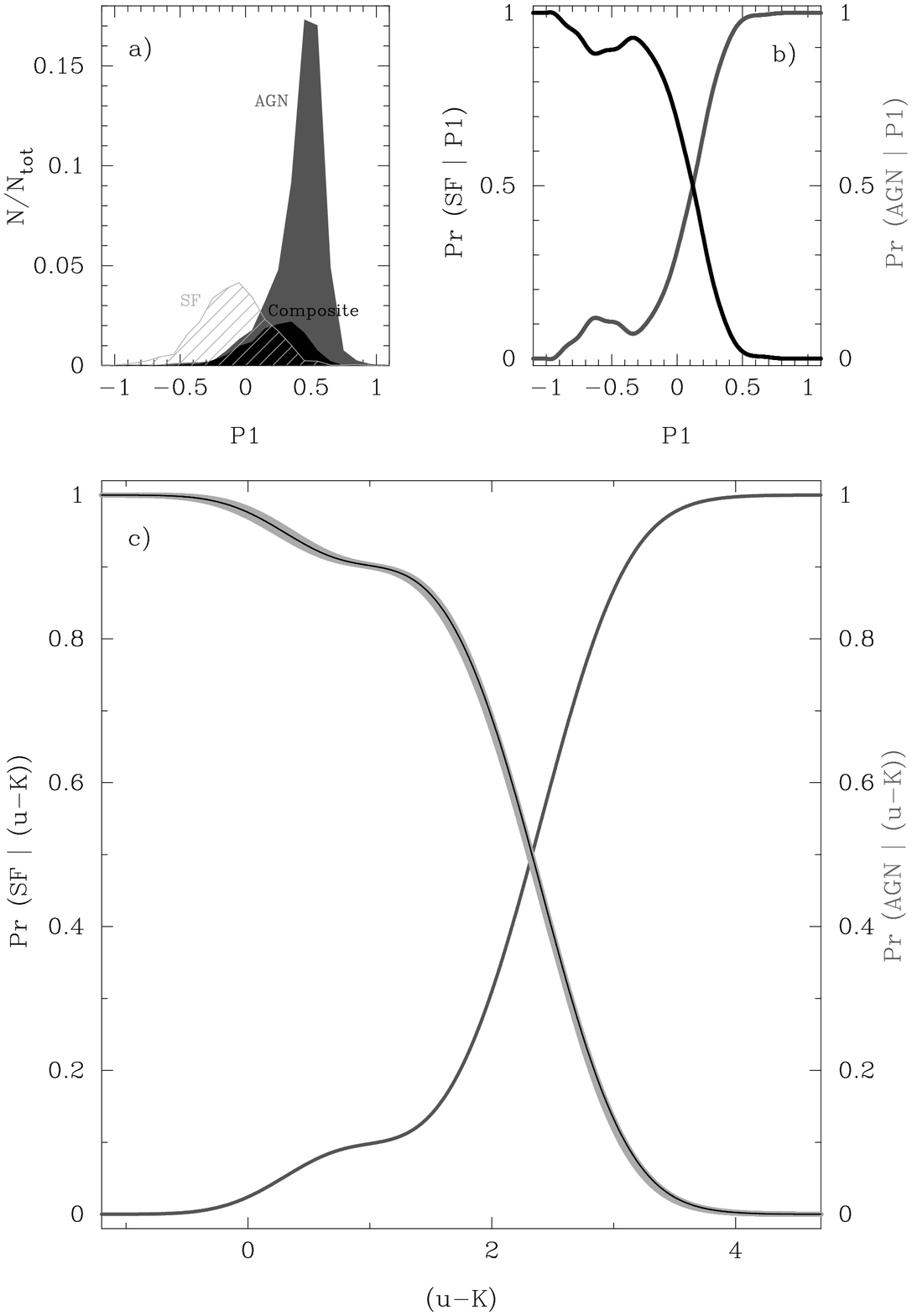}
\caption{{\it (a)} -- Distribution with respect to P1 colour of star forming (SF), AGN and composite systems in the SDSS/NVSS sample of \cite{smolcic08}. {\it (b)} -- Probability ${\rm Pr_{eff.}\,(SF\,|\,P1)}$ (cf. equation (\ref{eq:SFprob})) of correctly classifying a galaxy as star forming (black line) or AGN (dark grey line) at a given value of P1 colour based on panel {\it (a)} of this figure. {\it (c)} -- Probability ${\rm Pr_{\rm eff.}\,(SF)}$ (black curve), expressed as a function of rest frame ($u-K$) colour. The conversion from panel {\it (b)} is based on the average trend and the scatter of the relation shown in Fig. \ref{fig:P1equivcol}. The adjacent light grey area illustrates the error on the ${\rm Pr_{\rm eff.}\,(SF\,|\,}(u-K))$ curve induced by the uncertainty of the best-fit linear relation between P1 and ($u-K$). In dark grey the complementary probability of correct AGN-classification, ${\rm Pr_{\rm eff.}\,(AGN\,|\,}(u-K))$, is plotted (without the error which mirrors the one of the curve for SF systems). Equal probabilities of correctly classifying objects as SF or AGN, respectively, are reached at ($u-K$) = 2.36.\label{fig:uK_SFprob}}
\end{figure}

\begin{figure}
\centering
\includegraphics[scale=0.75]{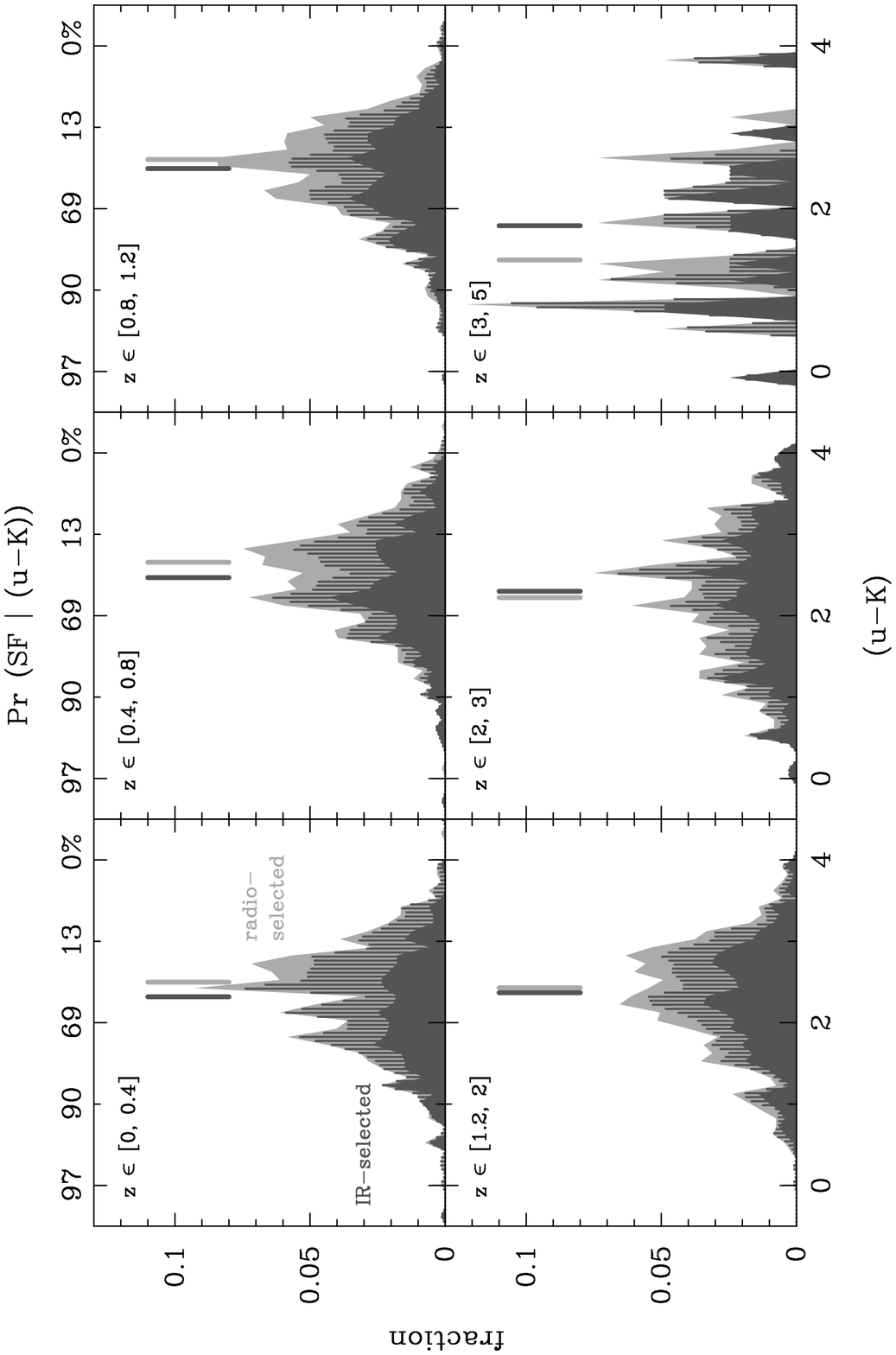} 
\caption{Distribution of rest frame ($u-K$) colours of galaxies in the radio-selected sample (light grey) and the IR-selected sample (limited to sources with $S_{\nu}(24\,{\mu}m)\geq$ 0.3 mJy; dark grey) in six different redshift slices. The hatched areas indicate those sources which are common to both samples. The median ($u-K$) colours in the radio- and IR-selected samples are marked by the vertical light and dark grey lines, respectively. Along the upper edge of the plot the probability ${\rm Pr\,(SF\,|\,}(u-K))$ of `SF-hood' associated with a given value of ($u-K$) colour is indicated.\label{fig:probcol_distr}}
\end{figure}

\begin{figure}
\epsscale{1}
\plotone{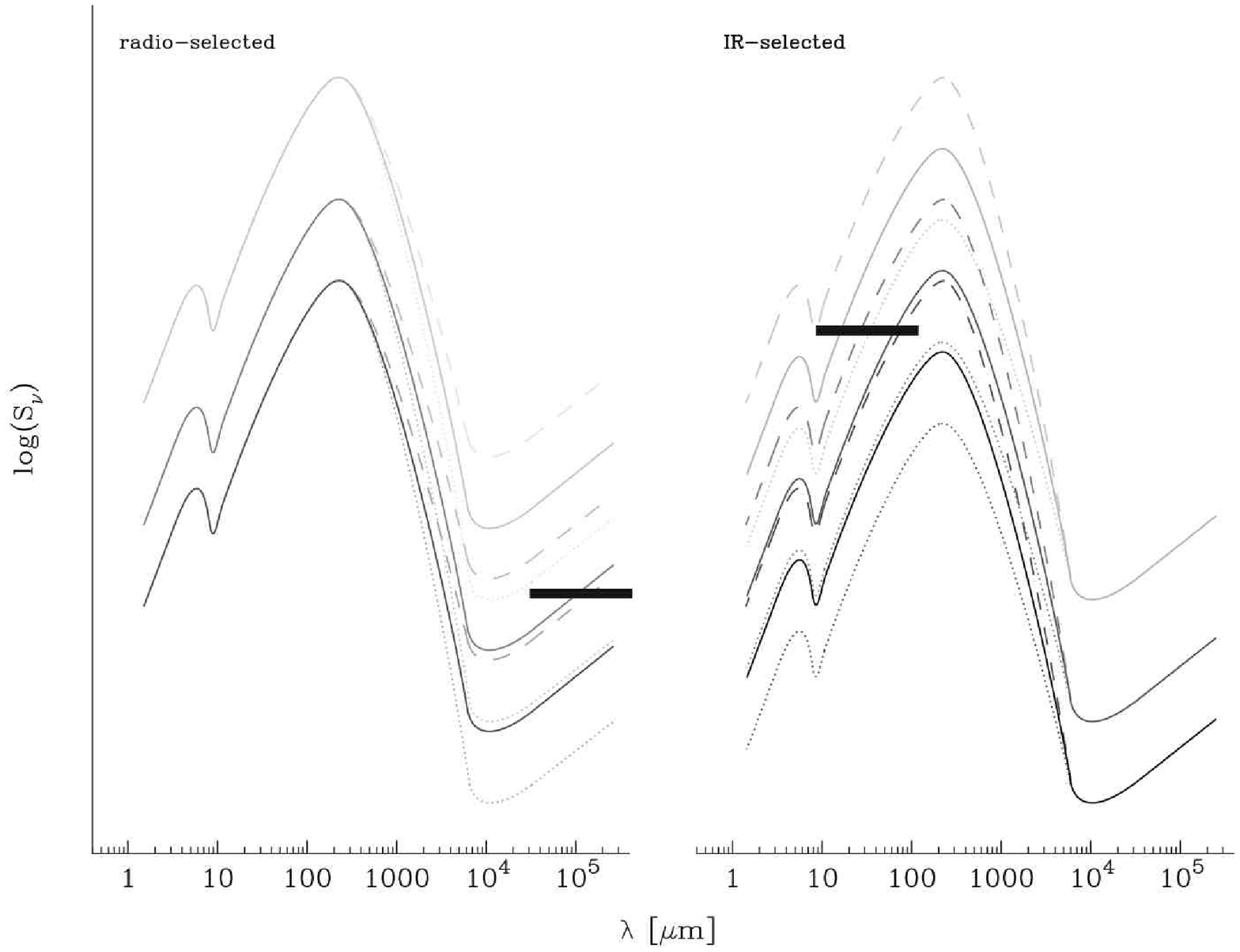} 
\caption{Schematic illustration of the origin of the difference $\Delta q_{\rm bias}$ between the average IR/radio ratio measured using an IR- or a radio-selected sample (see \S\,\ref{sect:selbias} for details). The vertically offset groups of curves show typical IR-radio SEDs of three objects with different observed bolometric flux. The increasing intensity (top to bottom) of the colour grey with which each of the three groups is drawn reflects that source counts rise with decreasing flux according to $\nicefrac{dN}{dS} \propto S^{-\beta}$. Within each group of curves with a similar observed flux density, the central one represents the SED of an object with an average  IR/radio ratio. The dashed (dotted) curve indicate +3\,$\sigma$ (-3\,$\sigma$) outliers to the IR-radio relation. Arbitrary observational limits in the radio (left) and MIR (right) window are marked with horizontal black bars. 
\label{fig:qselsketch}}
\end{figure}

\begin{figure}
\epsscale{.85}
\plotone{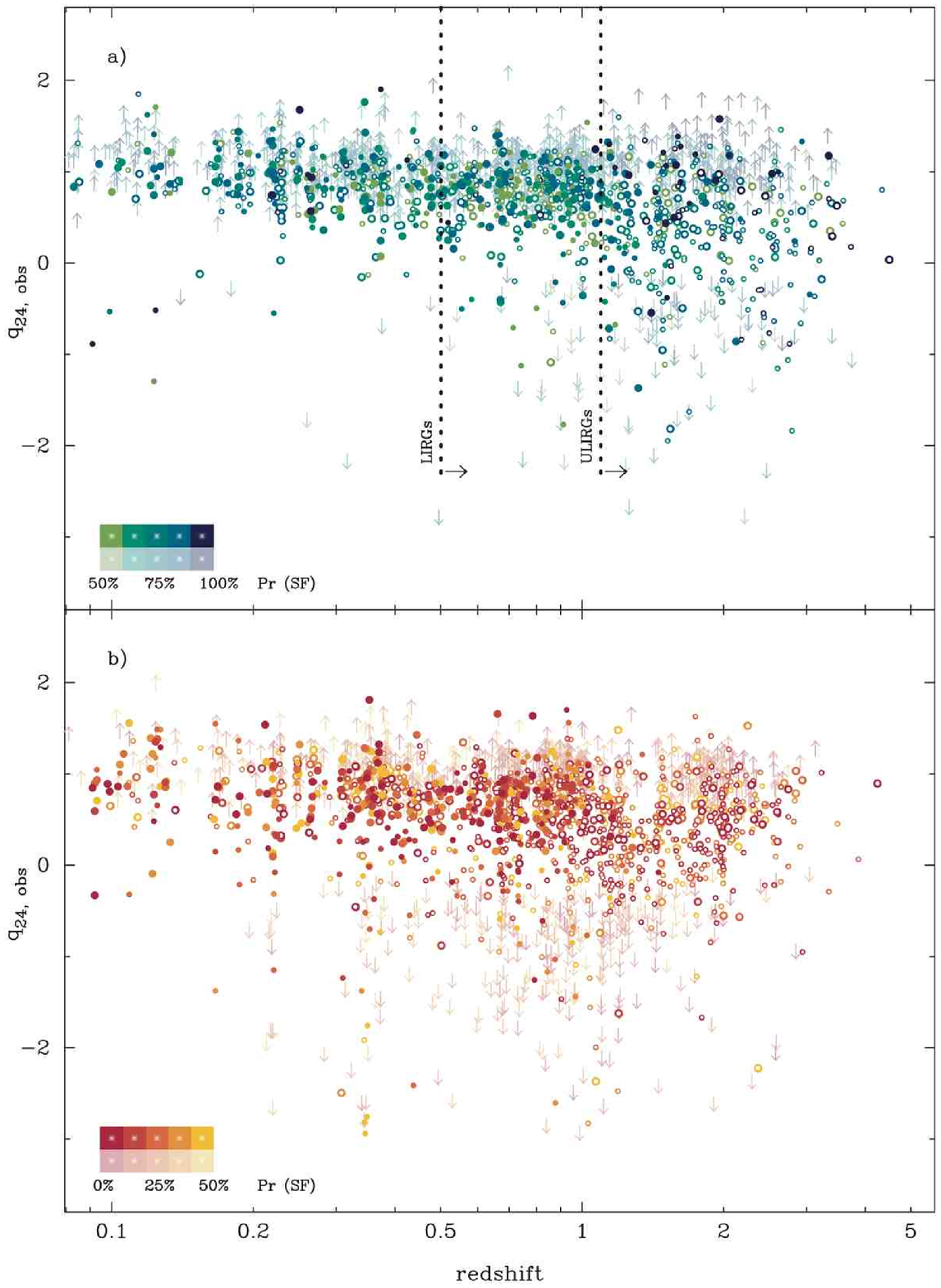} 
\caption{Ratio of observed (i.e. not $K$-corrected) 24\,$\mu$m and 1.4\,GHz fluxes, $q_{24}$, for galaxies classified as SF (top) and AGN (bottom). Sources with the highest probability of being star forming are plotted in dark blue, while those with the highest probability of being AGN are shown in red (see lower left corner of both panels). The dotted vertical lines mark the redshifts above which only (U)LIRGs remain in the sample. Large dots are used for sources that are in both the IR- and radio-selected sample; objects plotted with smaller dots are found in only one of the two samples. The style of the symbol reveals if a source has a spectroscopically (filled dot) or photometrically (open circle) measured redshift. This distinction between the quality of the redshift information is not made for lower and and upper limits (from the IR- and radio-selected sample, respectively) on $q_{24}$ which are reported with fainter colours. \label{fig:uncorrq24}}
\end{figure}

\begin{figure}
\epsscale{.85}
\plotone{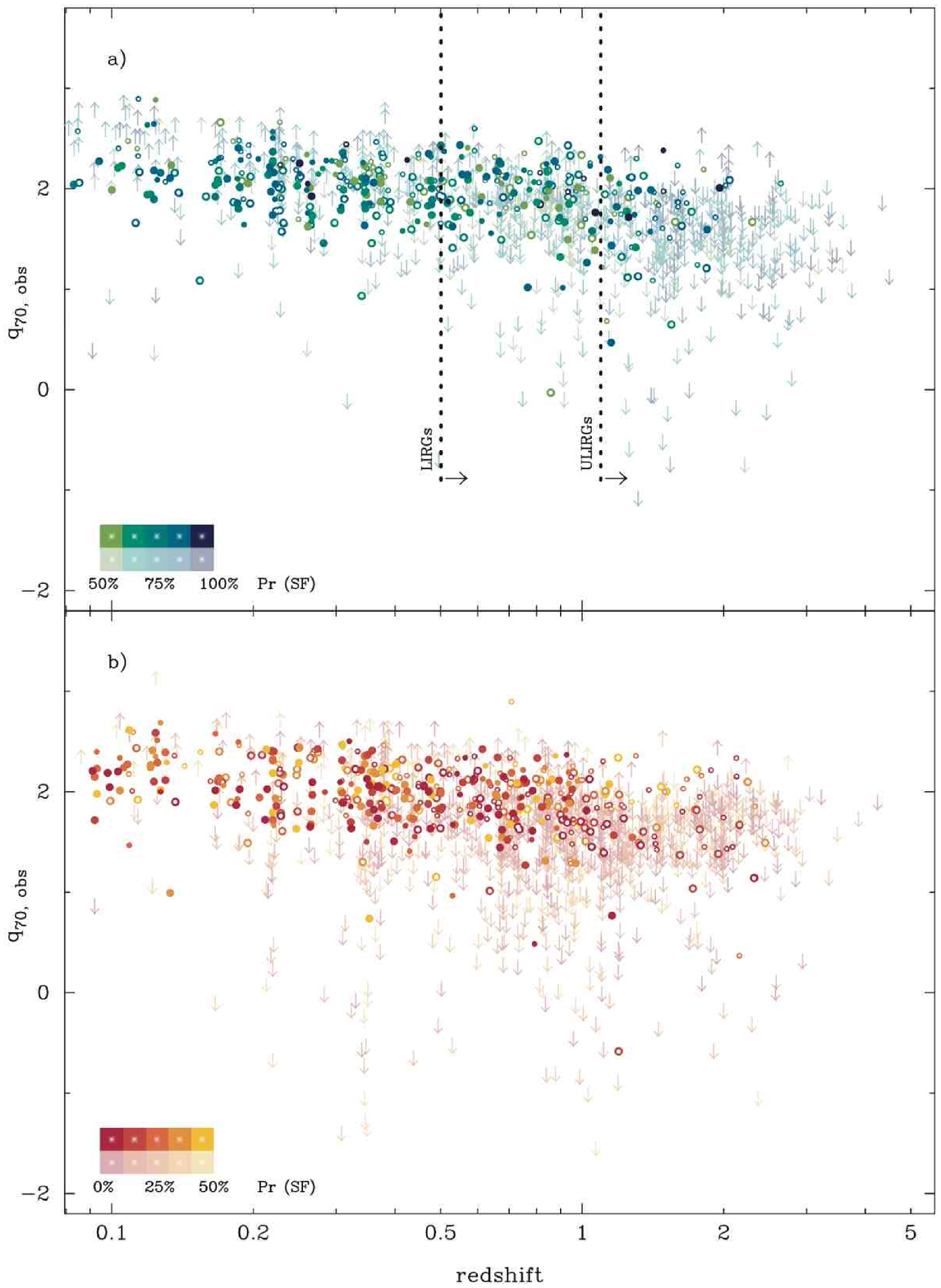} 
\caption{As for Fig. \ref{fig:uncorrq24} but for the ratio of observed 70\,$\mu$m and 1.4\,GHz flux, $q_{\rm 70, obs}$. Symbols and colours are identical to those used in Fig. \ref{fig:uncorrq24}. \label{fig:uncorrq70}}
\end{figure}

\begin{figure}
\epsscale{.9}
\plotone{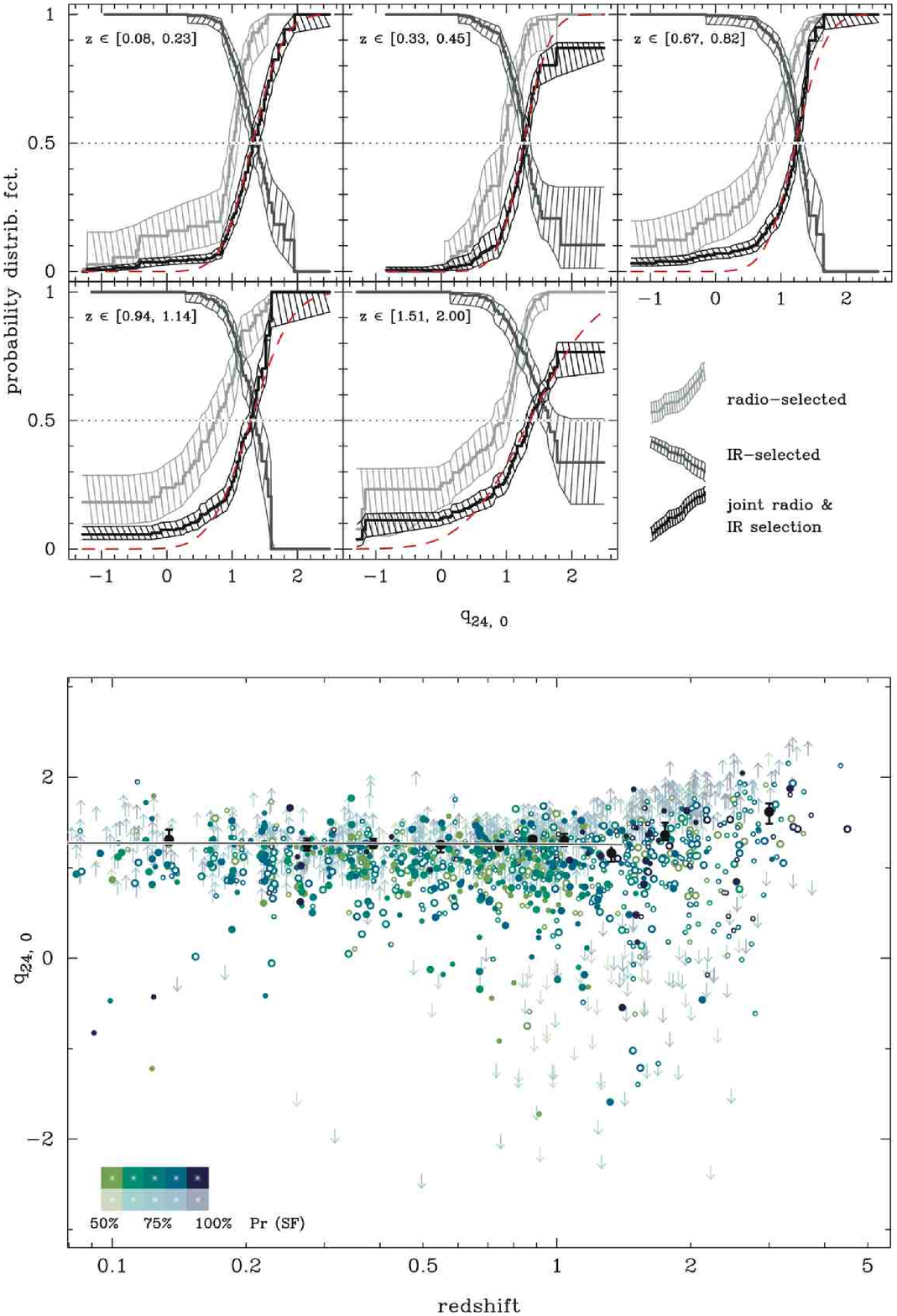} 
\caption{Evolution of the $K$-corrected 24\,$\mu$m/1.4\,GHz flux ratios, $q_{24,\,0}$, as a function of redshift. {\it Upper panel} -- Cumulative/probability distribution functions of $q_{24,\,0}$ in a radio- (light grey) and an IR-selected (dark grey) sample of SF galaxies, as well as in the union of the two (black curves). The panels shown here represent every second bin of a finer redshift sampling presented in full in the lower panel of the figure. The curves give the probability of finding a system with a smaller (or -- in the case of the IR-selected sample -- a larger) $q_{24,\,0}$ than a given value on the horizontal axis. Hatched areas span the 95\% confidence interval of the radio- and IR-selected distribution functions at fixed $q_{24,\,0}$. In each redshift bin the dashed red line traces the Gaussian distribution which fits the measured distribution function of the joint IR- and radio-selected samples best. The intersection of the black curve with the 50\% probability line (dotted horizontal line) defines the median value of $q_{24,\,0}$ in the three different samples (see also Table \ref{tab:q24info}).\newline
{\it Lower panel} -- $K$-corrected 24\,$\mu$m/1.4\,GHz flux ratios, $q_{24,\,0}$, as a function of redshift. Symbol colours and style are identical to those used in Fig. \ref{fig:uncorrq24}. The black dots and error bars mark the medians and associated 2\,$\sigma$ errors obtained from the distribution function of $q_{24,\,0}$ of the joint IR- and radio-selected sample (see black curves in the upper panel). Width and location of the redshift bins were chosen such as to always include the same number of objects. The black line represents the best-fitting evolutionary trend (fit as a function of linear redshift space) to the medians at $z<$ 1.4. \label{fig:SFcorrq24_medevo}}
\end{figure}

\begin{figure}
\epsscale{.8}
\plotone{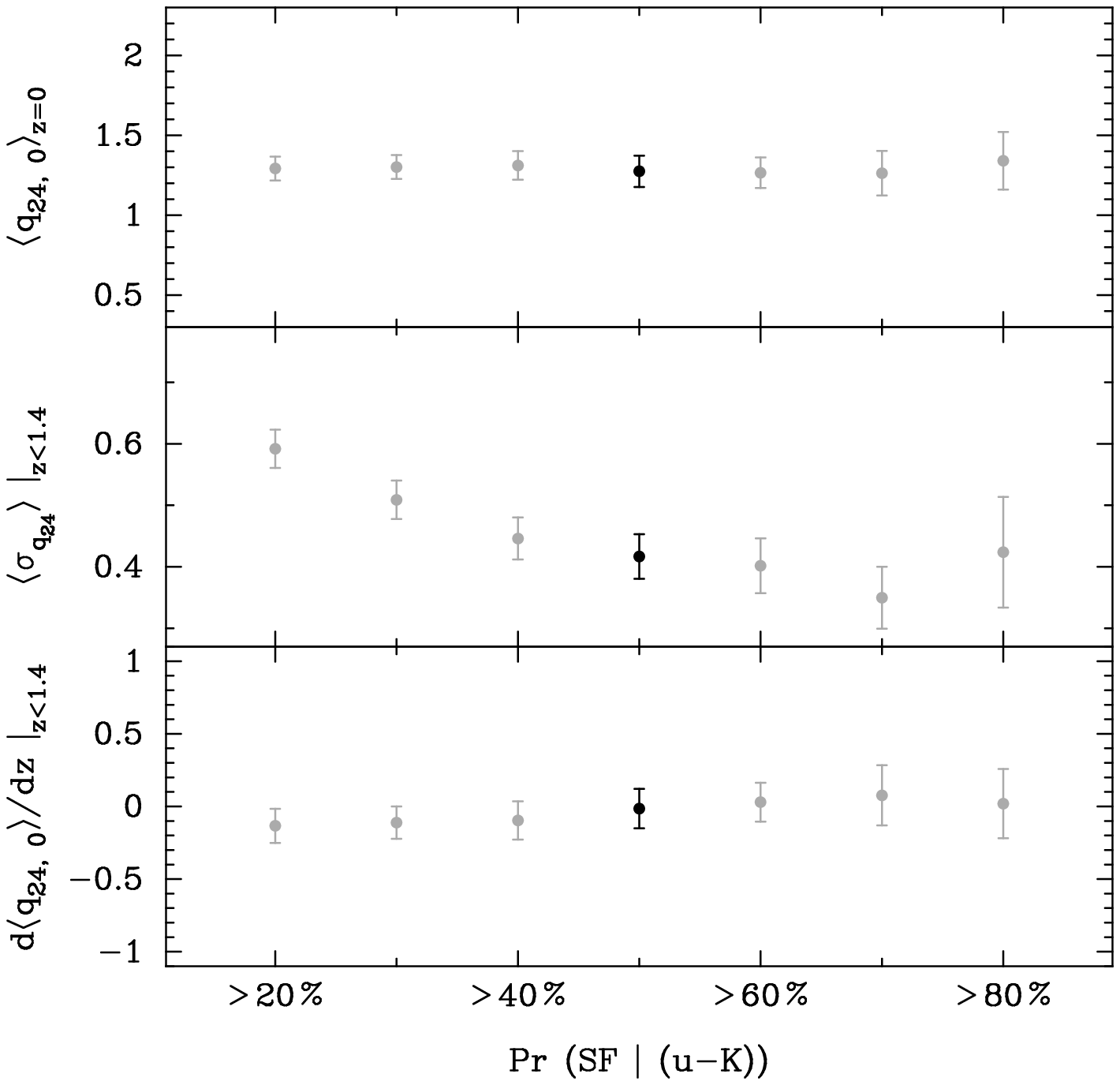} 
\caption{Variation as a function of the threshold used to select SF systems of: {\it (top)} -- the extrapolation to $z=$ 0 of the evolutionary trend line fitted to the medians $\langle q_{24,\,0}\rangle$ at $z<$ 1.4; {\it (centre)} -- the mean scatter in the population, averaged over all redshift slices at $z<$ 1.4; {\it (bottom)} -- the slope of the evolutionary trend of $q_{24, 0}$ at $z<$ 1.4.\newline
In each panel the black point highlights the measurement obtained if all objects with probability Pr\,(SF) $>$ 50\% are considered star forming (the convention used when plotting Figs. \ref{fig:uncorrq24} and \ref{fig:SFcorrq24_medevo}). The scatter decreases with increasing purity of the star forming sample, however no significant changes in the best-fitting evolutionary trend line are seen when samples consisting of sources with a probability of at least 20\%, 30\%, ... 80\% of being SF are constructed.\label{fig:SFcorrq24varevol}}
\end{figure}

\begin{figure}
\epsscale{.9}
\plotone{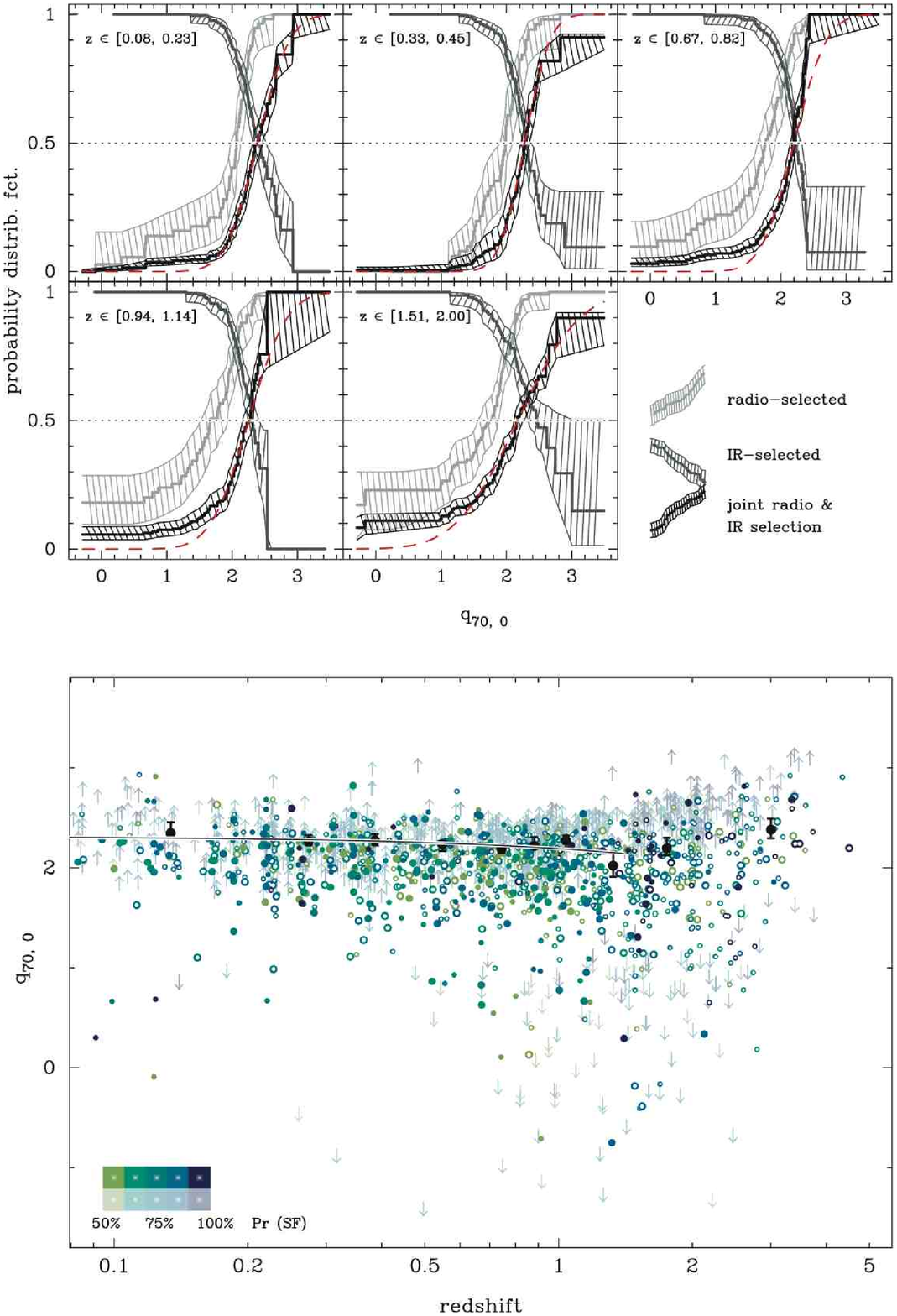} 
\caption{As for Fig. \ref{fig:SFcorrq24_medevo} but showing the distribution functions of the $K$-corrected IR/radio flux ratio $q_{70,\,0}$ ({\it top}) as well as the redshift evolution of the median ratio $\langle q_{70,\,0}\rangle$ at $z<$ 1.4 ({\it bottom}). \label{fig:SFcorrq70_medevo}}
\end{figure}

\begin{figure}
\epsscale{.8}
\plotone{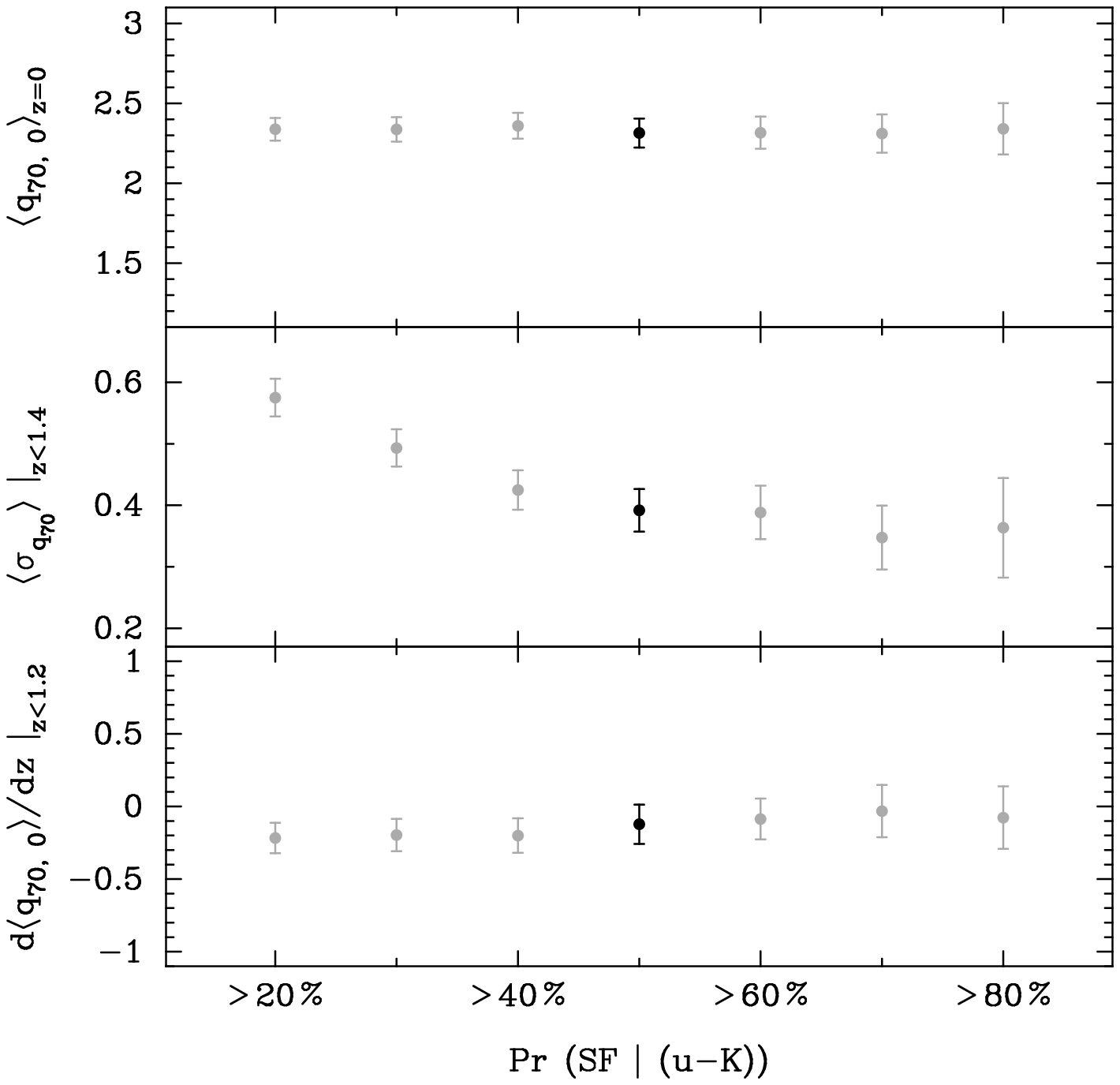} 
\caption{As for Fig. \ref{fig:SFcorrq24varevol} but investigating variations in the scatter and evolutionary trend of $q_{70,\,0}$ if different thresholds for the selection of a sample of SF galaxies are adopted. \label{fig:SFcorrq70varevol}}
\end{figure}

\begin{figure}
\epsscale{.9}
\plotone{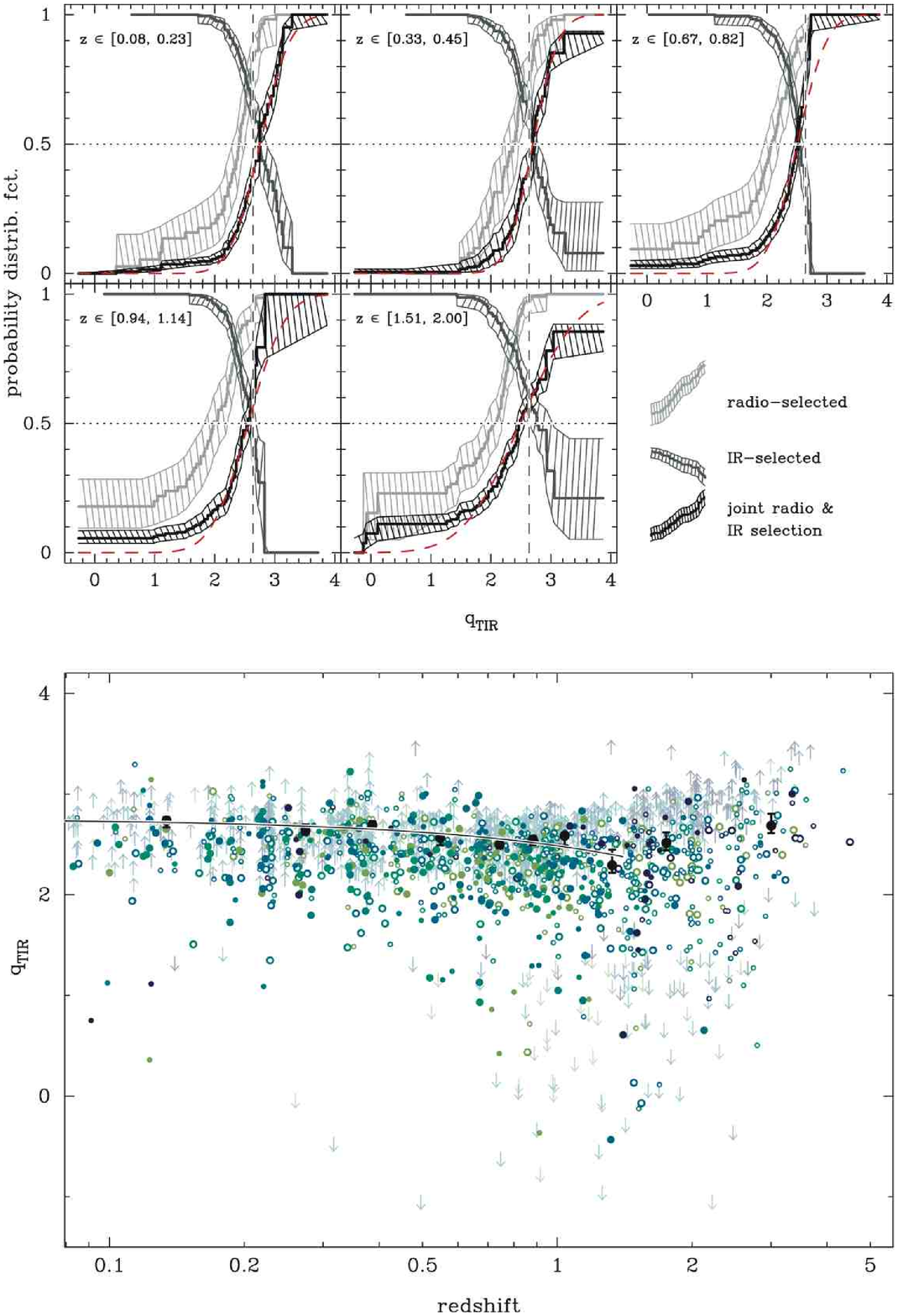} 
\caption{As for Figs. \ref{fig:SFcorrq24_medevo} and \ref{fig:SFcorrq70_medevo} but showing the distribution functions of the TIR/radio flux ratio $q_{\rm TIR}$ ({\it top}) as well as the redshift evolution of the median ratio $\langle q_{\rm TIR}\rangle$ at $z<$ 1.4 ({\it bottom}). The vertical dashed line marks the locally measured average TIR/radio ratio of 2.64 \citep{yun01, bell03}.\label{fig:SFqTIR_medevo}}
\end{figure}

\begin{figure}
\epsscale{.8}
\plotone{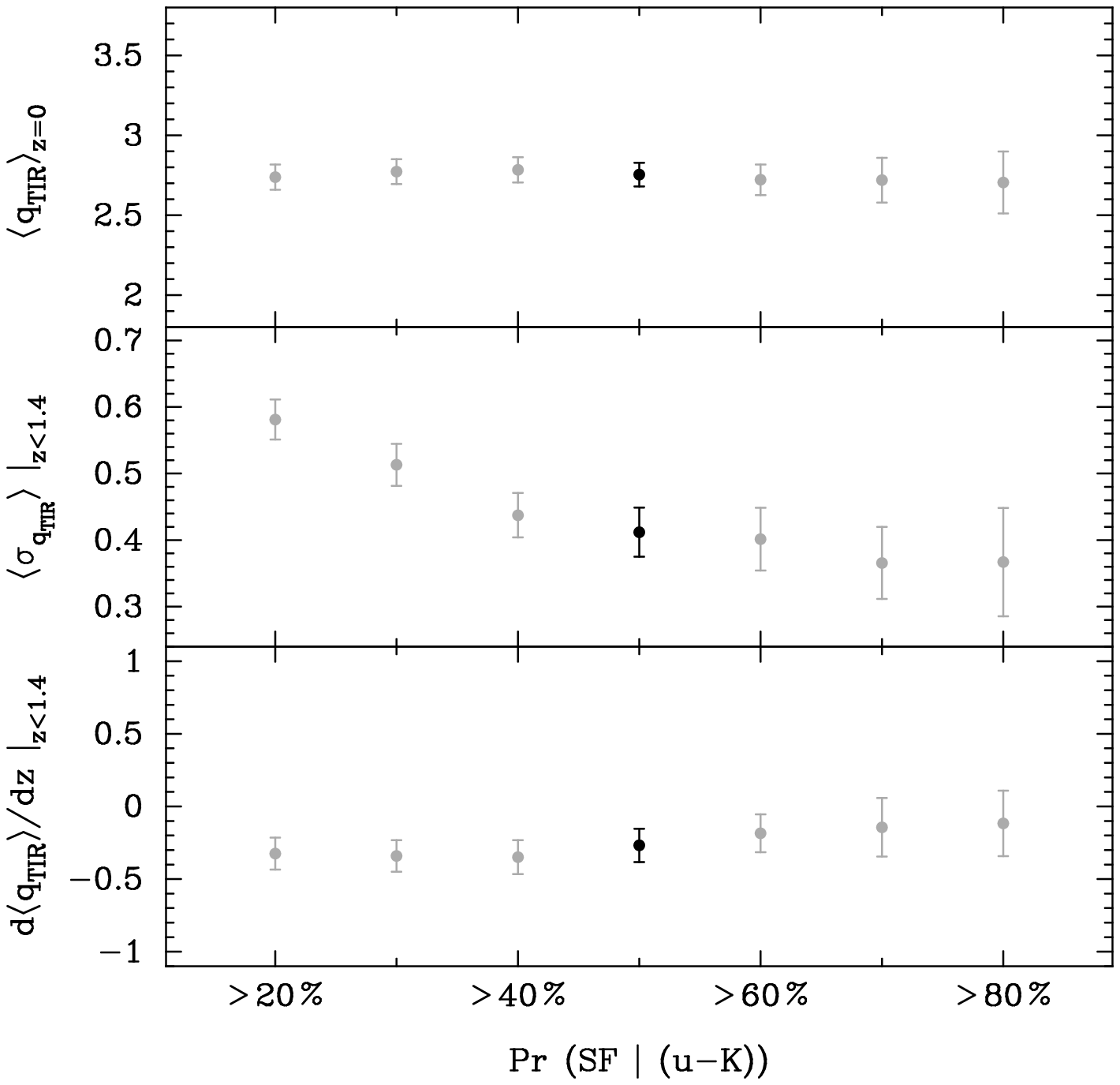} 
\caption{As for Figs. \ref{fig:SFcorrq24varevol} and \ref{fig:SFcorrq70varevol} but assessing the robustness of the best-fitting trend for the redshift evolution of $q_{\rm TIR}$ to changing the probability threshold for the selection of a sample of SF galaxies. Also shown are changes in the scatter of the TIR/radio correlation ({\it middle}) as a function of varying sample selection. \label{fig:SFqTIRvarevol}}
\end{figure}

\begin{figure}
\centering
\includegraphics[scale=0.65, angle=-90]{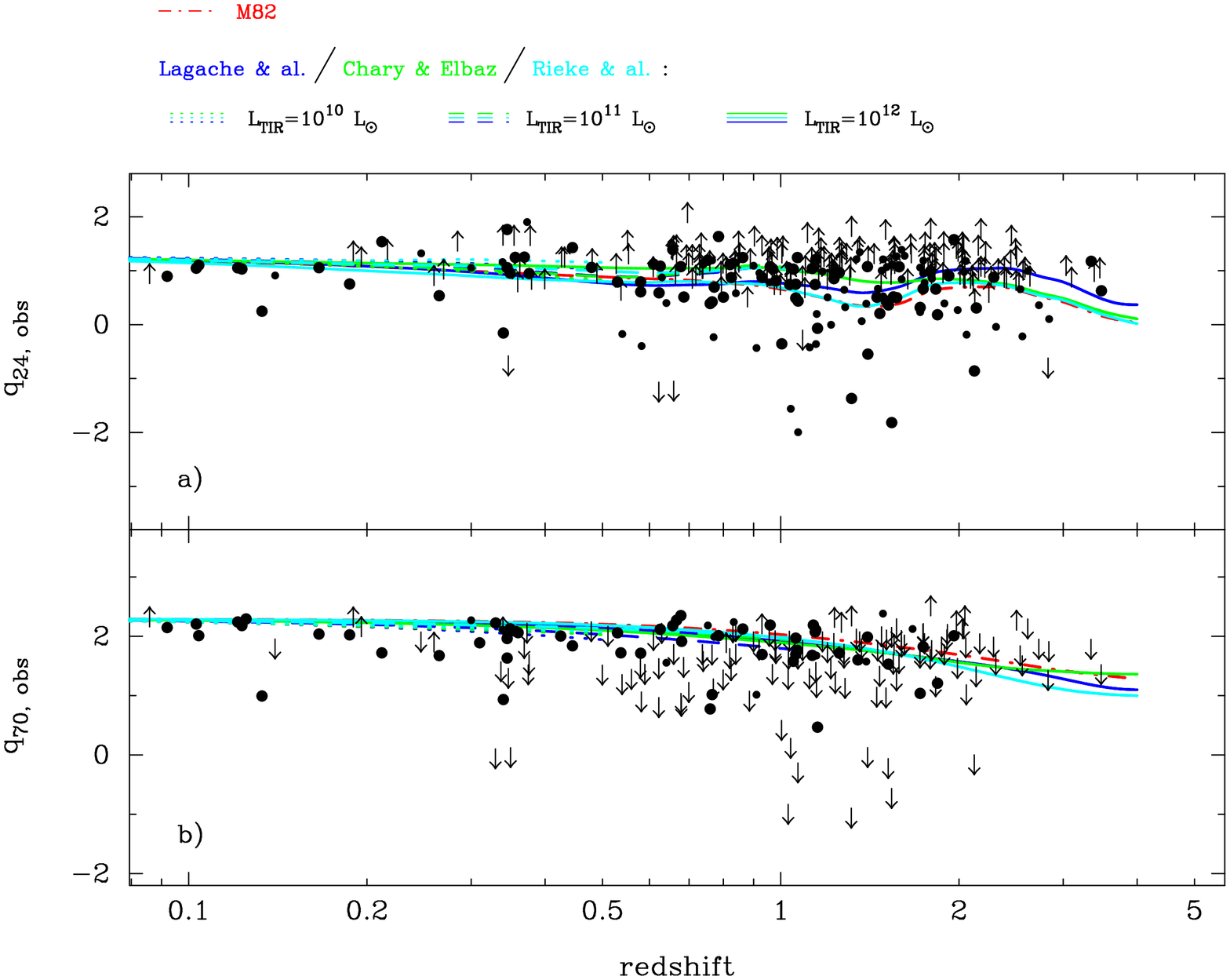}
\caption{Comparison of the IR/radio properties of model SF systems (coloured tracks; see legend along upper edge) with those of AGN-bearing sources detected in the {\it XMM-Newton} survey of the COSMOS field. {\it (a)} -- observed 24\,$\mu$m/1.4\,GHz flux density ratio $q_{\rm 24,\,obs}$; {\it (b)} -- observed 70\,$\mu$m/1.4\,GHz flux density ratio $q_{\rm 70,\,obs}$.\newline
Large symbols are used for sources that are found in both the IR- and radio-selected sample, small dots if a given source is only found in one of the two samples.\label{fig:XMMqs}}
\end{figure}

\begin{figure}
\epsscale{.85}
\plotone{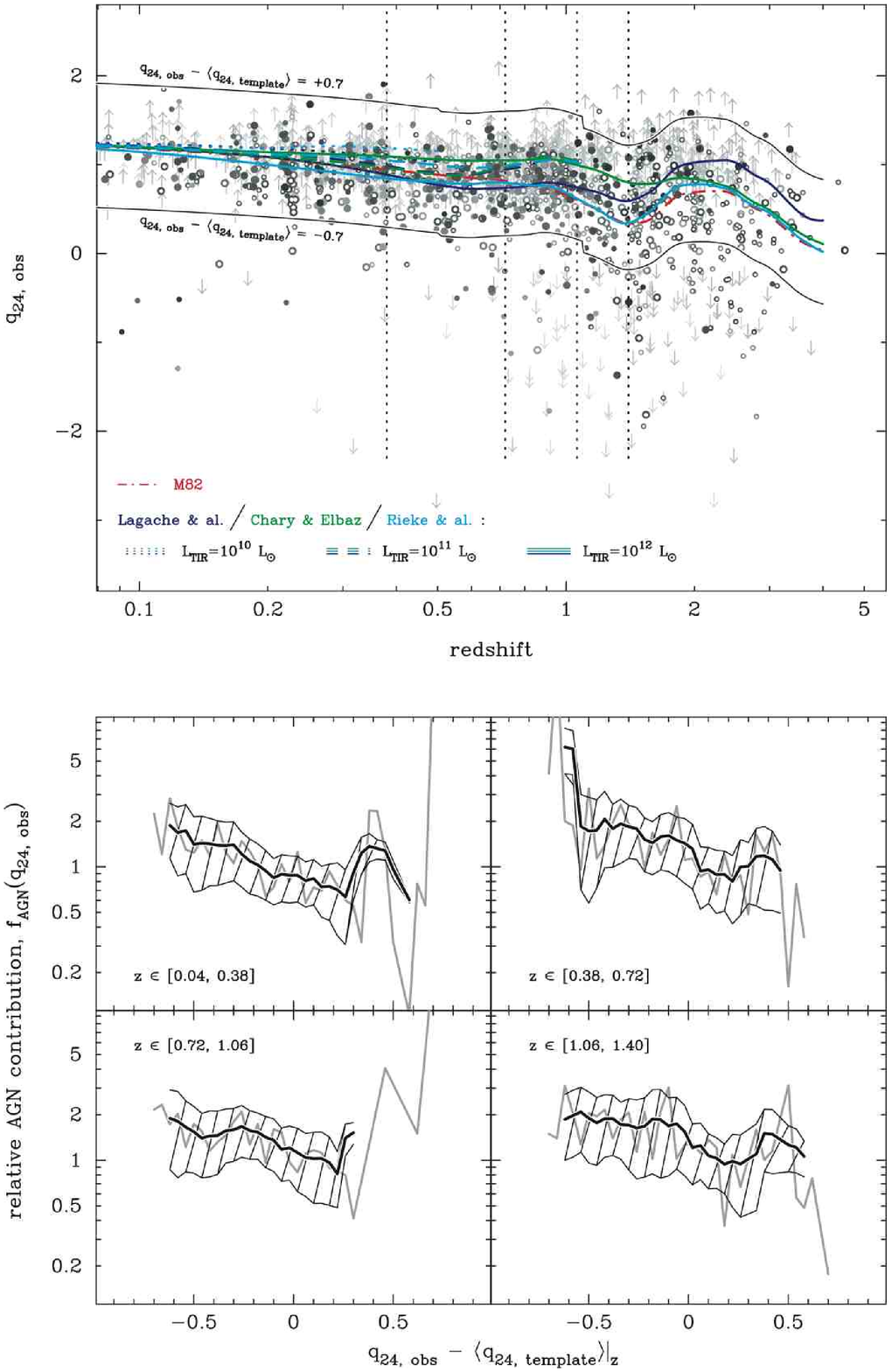} 
\caption{Assessment of the contribution of AGN-bearing sources, f$_{\rm AGN}(q_{\rm 24,\,obs})$, to the total number of objects lying on the IR/radio locus expected for SF systems. \newline
{\it Upper panel} -- Definition of the redshift-dependent locus of SFGs (delimited by the solid black lines marked with `$q_{\rm 24,\,obs} - \langle q_{\rm 24,\,template}\rangle = \pm 0.7$', respectively; see text for details) in which the relative frequency of AGN and SF sources is mapped. The coloured tracks show the evolution of the observed 24\,$\mu$m/1.4\,GHz flux ratios, $q_{\rm 24,\,obs}$, for different IR-SEDs from a variety of template libraries. The vertical dotted lines indicate the limits of the redshift bins used in the lower part of this figure.\newline
{\it Lower panel} -- Variation of the relative contribution of AGN and SF sources, f$_{\rm AGN}$, to the total population at a given value of $q_{\rm 24,\,obs}$. The changes are traced between $\pm2\,\sigma$ of the expected mean -- $\langle q_{\rm 24,\,template}\rangle|_z$ -- for local IR SEDs at different redshifts (see upper panel of the figure and details in text). The black line is a smoothed version of f$_{\rm AGN}(q_{\rm 24,\,obs})$, obtained by taking a 5-point running average of the finer and noisier mapping reported in light grey. The hatched area shows the associated $\pm2\sigma$ uncertainty region.\label{fig:q24AGNfrac} }
\end{figure}

\begin{figure}
\epsscale{.85}
\plotone{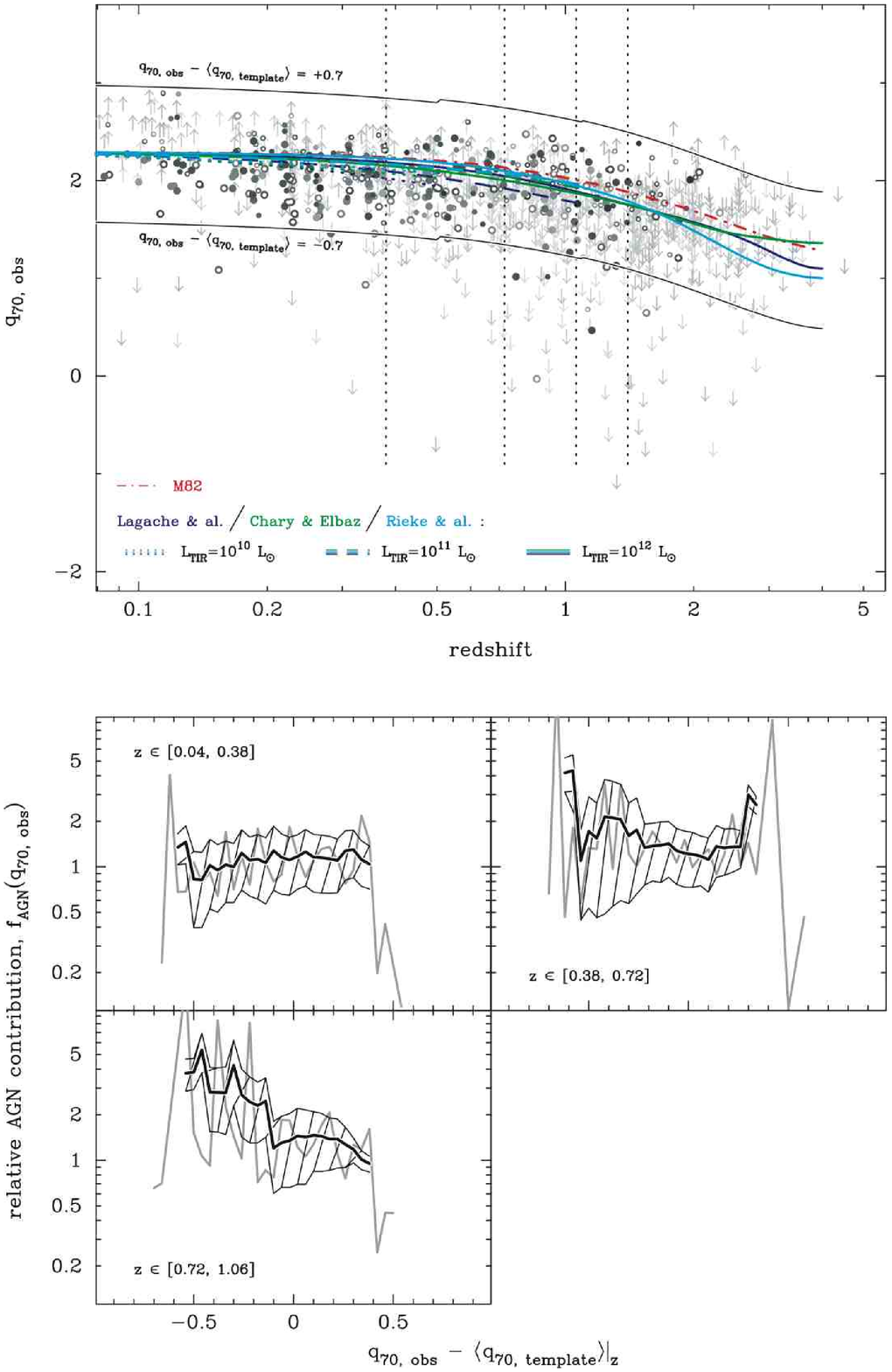} 
\caption{As for Fig. \ref{fig:q24AGNfrac} but measuring the relative frequency f$_{\rm AGN}(q_{\rm 70,\,obs})$ of AGN and SF galaxies in a $\pm2\,\sigma$ band (cf. black lines in upper panel) around the mean expected value of the observed 70\,$\mu$m/1.4\,GHz flux ratio $\langle q_{\rm 70,\,template}\rangle|_z$ based on template SEDs of star forming galaxies. Symbols and colours are identical to those used in Fig. \ref{fig:q24AGNfrac}. \label{fig:q70AGNfrac}}
\end{figure}

\begin{figure}
\epsscale{.85}
\plotone{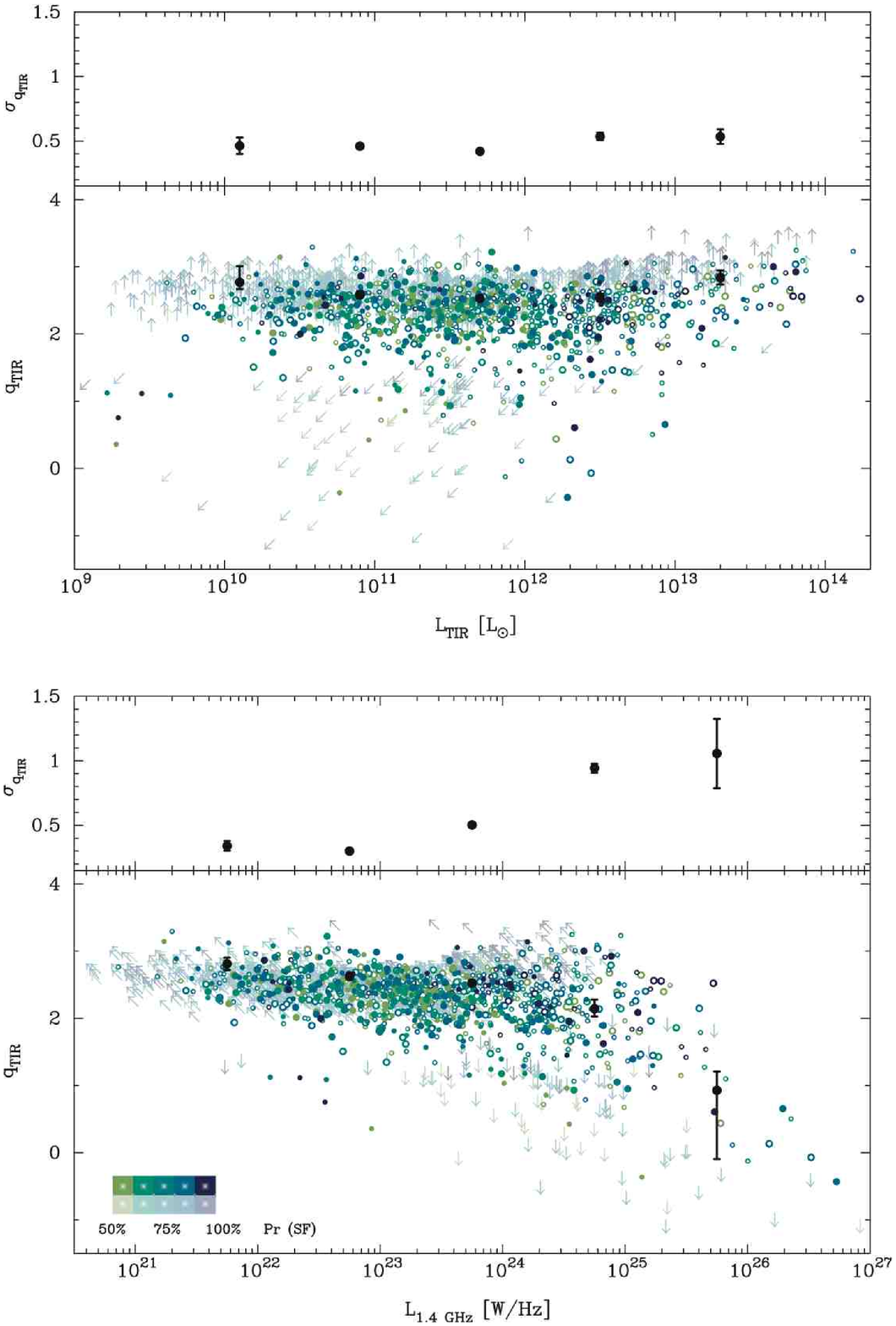} 
\caption{Dependence of the TIR/radio flux ratio $q_{\rm TIR}$ of star forming galaxies on total IR luminosity $L_{\rm TIR}$ ({\it top}) and 1.4\,GHz luminosity $L_{\rm 1.4\,GHz}$ ({\it bottom}). Medians and their associated errors derived with survival analysis are plotted in black. In the thinner two windows the scatter in the measurements of $q_{\rm TIR}$ is shown for different luminosity bins. A double off-set binning scheme was chosen in order to account for the uncertain $x$-axis value of upper limits in luminosity (diagonal arrows). \label{fig:SFqTIR_vs_lum}}
\end{figure}

\begin{figure}
\centering
\includegraphics[scale=0.6, angle=-90]{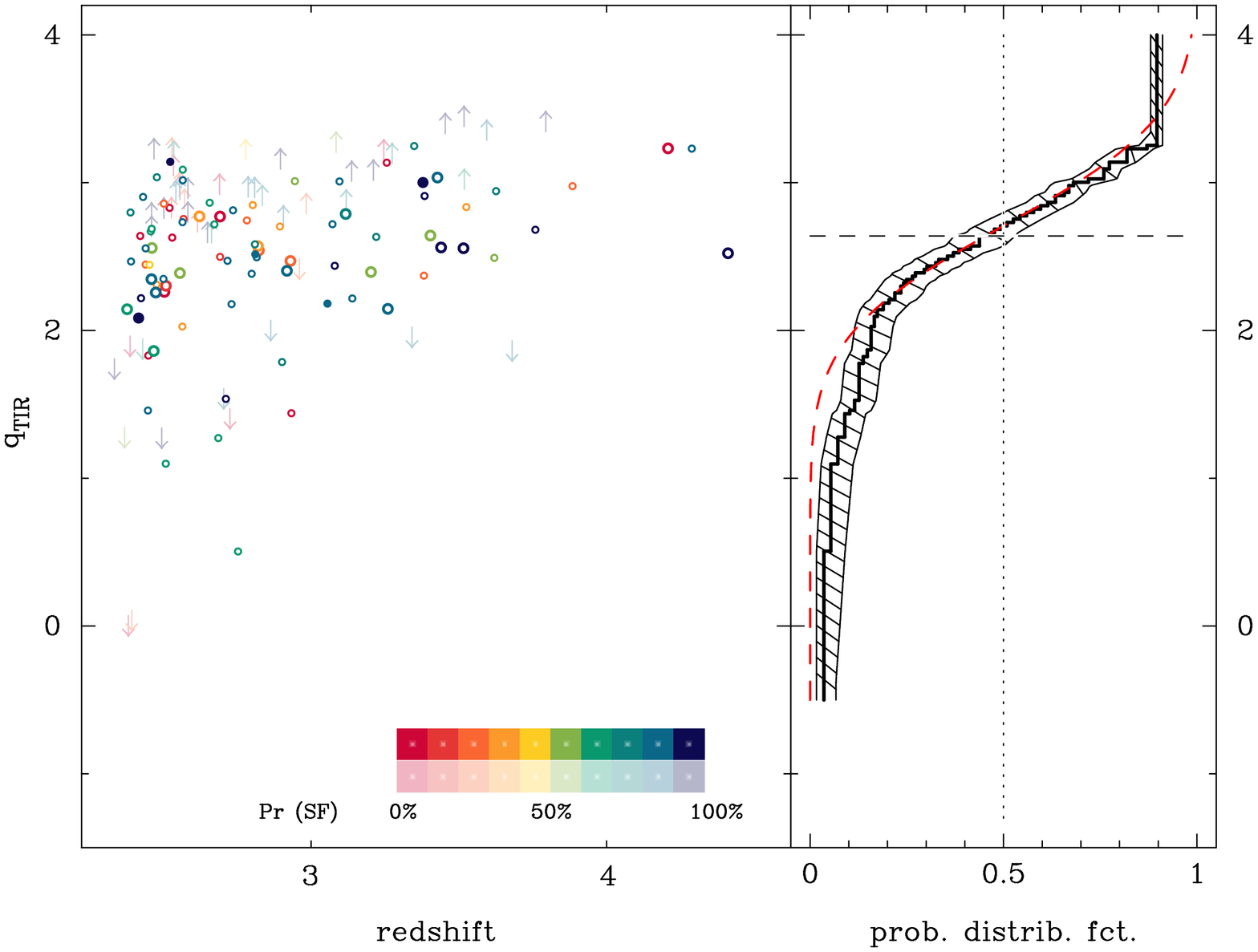} 
\caption{TIR/radio ratios of all IR- or radio-selected sources in the sample that lie at $z>$ 2.5 (sources are coloured according to their probability of `SF-hood'; see colour scale). On the right hand side of the plot the cumulative distribution function is plotted together with the best-fit Gaussian distribution (cf. Fig. \ref{fig:SFqTIR_medevo}). The dashed line marks the locally measured average TIR/radio ratio. \label{fig:SFqTIR_medevohighz}}
\end{figure}

\noindent The relative frequency of AGN at a given value of $q_{24\,[70]}$ (and in any of the considered redshift slices) is computed as
\begin{equation}
{\rm f_{AGN}}(q, z) = \frac{\sum_i\, {\rm Pr\,(AGN)}_i}{\sum_i\, {\rm Pr\,(SF)}_i}~,
\label{eq:agnfrac}
\end{equation}
where the summation is carried out over all objects $i$ in the given redshift bin that lie in the pertinent range of IR/radio ratios $q_{24\,[70]}$. Furthermore, we have by definition that Pr\,(AGN) = [1-Pr\,(SF)], as illustrated in Fig. \ref{fig:uK_SFprob}. Note that if the probabilities of SF- and AGN-`hood' are discretized (e.g. drawn from a set of integer probability values in per cent) the numerator and denominator in equation (\ref{eq:agnfrac}) could be rewritten as a summation over probabilities. In this case the denominator, for instance, takes the form:
\begin{equation}
\sum_i\, {\rm Pr\,(SF)}_i = \sum_{i=1}^{100}\,N_{i\%}\times i/100~.
\label{eq:altsum}
\end{equation}
Here $N_{i\%}$ stands for the number of objects with a probability of $i$ per cent of being a SFG. If a data set with only uncensored measurements is being studied the computation of ${\rm f_{AGN}}(q, z)$ is straightforward since the numbers $N_{i\%}$ can be directly determined by counting the suitable objects in each bin of $q$ and $z$. In the presence of censored data, however, $N_{i\%}$ should be regarded as an {\it effective} number of objects which accounts for the fact that ({\it i}) the true value of $q$ associated with a limit found in a specific bin, $j$, might lie outside that bin (if its actual value were determined with the help of better data), or, ({\it ii}) limits detected in neighbouring bins have a finite probability of ending up in the $j$-th bin (e.g. lower limits populating a bin with rank smaller than $j$ or upper limits in a bin centred on a larger value of $q$ than that of the $j$-th bin).\\
To compute this effective number of objects in the $j$-th bin, $N_{i\%,\, {\rm eff.}}(j)$, we derived the distribution function of objects with the desired probability of being SFGs following the techniques presented in Appendix \ref{appsect:doubcensetheory}. Due to the finite number of objects in each redshift bin it was necessary to work with a limited number of probability bins to ensure a well-behaved estimate of the distribution function $f_{i\%}(q)$. Since $f_{i\%}(q)$ specifies the fractional contribution of a bin in $q$ to the total number of measurements we have that
\begin{equation}
N_{i\%,\, {\rm eff.}}(j) = f_{j,\,i\%}\times N_{\rm tot, \,i\%}~,
\end{equation}
where $N_{\rm tot, \,i\%}$ is the total number of objects used to compute the distribution function. With these modifications the new expression for ${\rm f_{AGN}}(q, z)$ in the $j$-th bin (to be compared with equation (\ref{eq:agnfrac})) becomes:
\begin{equation}
{\rm f_{AGN}}(q, z) = \frac{\sum_{i=1}^{n}\,N_{i,\, {\rm eff.}}\times \langle {\rm 1-Pr\,(SF)}\rangle_i}{\sum_{i=1}^{n}\,N_{i,\, {\rm eff.}}\times \langle {\rm Pr\,(SF)}\rangle_i}~.
\end{equation}
Note that because of the coarse binning in probability space $i$ no longer defines a number in per cent in this case but a range of probabilities instead. Consequently, we also have introduced a new weighting factor, namely the median value of the probabilities in the grouped data, $\langle {\rm Pr\,(SF)}\rangle_i$. It replaces the probability of the individual objects, $\nicefrac{i}{100}$, used in equation (\ref{eq:altsum}). In practice the median usually does not differ much from the mean, i.e. it might take on a value of $\sim$0.16 for objects with probabilities in the range $0.1< {\rm Pr\,(SF)} < 0.2$.

\acknowledgments We gratefully acknowledges the contribution of the whole COSMOS collaboration and its more than 100 scientists around the globe. M.T.S thanks Ute Lisenfeld, Chris Carilli and George Helou for useful discussions, as well as Marco Scodeggio for comments on the manuscript.\\
The German DFG supported this research with grant SCHI 536/3-2. GZ acknowledges partial support from an INAF contract (PRIN-2007/1.06.10.08) and an ASI grant (ASI/COFIS/WP3110 I/026/07/0). This work is partly based on observations made with the {\it Spitzer Space Telescope}, which is operated by NASA/JPL/Caltech. The National Radio Astronomy Observatory is a facility of the National Science Foundation operated under cooperative agreement by Associated Universities, Inc.\\
 
\noindent {\it Facilities:} NRAO (VLA), Spitzer (IRAC, MIPS), ESO (VLT), Subaru (SuprimeCam)

\clearpage

\begin{deluxetable}{lccl}
\tabletypesize{\scriptsize}
\tablecaption{Summary of sample properties.\label{tab:zstats}}
\tablewidth{0pt}
\setlength{\tabcolsep}{0.04in}
\tablehead{
\colhead{category} &
\colhead{no. of src.} &
\colhead{\% of cat. src.} &
\colhead{remarks} }
\startdata
 {\bf radio-selected sample}& & &\\ [1ex]
\hline\\[-1.5ex]
all & 2901 & 100 & initial 1.4\,GHz sample\\
no redshift & 748 & 25.8 & includes 21 1.4\,GHz sources outside area covered by multi-$\lambda$ photometry\\
spec-z & 769 & 26.5 & \\
photo-z & 1232+114 & 46.4 & regular + AGN photo-zs\\
unreliable redshift & 38 & 1.3 & \\
& & & \\
useable & 2020 & 69.6 & no ambiguous counterparts, reliable redshift \& AGN/SF classification available\\
& & & \\
Pr\,(SF) $\geq$ 0.5 & 766 & 26.4 & optically selected SFGs with $(u-K) <$ 2.36 (38\% of `useable' sample)\\
Pr\,(SF) $<$ 0.5 & 1254 & 43.2 & optically selected AGN with $(u-K)\geq$ 2.36 (62\% of `useable' sample)\\ [1ex]
\hline\\[-1.5ex]
 {\bf IR-selected sample}& & &\\ [1ex]
\hline\\[-1.5ex]
all & 5301 & 100 & initial 24\,$\mu$m sample restricted to $S_{\nu}$(24\,$\mu$m)$\geq$ 0.3\,mJy\\
no redshift & 1003 & 18.9 & includes 304 24\,$\mu$m sources outside area covered by multi-$\lambda$ photometry\\
spec-z & 1254 & 23.7 & \\
photo-z & 2712 + 231 & 55.5 & regular + AGN photo-zs\\
unreliable redshift & 101 & 1.9 & \\
& & & \\
useable & 3259 & 61.5 & no ambiguous counterparts, reliable redshift \& AGN/SF classification available\\
& & & \\
Pr\,(SF) $\geq$ 0.5 & 1822 & 34.4 & optically selected SFGs with $(u-K) <$ 2.36 (56\% of `useable' sample)\\
Pr\,(SF) $<$ 0.5 & 1437 & 27.1 & optically selected AGN with $(u-K)\geq$ 2.36 (44\% of `useable' sample)\\ [1ex]
\hline\\[-1.5ex]
 \multicolumn{2}{l}{\bf jointly IR- and radio-selected sample} & &\\ [1ex]
\hline\\[-1.5ex]
all & 6863 & 100 &1560/3960 sources selected only at 1.4\,GHz/24\,$\mu$m; 1341 in both catalogs \\
no redshift & 1620 & 23.6 & includes 321 sources outside area covered by multi-$\lambda$ photometry\\
spec-z & 1486 & 21.7 & \\
photo-z & 3362 + 275 & 53.0 & regular + AGN photo-zs\\
unreliable redshift & 120 & 1.7 & \\
& & & \\
useable & 4454 & 64.9 & no ambiguous counterparts, reliable redshift \& AGN/SF classification available\\
& & & \\
Pr\,(SF) $\geq$ 0.5 & 2215 & 32.3 & optically selected SFGs with $(u-K) <$ 2.36 (49.7\% of `useable' sample)\\
Pr\,(SF) $<$ 0.5 & 2239 & 32.6 & optically selected AGN with $(u-K)\geq$ 2.36 (50.3\% of `useable' sample)\\ [1ex]
\enddata
\end{deluxetable}

\clearpage

\begin{landscape}
\begin{deluxetable}{lcrccc}
\tabletypesize{\scriptsize}
\tablecaption{Representative studies (ordered according to the investigated IR/radio parameter) on the spatially unresolved IR-radio correlation using IR and 1.4\,GHz data. \label{tab:prevwork}}
\tablewidth{0pt}
\setlength{\tabcolsep}{0.04in}
\tablehead{
\colhead{survey/field} &
\colhead{$z$} &
\colhead{reference} &
\colhead{selection band} &
\colhead{other flux limits} &
\colhead{$\langle q\rangle$}}
\startdata
HDF-N & -- & \cite{beswick08}\tablenotemark{$\dagger$} & IR; $S_{\nu}(24\,\mu{\rm m})>$ 0.08\,mJy & $\sigma_{\rm 1.4\,GHz}\sim$ 0.004\,mJy/bm & $q_{24}$ = 0.52-0.7 ($S_{\nu}(24\,\mu{\rm m})<$ 1\,mJy) \\ [1 ex]
SWIRE & -- & \cite{boyle07}\tablenotemark{$\dagger$} & IR; $S_{\nu}(24\,\mu{\rm m})>$ 0.1\,mJy & $\sigma_{\rm 1.4\,GHz}\sim$ 0.03\,mJy/bm & $q_{24} =$ 1.39 \\ [1 ex]
xFLS/SWIRE & -- & \cite{garn09a}\tablenotemark{$\dagger$}  & IR; $S_{\nu}(24\,\mu{\rm m})>$ 0.15\,mJy & -- & $q_{24}=$ 0.92-1.02 ($S_{\nu}(24\,\mu{\rm m})<$ 1\,mJy) \\ [1 ex]
\multirow{4}{*}{{\it IRAS}} & \multirow{4}{*}{local} & \multirow{4}{*}{\cite{rieke09}} & \multirow{4}{*}{IR; $S_{\nu}(60\,\mu{\rm m})>$ 2 Jy} & \multirow{4}{*}{--} & $q_{24}=1.22\pm0.24$ \\
& & & & & \qquad\qquad\qquad\qquad\qquad\qquad\qquad(log$(\nicefrac{L_{\rm TIR}}{L_{\sun}})\leq$ 11), \\
& & & & & $q_{24}=(-1.28\pm0.76)+(0.22\pm0.07)\cdot {\rm log}(\nicefrac{L_{\rm TIR}}{L_{\sun}})$ \\
& & & & & \qquad\qquad\qquad\qquad\qquad\qquad\qquad(log$(\nicefrac{L_{\rm TIR}}{L_{\sun}})>$ 11) \\ [1 ex]
Subaru--{\it XMM-New-} & \multirow{2}{*}{$z\lesssim$ 1} & \multirow{2}{*}{\cite{ibar08}} & \multirow{2}{*}{radio; $S_{\nu}({\rm 1.4\,GHz})>$ 0.035\,mJy} & \multirow{2}{*}{$S_{\nu}(24\,\mu{\rm m})>$ 0.2\,mJy} & \multirow{2}{*}{$q_{24}$=(0.94$\pm$0.01)-(0.01$\pm$0.01)$\cdot z$} \\
{\it ton} Deep Field\\ [1 ex]
\multirow{2}{*}{xFLS} & \multirow{2}{*}{$z\lesssim$ 2} & \multirow{2}{*}{\cite{appleton04}} & \multirow{2}{*}{radio; $S_{\nu}({\rm 1.4\,GHz})>$ 0.09\,mJy} & $S_{\nu}(24\,\mu{\rm m})>$ 0.5\,mJy, & $q_{24}$ = 0.94-1$\pm$0.25, \\
& & & & $S_{\nu}(70\,\mu{\rm m})>$ 30 mJy & $q_{70}$ = 2.15$\pm$0.16 \\ [1ex]
xFLS & $z<$ 1 & \cite{frayer06} & radio: $S_{\nu}({\rm 1.4\,GHz})>$ 0.09\,mJy & $S_{\nu}(70\,\mu{\rm m})>$ 15\,mJy & $q_{70}=2.10\pm0.16$ ($z\simeq$ 0.2)\\ [1 ex]
13$^{\rm H}${\it XMM-Newton}/& \multirow{2}{*}{$z<$ 3} & \multirow{2}{*}{\cite{seymour09}} & \multirow{2}{*}{radio: $S_{\nu}({\rm 1.4\,GHz})>$ 0.03\,mJy} & \multirow{2}{*}{$S_{\nu}(70\,\mu{\rm m})>$ 6 mJy} & \multirow{2}{*}{$q_{70}=2.14\pm0.10-0.75\pm0.32\cdot{\rm log}(1+z)$}\\
{\it Chandra} Deep Field \\ [1 ex]
{\it IRAS} & $z\lesssim$ 0.16 & \cite{yun01} & radio-matched w/ $S_{\nu}(60\,\mu{\rm m})>$ 2 Jy & -- & $q_{\rm FIR}$ = 2.34$\pm$0.26 \\ [1ex]
HDF-N & $z\lesssim$ 1.4 & \cite{garrett02} & radio; 5$\sigma$ WSRT srcs. & -- & $q_{\rm FIR}\simeq$ 2 \\ [1 ex]
var. survey fields & $z\in$ [1,\,3] & \cite{kovacs06} & sub-mm/radio & -- & $q_{\rm FIR}=$ 2.07$\pm$0.08 \\ [1 ex]
GOODS-N & $z\in$ [0.6,\,2.6] & \cite{murphy09b} & -- & $S_{\nu}(24\,\mu{\rm m})>$ 0.2\,mJy & $q_{\rm TIR}=$ 2.41$\pm$0.39 \\ [1 ex]
xFLS & $z\in$ [0.5,\,3] & \cite{sajina08} & IR; $S_{\nu}(24\,\mu{\rm m})>$ 0.9\,mJy & $S_{\nu}({\rm 1.4\,GHz})>$ 0.09\,mJy & $q_{\rm FIR}=$ 2.07/2.21 \\ [1 ex]
EGS/FIDEL & $z\sim$ 2 & \cite{younger09} & IR; $S_{\nu}(24\,\mu{\rm m})>$ 0.5\,mJy & -- & $q_{\rm FIR}=$ 2.23$\pm$0.19 \\ [1 ex]
\multirow{2}{*}{misc. literature} & \multirow{2}{*}{local} & \multirow{2}{*}{\cite{bell03}} & \multirow{2}{*}{FUV (\& IR)} & \multirow{2}{*}{--} & $q_{\rm FIR}$ = 2.36$\pm$0.26, \\
& & & & & $q_{\rm TIR}$ = 2.64$\pm$0.26 \\ [1 ex]
\enddata
\tablenotetext{$\dagger$}{Radio stacking of 24\,$\mu$m sources.}
\end{deluxetable}
\clearpage
\end{landscape}

\begin{deluxetable}{lc|c|cc}
\tabletypesize{\scriptsize}
\tablecaption{Median of $K-$corrected logarithmic 24\,$\mu$m/1.4\,GHz flux ratios, $q_{24,\,0}$, as a function of redshift (cf. also Fig. \ref{fig:SFcorrq24_medevo}). The average IR/radio properties are given for three different samples of SFGs: a radio-selected sample, an IR-selected sample and the combination of the two of these (from left to right). For the jointly selected sample we also give the scatter in the relation (right-most column). All errors are 2\,$\sigma$ uncertainties. \label{tab:q24info}}
\tablewidth{0pt}
\setlength{\tabcolsep}{0.04in}
\tablehead{
\colhead{redshift} &
\colhead{$\langle q_{24,\,0}\rangle$} &
\colhead{$\langle q_{24,\,0}\rangle$} &
\colhead{$\langle q_{24,\,0}\rangle$} &
\colhead{$\sigma_{q_{24}}$} }
\startdata
& {\bf radio-selected sample} & {\bf IR-selected sample} & \multicolumn{2}{c}{\bf jointly selected sample}\\
\\ [-2ex]
\hline \\ [-2ex]
0.08 $\leq z <$ 0.23 & 1.02$^{+0.08}_{-0.07}$ & 1.35$^{+0.08}_{-0.12}$ & 1.31$^{+0.10}_{-0.05}$ & 0.37$\pm$0.03\\ [1ex]
0.23 $\leq z <$ 0.33 & 0.88$^{+0.16}_{-0.10}$ & 1.28$^{+0.07}_{-0.07}$ & 1.26$^{+0.07}_{-0.06}$ & 0.25$\pm$0.03\\ [1ex]
0.33 $\leq z <$ 0.45 & 0.94$^{+0.11}_{-0.04}$ & 1.29$^{+0.04}_{-0.08}$ & 1.27$^{+0.08}_{-0.04}$ & 0.31$\pm$0.04\\ [1ex]
0.45 $\leq z <$ 0.67 & 0.97$^{+0.08}_{-0.14}$ & 1.26$^{+0.05}_{-0.14}$ & 1.26$^{+0.02}_{-0.08}$ & 0.35$\pm$0.03\\ [1ex]
0.67 $\leq z <$ 0.82 & 0.83$^{+0.18}_{-0.16}$ & 1.29$^{+0.06}_{-0.06}$ & 1.23$^{+0.06}_{-0.03}$ & 0.36$\pm$0.03\\ [1ex]
0.82 $\leq z <$ 0.94 & 0.78$^{+0.20}_{-0.17}$ & 1.32$^{+0.03}_{-0.09}$ & 1.31$^{+0.03}_{-0.06}$ & 0.35$\pm$0.03\\ [1ex]
0.94 $\leq z <$ 1.14 & 0.77$^{+0.18}_{-0.26}$ & 1.36$^{+0.07}_{-0.10}$ & 1.29$^{+0.07}_{-0.09}$ & 0.49$\pm$0.04\\ [1ex]
1.14 $\leq z <$ 1.51 & 0.75$^{+0.12}_{-0.15}$ & 1.47$^{+0.11}_{-0.07}$ & 1.16$^{+0.09}_{-0.10}$ & 0.91$\pm$0.06\\ [1ex]
1.51 $\leq z <$ 2.00 & 0.96$^{+0.09}_{-0.21}$ & 1.63$^{+0.12}_{-1.01}$ & 1.35$^{+0.10}_{-0.06}$ & 0.79$\pm$0.04\\ [1ex]
2.00 $\leq z <$ 4.50 & 1.06$^{+0.11}_{-0.28}$ & 1.93$^{+0.11}_{-0.21}$ & 1.62$^{+0.10}_{-0.12}$ & 0.88$\pm$0.06\\ [1ex]
\enddata
\end{deluxetable}

\begin{deluxetable}{lccc}
\tabletypesize{\scriptsize}
\tablecaption{Slope (column 1) and $y$-axis intercept (column 2) of the best-fitting linear trend to the evolution of the three investigated IR/radio parameters $q_{24,\,0}$, $q_{70,\,0}$ and $q_{\rm TIR}$. The linear fit was performed using the median logarithmic IR/radio ratios in all redshift slices at $z<$ 1.4. The third column gives the average scatter in the relation measured over this redshift range. The states errors are 1\,$\sigma$ uncertainties. \label{tab:medqevo}}
\tablewidth{0pt}
\setlength{\tabcolsep}{0.04in}
\tablehead{
\colhead{IR/radio parameter} &
\colhead{$d\langle q\rangle/dz |_{z<1.4}$} &
\colhead{$\langle q\rangle_{z=0}$} &
\colhead{$\langle \sigma_{q}\rangle |_{z<1.4}$}}
\startdata
$q_{24,\,0}$ & -0.015$\pm$0.136 & 1.275$\pm$0.098 & 0.417$\pm$0.038\\
$q_{70,\,0}$  & -0.123$\pm$0.135 & 2.314$\pm$0.091 & 0.392$\pm$0.035\\
$q_{\rm TIR}$  & -0.268$\pm$0.115 & 2.754$\pm$0.074 & 0.412$\pm$0.037\\
\enddata
\end{deluxetable}

\begin{deluxetable}{lc|c|cc}
\tabletypesize{\scriptsize}
\tablecaption{As for Table \ref{tab:q24info} but for the $K-$corrected logarithmic 70\,$\mu$m/1.4\,GHz flux ratio $q_{70,\,0}$. \label{tab:q70info}}
\tablewidth{0pt}
\setlength{\tabcolsep}{0.04in}
\tablehead{
\colhead{redshift} &
\colhead{$\langle q_{70,\,0}\rangle$} &
\colhead{$\langle q_{70,\,0}\rangle$} &
\colhead{$\langle q_{70,\,0}\rangle$} &
\colhead{$\sigma_{q_{70}}$} }
\startdata
& {\bf radio-selected sample} & {\bf IR-selected sample} & \multicolumn{2}{c}{\bf jointly selected sample}\\
\\ [-2ex]
\hline \\ [-2ex]
0.08 $\leq z <$ 0.23 & 2.07$^{+0.06}_{-0.09}$ & 2.41$^{+0.09}_{-0.11}$ & 2.39$^{+0.08}_{-0.08}$ & 0.39$\pm$0.04\\ [1ex]
0.23 $\leq z <$ 0.33 & 1.91$^{+0.16}_{-0.10}$ & 2.27$^{+0.03}_{-0.05}$ & 2.26$^{+0.06}_{-0.04}$ & 0.25$\pm$0.03\\ [1ex]
0.33 $\leq z <$ 0.45 & 1.99$^{+0.10}_{-0.08}$ & 2.30$^{+0.05}_{-0.08}$ & 2.28$^{+0.07}_{-0.06}$ & 0.30$\pm$0.04\\ [1ex]
0.45 $\leq z <$ 0.67 & 1.94$^{+0.10}_{-0.15}$ & 2.23$^{+0.03}_{-0.08}$ & 2.24$^{+0.02}_{-0.07}$ & 0.36$\pm$0.03\\ [1ex]
0.67 $\leq z <$ 0.82 & 1.77$^{+0.14}_{-0.13}$ & 2.23$^{+0.04}_{-0.08}$ & 2.20$^{+0.02}_{-0.05}$ & 0.39$\pm$0.03\\ [1ex]
0.82 $\leq z <$ 0.94 & 1.79$^{+0.18}_{-0.20}$ & 2.29$^{+0.07}_{-0.05}$ & 2.22$^{+0.06}_{-0.02}$ & 0.28$\pm$0.03\\ [1ex]
0.94 $\leq z <$ 1.14 & 1.67$^{+0.26}_{-0.15}$ & 2.32$^{+0.06}_{-0.10}$ & 2.25$^{+0.06}_{-0.10}$ & 0.46$\pm$0.04\\ [1ex]
1.14 $\leq z <$ 1.51 & 1.67$^{+0.12}_{-0.14}$ & 2.33$^{+0.08}_{-0.13}$ & 2.02$^{+0.14}_{-0.11}$ & 0.83$\pm$0.05\\ [1ex]
1.51 $\leq z <$ 2.00 & 1.78$^{+0.10}_{-0.22}$ & 2.45$^{+0.12}_{-1.66}$ & 2.17$^{+0.08}_{-0.06}$ & 0.74$\pm$0.04\\ [1ex]
2.00 $\leq z <$ 4.50 & 1.85$^{+0.10}_{-0.28}$ & 2.71$^{+0.07}_{-0.18}$ & 2.38$^{+0.11}_{-0.08}$ & 0.89$\pm$0.06\\ [1ex]
\enddata
\end{deluxetable}

\begin{deluxetable}{lc|c|cc}
\tabletypesize{\scriptsize}
\tablecaption{As for Tables \ref{tab:q24info} and \ref{tab:q70info} but for the logarithmic TIR/1.4\,GHz flux ratio $q_{\rm TIR}$. \label{tab:qTIRinfo}}
\tablewidth{0pt}
\setlength{\tabcolsep}{0.04in}
\tablehead{
\colhead{redshift} &
\colhead{$\langle q_{\rm TIR}\rangle$} &
\colhead{$\langle q_{\rm TIR}\rangle$} &
\colhead{$\langle q_{\rm TIR}\rangle$} &
\colhead{$\sigma_{q_{\rm TIR}}$} }
\startdata
& {\bf radio-selected sample} & {\bf IR-selected sample} & \multicolumn{2}{c}{\bf jointly selected sample}\\
\\ [-2ex]
\hline \\ [-2ex]
0.08 $\leq z <$ 0.23 & 2.41$^{+0.10}_{-0.12}$ & 2.75$^{+0.05}_{-0.11}$ & 2.76$^{+0.02}_{-0.08}$ & 0.39$\pm$0.03\\ [1ex]
0.23 $\leq z <$ 0.33 & 2.29$^{+0.23}_{-0.09}$ & 2.63$^{+0.02}_{-0.07}$ & 2.64$^{+0.05}_{-0.04}$ & 0.29$\pm$0.04\\ [1ex]
0.33 $\leq z <$ 0.45 & 2.37$^{+0.10}_{-0.16}$ & 2.71$^{+0.04}_{-0.07}$ & 2.70$^{+0.05}_{-0.06}$ & 0.34$\pm$0.04\\ [1ex]
0.45 $\leq z <$ 0.67 & 2.28$^{+0.09}_{-0.12}$ & 2.61$^{+0.07}_{-0.12}$ & 2.56$^{+0.05}_{-0.08}$ & 0.39$\pm$0.04\\ [1ex]
0.67 $\leq z <$ 0.82 & 2.09$^{+0.12}_{-0.14}$ & 2.54$^{+0.04}_{-0.08}$ & 2.51$^{+0.03}_{-0.05}$ & 0.39$\pm$0.03\\ [1ex]
0.82 $\leq z <$ 0.94 & 2.07$^{+0.20}_{-0.16}$ & 2.56$^{+0.04}_{-0.07}$ & 2.51$^{+0.06}_{-0.02}$ & 0.28$\pm$0.03\\ [1ex]
0.94 $\leq z <$ 1.14 & 1.98$^{+0.23}_{-0.16}$ & 2.61$^{+0.06}_{-0.08}$ & 2.56$^{+0.06}_{-0.12}$ & 0.48$\pm$0.05\\ [1ex]
1.14 $\leq z <$ 1.51 & 1.97$^{+0.10}_{-0.15}$ & 2.64$^{+0.09}_{-0.11}$ & 2.29$^{+0.16}_{-0.08}$ & 0.78$\pm$0.05\\ [1ex]
1.51 $\leq z <$ 2.00 & 2.10$^{+0.09}_{-0.22}$ & 2.77$^{+0.12}_{-0.26}$ & 2.49$^{+0.08}_{-0.06}$ & 0.75$\pm$0.04\\ [1ex]
2.00 $\leq z <$ 4.50 & 2.17$^{+0.10}_{-0.28}$ & 3.04$^{+0.08}_{-0.19}$ & 2.72$^{+0.10}_{-0.10}$ & 0.85$\pm$0.06\\ [1ex]
\enddata
\end{deluxetable}

\clearpage

\clearpage

\clearpage

\clearpage

\clearpage

\clearpage

\clearpage

\clearpage

\clearpage

\clearpage

\clearpage

\clearpage

\clearpage

\clearpage

\clearpage

\clearpage

\clearpage

\clearpage

\clearpage

\clearpage

\clearpage

\clearpage

\clearpage

\begin{thebibliography}{}
\bibitem[Appleton et al.(2004)]{appleton04} Appleton, P.~N., et al.\ 2004, \apjs, 154, 147
\bibitem[Avni et al.(1980)]{avni80} Avni, Y., Soltan, A., Tananbaum, H., \& Zamorani, G.\ 1980, \apj, 238, 800
\bibitem[Baldwin et al.(1981)]{baldwin81} Baldwin, J.~A., Phillips, M.~M., \& Terlevich, R.\ 1981, \pasp, 93, 5
\bibitem[Bally \& Thronson(1989)]{bally89} Bally, J., \& Thronson, H.~A., Jr.\ 1989, \aj, 97, 69
\bibitem[Barthel(2006)]{barthel06} Barthel, P.~D.\ 2006, \aap, 458, 107
\bibitem[Beck \& Golla(1988)]{beckgolla88} Beck, R., \& Golla, G.\ 1988, \aap, 191, L9
\bibitem[Beck(2005)]{beck05} Beck, R.\ 2005, In: The Magnetized Plasma in Galaxy Evolution, 193
\bibitem[Beelen et al.(2006)]{beelen06} Beelen, A., Cox, P., Benford, D.~J., Dowell, C.~D., Kov{\'a}cs, A., Bertoldi, F., Omont, A., \& Carilli, C.~L.\ 2006, \apj, 642, 694
\bibitem[Bell(2003)]{bell03} Bell, E.~F.\ 2003, \apj, 586, 794
\bibitem[Bertin \& Arnouts(1996)]{bertin96} Bertin, E., \& Arnouts, S.\ 1996, A\&AS, 117, 393
\bibitem[Beswick et al.(2008)]{beswick08} Beswick, R.~J., Muxlow, T.~W.~B., Thrall, H., Richards, A.~M.~S., \& Garrington, S.~T.\ 2008, \mnras, 385, 1143
\bibitem[Bettens et al.(1993)]{bettens93} Bettens, R.~P.~A., Brown, R.~D., Cragg, D.~M., Dickinson, C.~J., \& Godfrey, P.~D.\ 1993, \mnras, 263, 93
\bibitem[Bicay \& Helou(1990)]{bicayhelou90} Bicay, M.~D., \& Helou, G.\ 1990, \apj, 362, 59
\bibitem[Bondi et al.(2008)]{bondi08} Bondi, M., Ciliegi, P., Schinnerer, E., Smol{\v c}i{\'c}, V., Jahnke, K., Carilli, C., \& Zamorani, G.\ 2008, \apj, 681, 1129
\bibitem[Boyle et al.(2007)]{boyle07} Boyle, B.~J., Cornwell, T.~J., Middelberg, E., Norris, R.~P., Appleton, P.~N., \& Smail, I.\ 2007, \mnras, 376, 1182
\bibitem[Bressan et al.(2002)]{bressan02} Bressan, A., Silva, L., \& Granato, G.~L.\ 2002, \aap, 392, 377
\bibitem[Campbell(1981)]{campbell81} Campbell, G.\ 1981, Biometrika, 68, 417
\bibitem[Capak et al.(2007)]{capak07} Capak, P., et al.\ 2007, \apjs, 172, 99
\bibitem[Capak et al.(2008)]{capak08} Capak, P., et al.\ 2008, VizieR Online Data Catalog, 2284, 0
\bibitem[Cappelluti et al.(2007)]{cappelluti07} Cappelluti, N., et al.\ 2007, \apjs, 172, 341
\bibitem[Cappelluti et al.(2009)]{cappelluti09} Cappelluti, N., et al.\ 2009, \aap, 497, 635
\bibitem[Carilli \& Yun(1999)]{carilliyun99} Carilli, C.~L., \& Yun, M.~S.\ 1999, \apjl, 513, L13
\bibitem[Carilli et al.(2008)]{carilli08} Carilli, C.~L., et al.\ 2008, \apj, 689, 883
\bibitem[Chary \& Elbaz(2001)]{charyelbaz01} Chary, R., \& Elbaz, D.\ 2001, \apj, 556, 562
\bibitem[Chary et al.(2004)]{chary04} Chary, R., et al.\ 2004, \apjs, 154, 80
\bibitem[Colina \& Perez-Olea(1995)]{colina95} Colina, L., \& Perez-Olea, D.~E.\ 1995, \mnras, 277, 845
\bibitem[Condon et al.(1982)]{condon82} Condon, J.~J., Condon, M.~A., Gisler, G., \& Puschell, J.~J.\ 1982, \apj, 252, 102
\bibitem[Condon(1984)]{condon84} Condon, J.~J.\ 1984, \apj, 287, 461
\bibitem[Condon(1992)]{condon92} Condon, J.~J.\ 1992, \araa, 30, 575
\bibitem[Condon(1997)]{condon97} Condon, J.~J.\ 1997, \pasp, 109, 166
\bibitem[Condon et al.(1998)]{condon98} Condon, J.~J., Cotton, W.~D., Greisen, E.~W., Yin, Q.~F., Perley, R.~A., Taylor, G.~B., \& Broderick, J.~J.\ 1998, \aj, 115, 1693
\bibitem[Condon et al.(2003)]{condon03} Condon, J.~J., Cotton, W.~D., Yin, Q.~F., Shupe, D.~L., Storrie-Lombardi, L.~J., Helou, G., Soifer, B.~T., \& Werner, M.~W.\ 2003, \aj, 125, 2411
\bibitem[Dale \& Helou(2002)]{dalehelou02} Dale, D.~A., \& Helou, G.\ 2002, \apj, 576, 159
\bibitem[de Jong et al.(1985)]{dejong85} de Jong, T., Klein, U., Wielebinski, R., \& Wunderlich, E.\ 1985, \aap, 147, L6
\bibitem[Dickey \& Salpeter(1984)]{dickeysalpeter84} Dickey, J.~M., \& Salpeter, E.~E.\ 1984, \apj, 284, 461
\bibitem[Domingue et al.(2005)]{domingue05} Domingue, D.~L., Sulentic, J.~W., \& Durbala, A.\ 2005, \aj, 129, 2579
\bibitem[Donley et al.(2005)]{donley05} Donley, J.~L., Rieke, G.~H., Rigby, J.~R., \& P{\'e}rez-Gonz{\'a}lez, P.~G.\ 2005, \apj, 634, 169
\bibitem[Dunkley et al.(2009)]{dunkley09} Dunkley, J., et al.\ 2009, \apjs, 180, 306
\bibitem[Dunne et al.(2000)]{dunne00} Dunne, L., Clements, D.~L., \& Eales, S.~A.\ 2000, \mnras, 319, 813
\bibitem[Dunne et al.(2009)]{dunne09} Dunne, L., et al.\ 2009, \mnras, 394, 3
\bibitem[Elbaz et al.(2002)]{elbaz02} Elbaz, D., Cesarsky, C.~J., Chanial, P., Aussel, H., Franceschini, A., Fadda, D., \& Chary, R.~R.\ 2002, \aap, 384, 848
\bibitem[Elvis et al.(2009)]{elvis09} Elvis, M., et al.\ 2009, \apjs, 184, 158
\bibitem[Engelbracht et al.(2007)]{engelbracht07} Engelbracht, C.~W., et al.\ 2007, \pasp, 119, 994
\bibitem[Feigelson \& Nelson(1985)]{feigelson85} Feigelson, E.~D., \& Nelson, P.~I.\ 1985, \apj, 293, 192
\bibitem[Feldmann et al.(2006)]{feldmann06} Feldmann, R., et al.\ 2006, \mnras, 372, 565
\bibitem[Fomalont et al.(2006)]{fomalont06} Fomalont, E.~B., Kellermann, K.~I., Cowie, L.~L., Capak, P., Barger, A.~J., Partridge, R.~B., Windhorst, R.~A., \& Richards, E.~A.\ 2006, \apjs, 167, 103
\bibitem[Francis(1993)]{francis93} Francis, P.~J.\ 1993, \apj, 407, 519
\bibitem[Frayer et al.(2006)]{frayer06} Frayer, D.~T., et al.\ 2006, \aj, 131, 250
\bibitem[Frayer et al.(2009)]{frayer09} Frayer, D.~T., et al.\ 2009, arXiv:0902.3273
\bibitem[Garn \& Alexander(2009)]{garn09a} Garn, T., Green, D.~A., Riley, J.~M., \& Alexander, P.\ 2009, arXiv:0905.1218
\bibitem[Garn et al.(2009)]{garn09b} Garn, T., \& Alexander, P.\ 2009, \mnras, 394, 105
\bibitem[Garrett(2002)]{garrett02} Garrett, M.~A.\ 2002, \aap, 384, L19
\bibitem[Gordon et al.(2007)]{gordon07} Gordon, K.~D., et al.\ 2007, \pasp, 119, 1019
\bibitem[Glazebrook \& Economou(1997)]{glazebrook97} Glazebrook, K. \& Economou, F.\ 1997, The Perl Journal, 5, 5
\bibitem[Gonzalez et al.(2009)]{gonzalez09} Gonzalez, A.~H., Clowe, D., Brada{\v c}, M., Zaritsky, D., Jones, C., \& Markevitch, M.\ 2009, \apj, 691, 525
\bibitem[Greisen(2003)]{greisen03} Greisen, E.~W.\ 2003, Astrophysics and Space Science Library, 285, 109
\bibitem[Groves et al.(2003)]{groves03} Groves, B.~A., Cho, J., Dopita, M., \& Lazarian, A.\ 2003, Publications of the Astronomical Society of Australia, 20, 252
\bibitem[Haarsma \& Partridge(1998)]{haarsmapartridge98} Haarsma, D.~B., \& Partridge, R.~B.\ 1998, \apjl, 503, L5
\bibitem[Haarsma et al.(2000)]{haarsma00} Haarsma, D.~B., Partridge, R.~B., Windhorst, R.~A., \& Richards, E.~A.\ 2000, \apj, 544, 641
\bibitem[Harwit \& Pacini(1975)]{harwitpacini75} Harwit, M., \& Pacini, F.\ 1975, \apjl, 200, L127
\bibitem[Hasinger et al.(2007)]{hasinger07} Hasinger, G., et al.\ 2007, \apjs, 172, 29
\bibitem[Helou et al.(1985)]{helou85} Helou, G., Soifer, B.~T., \& Rowan-Robinson, M.\ 1985, \apjl, 298, L7
\bibitem[Helou(1991)]{helou91} Helou, G.\ 1991, The Interpretation of Modern Synthesis Observations of Spiral Galaxies, 18, 125
\bibitem[Helou \& Bicay(1993)]{helou93} Helou, G., \& Bicay, M.~D.\ 1993, \apj, 415, 93
\bibitem[Hummel et al.(1988)]{hummel88} Hummel, E., Davies, R.~D., Pedlar, A., Wolstencroft, R.~D., \& van der Hulst, J.~M.\ 1988, \aap, 199, 91
\bibitem[Ibar et al.(2008)]{ibar08} Ibar, E., et al.\ 2008, \mnras, 386, 953
\bibitem[Ilbert et al.(2009a)]{ilbert09a} Ilbert, O., et al.\ 2009, \apj, 690, 1236
\bibitem[Ilbert et al.(2009b)]{ilbert09b} Ilbert, O., et al.\ 2009, arXiv:0903.0102
\bibitem[Isobe et al.(1990)]{isobe90} Isobe, T., Feigelson, E.~D., Akritas, M.~G., \& Babu, G.~J.\ 1990, \apj, 364, 104
\bibitem[Jarrett et al.(2000)]{jarrett00} Jarrett, T.~H., Chester, T., Cutri, R., Schneider, S., Skrutskie, M., \& Huchra, J.~P.\ 2000, \aj, 119, 2498
\bibitem[Kaplan \& Meier(1958)]{kaplanmeier58} Kaplan, E.~L. \& Meier, P.\ 1958, J. Am. Statist. Assoc., 53, 457
\bibitem[Kauffmann et al.(2003)]{kauffmann03} Kauffmann, G., et al.\ 2003, \mnras, 346, 1055
\bibitem[Kellermann(1964)]{kellermann64} Kellermann, K.~I.\ 1964, \apj, 140, 969
\bibitem[Kennicutt(1998)]{kennicutt98} Kennicutt, R.~C., Jr.\ 1998, \araa, 36, 189
\bibitem[Kewley et al.(2006)]{kewley06} Kewley, L.~J., Groves, B., Kauffmann, G., \& Heckman, T.\ 2006, \mnras, 372, 961
\bibitem[Koekemoer et al.(2007)]{koekemoer07} Koekemoer, A.~M., et  al.\ 2007, \apjs, 172, 196
\bibitem[Kov{\'a}cs et al.(2006)]{kovacs06} Kov{\'a}cs, A., Chapman, S.~C., Dowell, C.~D., Blain, A.~W., Ivison, R.~J., Smail, I., \& Phillips, T.~G.\ 2006, \apj, 650, 592
\bibitem[Krist(2002)]{krist02} Krist, J.\ 2002, Tiny Tim/{\it SIRTF} User's Guide (Pasadena: SSC)
\bibitem[Lagache et al.(2003)]{lagache03} Lagache, G., Dole, H., \& Puget, J.-L.\ 2003, \mnras, 338, 555
\bibitem[LeFloc'h et al.(2009)]{lefloch09} LeFloc'h, E., et al.\ 2009, \apj, 703, 222
\bibitem[Lauer et al.(2007)]{lauer07} Lauer, T.~R., Tremaine, S., Richstone, D., \& Faber, S.~M.\ 2007, \apj, 670, 249
\bibitem[Lilly et al.(2007)]{lilly07} Lilly, S.~J., et al.\ 2007, \apjs, 172, 70
\bibitem[Lilly et al.(2009)]{lilly09} Lilly, S.~J., et al.\ 2009, \apjs, 184, 218
\bibitem[Lisenfeld et al.(1996)]{lisenfeld96} Lisenfeld, U., V\"olk, H.~J., \& Xu, C.\ 1996, \aap, 306, 677
\bibitem[Magnelli et al.(2009)]{magnelli09} Magnelli, B., Elbaz, D., Chary, R.~R., Dickinson, M., Le Borgne, D., Frayer, D.~T., \& Willmer, C.~N.~A.\ 2009, \aap, 496, 57
\bibitem[Makovoz \& Khan(2005)]{makovoz05a} Makovoz, D., \& Khan, I.\ 2005, Astronomical Data Analysis Software and Systems XIV, 347, 81
\bibitem[Makovoz \& Marleau(2005)]{makovoz05b} Makovoz, D., \& Marleau, F.~R.\ 2005, \pasp, 117, 1113
\bibitem[Marx et al.(1994)]{marx94} Marx, M., Kruegel, E., Klein, U., \& Wielebinski, R.\ 1994, \aap, 281, 718
\bibitem[Mauch \& Sadler(2007)]{mauchsadler07} Mauch, T., \& Sadler, E.~M.\ 2007, \mnras, 375, 931
\bibitem[Miller \& Owen(2001)]{millerowen01} Miller, N.~A., \& Owen, F.~N.\ 2001, \aj, 121, 1903
\bibitem[Murgia et al.(2005)]{murgia05} Murgia, M., Helfer, T.~T., Ekers, R., Blitz, L., Moscadelli, L., Wong, T., \& Paladino, R.\ 2005, \aap, 437, 389
\bibitem[Murphy et al.(2008)]{murphy08} Murphy, E.~J., Helou, G., Kenney, J.~D.~P., Armus, L., \& Braun, R.\ 2008, \apj, 678, 828
\bibitem[Murphy et al.(2009a)]{murphy09a} Murphy, E.~J., Kenney, J.~D.~P., Helou, G., Chung, A., \& Howell, J.~H.\ 2009, \apj, 694, 1435
\bibitem[Murphy et al.(2009b)]{murphy09b} Murphy, E.~J., Chary, R.-R., Alexander, D.~M., Dickinson, M., Magnelli, B., Morrison, G., Pope, A., \& Teplitz, H.~I.\ 2009, \apj, 698, 1380
\bibitem[Niklas \& Beck(1997)]{niklasbeck97} Niklas, S., \& Beck, R.\ 1997, \aap, 320, 54
\bibitem[Obri{\'c} et al.(2006)]{obric06} Obri{\'c}, M., et al.\ 2006, \mnras, 370, 1677
\bibitem[Odell et al.(2002)]{odell02} Odell, A.~P., Schombert, J., \& Rakos, K.\ 2002, \aj, 124, 3061
\bibitem[Oke(1974)]{oke74} Oke, J.~B.\ 1974, \apjs, 27, 21
\bibitem[Papovich et al.(2004)]{papovich04} Papovich, C., et al.\ 2004, \apjs, 154, 70
\bibitem[Park et al.(2008)]{park08} Park, S.~Q., et al.\ 2008, \apj, 678, 744
\bibitem[Pierini et al.(2003)]{pierini03} Pierini, D., Popescu, C.~C., Tuffs, R.~J., V\"olk, H.~J.\ 2003, \aap, 409, 907
\bibitem[Pohl(1994)]{pohl94} Pohl, M.\ 1994, \aap, 287, 453
\bibitem[Prentice \& Marek(1979)]{prenticemarek79} Prentice, R.~L., \& Marek, P.\ 1979, Biometrics, 35, 861
\bibitem[Price \& Duric(1992)]{priceduric92} Price, R., \& Duric, N.\ 1992, \apj, 401, 81
\bibitem[Reddy \& Yun(2004)]{reddy04} Reddy, N.~A., \& Yun, M.~S.\ 2004, \apj, 600, 695
\bibitem[Richards(2000)]{richards00} Richards, E.~A.\ 2000, \apj, 533, 611
\bibitem[Richards et al.(2006)]{richards06} Richards, G.~T., et al.\ 2006, \aj, 131, 2766
\bibitem[Rieke et al.(2009)]{rieke09} Rieke, G.~H., Alonso-Herrero, A., Weiner, B.~J., P{\'e}rez-Gonz{\'a}lez, P.~G., Blaylock, M., Donley, J.~L., \& Marcillac, D.\ 2009, \apj, 692, 556 
\bibitem[Rigby et al.(2008)]{rigby08} Rigby, J.~R., et al.\ 2008, \apj, 675, 262
\bibitem[Roy et al.(1998)]{roy98} Roy, A.~L., Norris, R.~P., Kesteven, M.~J., Troup, E.~R., \& Reynolds, J.~E.\ 1998, \mnras, 301, 1019
\bibitem[Sajina et al.(2008)]{sajina08} Sajina, A., et al.\ 2008, \apj, 683, 659
\bibitem[Salvato et al.(2009)]{salvato09} Salvato, M., et al.\ 2009, \apj, 690, 1250
\bibitem[Sanders et al.(1989)]{sanders89} Sanders, D.~B., Phinney, E.~S., Neugebauer, G., Soifer, B.~T., \& Matthews, K.\ 1989, \apj, 347, 29
\bibitem[Sanders \& Mirabel(1996)]{sandersmirabel96} Sanders, D.~B., \& Mirabel, I.~F.\ 1996, \araa, 34, 749
\bibitem[Sanders et al.(2007)]{sanders07} Sanders, D.~B., et al.\ 2007, \apjs, 172, 86
\bibitem[Schinnerer et al.(2004)]{vlacos1} Schinnerer, E., et al.\ 2004, \aj, 128, 1974
\bibitem[Schinnerer et al.(2007)]{vlacos2} Schinnerer, E., et al.\ 2007, \apjs, 172, 46
\bibitem[Schmitt(1985)]{schmitt85} Schmitt, J.~H.~M.~M.\ 1985, \apj, 293, 178
\bibitem[Schmitt et al.(1993)]{schmitt93} Schmitt, J.~H.~M.~M., Kahabka, P., Stauffer, J., \& Piters, A.~J.~M.\ 1993, \aap, 277, 114
\bibitem[Scoville et al.(2007)]{scoville07} Scoville, N., et al.\ 2007, \apjs, 172, 1
\bibitem[Seymour et al.(2008)]{seymour08} Seymour, N., et al.\ 2008, \mnras, 386, 1695
\bibitem[Seymour et al.(2009)]{seymour09} Seymour, N., Huynh, M., Dwelly, T., Symeonidis, M., Hopkins, A., McHardy, I.~M., Page, M.~J., \& Rieke, G.\ 2009, \mnras, 1077
\bibitem[Siana et al.(2008)]{siana08} Siana, B., Teplitz, H.~I., Chary, R.-R., Colbert, J., \& Frayer, D.~T.\ 2008, \apj, 689, 59
\bibitem[Silverman et al.(2009)]{silverman09} Silverman, J.~D., et al.\ 2009, \apj, 696, 396
\bibitem[Smol{\v c}i{\'c} et al.(2006)]{smolcic06} Smol{\v c}i{\'c}, V., et al.\ 2006, \mnras, 371, 121
\bibitem[Smol{\v c}i{\'c} et al.(2008)]{smolcic08} Smol{\v c}i{\'c}, V., et al.\ 2008, \apjs, 177, 14
\bibitem[Smol{\v c}i{\'c} et al.(2009)]{smolcic09a} Smol{\v c}i{\'c}, V., et al.\ 2009, \apj, 690, 610
\bibitem[Sopp \& Alexander(1991)]{sopp91} Sopp, H.~M., \& Alexander, P.\ 1991, \mnras, 251, 14P
\bibitem[Stetson(1987)]{stetson87} Stetson, P.~B.\ 1987, \pasp, 99, 191
\bibitem[Symeonidis et al.(2008)]{symeonidis08} Symeonidis, M., Willner, S.~P., Rigopoulou, D., Huang, J.-S., Fazio, G.~G., \& Jarvis, M.~J.\ 2008, \mnras, 385, 1015
\bibitem[Taniguchi et al.(2007)]{taniguchi07} Taniguchi, Y., et al.\ 2007, \apjs, 172, 9
\bibitem[Thompson et al.(2006)]{thompson06} Thompson, T.~A., Quataert, E., Waxman, E., Murray, N., \& Martin, C.~L.\ 2006, \apj, 645, 186
\bibitem[Thompson et al.(2007)]{thompson07} Thompson, T.~A., Quataert, E., \& Waxman, E.\ 2007, \apj, 654, 219
\bibitem[Trump et al.(2007)]{trump07} Trump, J.~R., et al.\  2007, \apjs, 172, 383
\bibitem[Trump et al.(2009)]{trump09} Trump, J.~R., et al.\ 2009, \apj, 696, 1195
\bibitem[Turnbull(1974)]{turnbull74} Turnbull, B.~W.\ 1974, J. Am. Statist. Assoc., 69, 169
\bibitem[van der Kruit(1973)]{vanderkruit73} van der Kruit, P.~C.\ 1973, \aap, 29, 263
\bibitem[Vlahakis et al.(2007)]{vlahakis07} Vlahakis, C., Eales, S., \& Dunne, L.\ 2007, \mnras, 379, 1042
\bibitem[V\"olk(1989)]{voelk89} V\"olk, H.~J.\ 1989, \aap, 218, 67
\bibitem[White et al.(1997)]{white97} White, R.~L., Becker, R.~H., Helfand, D.~J., \& Gregg, M.~D.\ 1997, \apj, 475, 479
\bibitem[Wolf et al.(2003)]{wolf03} Wolf, C., Wisotzki, L., Borch, A., Dye, S., Kleinheinrich, M., \& Meisenheimer, K.\ 2003, \aap, 408, 499
\bibitem[Wrobel \& Heeschen(1988)]{wrobel88} Wrobel, J.~M., \& Heeschen, D.~S.\ 1988, \apj, 335, 677
\bibitem[Wunderlich et al.(1987)]{wunderlich87} Wunderlich, E., Wielebinski, R., \& Klein, U.\ 1987, \aaps, 69, 487
\bibitem[York et al.(2000)]{york00} York, D.~G., et al.\ 2000, \aj, 120, 1579
\bibitem[Younger et al.(2009)]{younger09} Younger, J.~D., et al.\ 2009, \mnras, 394, 1685
\bibitem[Yun et al.(2001)]{yun01} Yun, M.~S., Reddy, N.~A., \& Condon, J.~J.\ 2001, \apj, 554, 80
\bibitem[Zhu \& Sun(2007)]{zhusun07} Zhu, C., \& Sun, J.\ 2007, In: Advances in statistical methods for the health sciences, 225
\end{thebibliography}
\end{document}